\documentclass[twocolumn,aps,prb,floats]{revtex4} 
\usepackage{amssymb,amsbsy,amsmath,mathrsfs}
\usepackage{epsfig,psfrag,subfigure}
\usepackage{graphicx,float}
\usepackage{times}
\usepackage{color}
\usepackage{subfigure}
\usepackage{setspace}
\usepackage{bm} % bold math.
\usepackage{xr}
\usepackage{stackengine}
\usepackage{natbib,url,hyperref}
\hypersetup{colorlinks,allcolors=blue}
\usepackage{filecontents}
\usepackage{comment}
\usepackage{hyperref}
\usepackage{tikz}
\newcommand\bea{\begin{eqnarray}}
\newcommand\eea{\end{eqnarray}}
\newcommand\beq{\begin{equation}}
\newcommand\eeq{\end{equation}}
\newcommand\bib{\bibitem}

\newcommand{\noi}{\noindent}
\newcommand{\non}{\nonumber}
\newcommand{\al}{\alpha}

\newcommand{\ga}{\gamma}

\newcommand{\la}{\langle}
\newcommand{\ra}{\rangle}

\newcommand{\bra}[1]{\langle #1|}
\newcommand{\ket}[1]{|#1\rangle}

\begin{document}
\title{Aspects of Hilbert space fragmentation in the quantum East model: 
fragmentation, subspace-restricted quantum scars, 
and effects of density-density interactions}
%Absence of thermalization in a correlated-hopping model 
%due to strong Hilbert space fragmentation and its characterization by irreducible strings \\

\author{Maitri Ganguli$^1\footnote{These authors contributed equally to this 
work.}$, Sreemayee Aditya$^{2*}$, and Diptiman Sen$^2$}
\affiliation{$^1$Department of Physics, Indian Institute of Science, Bengaluru
560012, India \\
$^2$Center for High Energy Physics, Indian Institute of Science, Bengaluru 560012, India}

\begin{abstract}
We investigate a one-dimensional correlated-hopping model of spinless fermions with an East constraint.
We first analytically unravel the complete fragmentation structure of this model by labeling each fragment by a unique root configuration and utilizing the transfer matrix method. We show that the growth of the
size of each fragment of the model follows the widely studied Dyck sequence, and
is therefore analytically tractable with the help of Catalan triangles. 
While the eigenstate thermalization hypothesis (ETH) does not hold within the full Hilbert space which exhibits Poisson statistics of the 
energy level spacing, an examination of various quantities restricted to the largest fragments shows that a weaker version of the subspace-restricted thermalization holds. This weaker violation of the ETH within the largest fragments is supported by the presence of subspace-restricted quantum many-body scars due to quantum fragmentation. Our analysis further 
reveals that the filling fraction at which the model has the largest fragment shifts with increasing system sizes from half-filling due to the distinct fragmentation structure. %In order to stabilize the ground state at a particular filling fraction, we examine the effect of perturbations, namely, a nearest-neighbor density-density interaction with strength $V$. 
Next, we show that the inclusion of a nearest-neighbor density-density interaction with strength $V$ induces a spectral transition within the largest fragment from a weakly ETH-violating phase containing scars to a  statistical bubble localized 
phase as $V$ increases. In particular, the $V\to \infty$ limit produces an integrable model. We find that the addition of finite-$V$ stabilizes the ground state near half-filling while keeping intact the fragmentation structure of the East model. However,
this behavior abruptly changes exactly at $V = \infty$ due to the emergence of a distinct fragmentation structure; we analytically explore this in a comprehensive manner.
The infinite-$V$ model has many interesting properties, among which the appearance of the ground state and the largest fragment at two different filling fractions is specially noteworthy. Finally, we propose an experimental setup to realize the infinite-$V$ model as a particular
limit of a special kind of $t-V$ model with an on-site potential.

\end{abstract}

\maketitle

\section{Introduction}

Quantum many-body systems far from equilibrium have been extensively studied for many years across a variety of platforms, ranging from optical lattices through 
Rydberg chains with tunable interactions to superconducting processors~\cite{exp1,exp2,exp3,exp4,exp5}. In this direction, the thermalization of 
an isolated many-body quantum system has been comprehensively explored over the past decade. The notion of thermalization in such systems has been formally comprehended 
with the help of the eigenstate thermalization hypothesis (ETH)~\cite{ETH1,ETH2,ETH3,ETH4},
which states that each eigenstate of a generic many-body strongly interacting quantum system implicitly carries the information of a thermal ensemble. As a 
consequence, these eigenstates self-thermalize on their own under time evolution.

The quest for finding novel forms of non-equilibrium universality has also shed light on various ergodicity-breaking mechanisms in many-body quantum 
systems~\cite{rev1,rev2,rev3,rev4}, which go beyond the ETH-satisfying family of models. The key quantity that plays a pivotal role in pinpointing the 
characteristics of such out-of-equilibrium many-body phases is quantum entanglement, which is also a central quantity from the viewpoint of quantum information 
processing~\cite{preskill}. These investigations thus lie at the frontier between quantum many-body physics and quantum information science, which is an active field 
of research in recent times due to the experimental advancements in diverse platforms~\cite{exp1,exp2,exp3,exp4,exp5}.

It is known that models with kinetic constraints can give rise to several intriguing phenomena where the central mechanisms for the breakdown of quantum thermalization can range from
quantum many-body scars ~\cite{rev1,QMBS1, QMBS2,QMBS5,QMBS6} to Hilbert space fragmentation \cite{rev1,HSF1_moudgalya,HSF2_moudgalya_2020,HSF3_sala_2020,HSF4_khemani,moudgalya_commu_2022,moudgalya_2021,brighi_2023,zadnik_2021,pozsgay_2023}. These quantum models 
also have received attention since considerable effort has been devoted in the past to investigating their classical counterparts, e.g., classical glasses 
undergoing a Markov evolution with kinetic constraints~\cite{menon_1995,menon_1997,barma_1994}. 

The specific mechanism on which we will focus in this paper is Hilbert space fragmentation (HSF)~\cite{rev1}, namely, a fracturing of the full Hilbert space into a number of fragments which grows exponentially with the system size; these different fragments may be 
integrable or non-integrable. Another fascinating feature of these 
fragmented systems is the presence of frozen eigenstates and blockades, which significantly impact the static and dynamical measures of thermalization in such 
systems. The frozen states are zero-energy eigenstates of the Hamiltonian, which do not participate in the dynamics. On the other hand, the presence of blockades 
forbids the dynamically active regions from interacting with one another.
%, thus yielding rich attributes in these fragmented systems. 
HSF can further be organized into two classes based on the growth of the largest fragment compared to the full Hilbert space size. Strong 
fragmentation~\cite{rev1,moudgalya_2021,aditya_2024} refers to the case when the sizes of the largest fragments are exponentially smaller than the size of the full 
Hilbert space in the thermodynamic limit, whereas in weakly fragmented systems,
the sizes of the largest fragments are smaller than the size of the full Hilbert
space by only a power of the system size~\cite{brighi_2023,weak1,weak2}. There is an alternative classification possible in terms of the basis states, which defines the fragmentation structure of the full Hilbert space. Classical 
fragmentation ~\cite{rev1,moudgalya_2021,moudgalya_commu_2022} occurs when the fragmentation 
structure is defined in terms of the product of local basis states. When the
fragmentation structure is defined in terms of entangled basis states, it is dubbed as quantum 
fragmentation~\cite{moudgalya_commu_2022,brighi_2023}. The different mechanisms can severely impact various measures of thermalization in a profound manner, yielding 
subspace- or Krylov-restricted thermalization ~\cite{rev1,moudgalya_2021,HSF1_moudgalya,HSF2_moudgalya_2020} to slow subdiffusive 
transport~\cite{trans1,trans2} in such systems. In addition, there have been 
studies that investigate the prethermal HSF in driven systems which have no 
static counterparts~\cite{dynloc,ghosh_2023}.

In this paper, we will investigate a one-dimensional model of spinless fermions in which the Hamiltonian has terms involving three consecutive sites on a lattice. 
The model allows nearest-neighbor hoppings 
between sites $j$ and $j+1$ only when the left neighboring site $j-1$ is occupied; such a constraint is called the East constraint~\cite{brighi_2023,weak2}. 
This model thus conserves the total particle number, but it breaks inversion symmetry. Furthermore, it is the most simple member of a recently introduced and
a more general family of particle-conserving quantum East models~\cite{brighi_2023}. 
Our model is also known to exhibit a freezing transition~\cite{weak1,weak2,freezing_2024} as the filling is varied. The
transition occurs at half-filling; there is strong fragmentation below 
half-filling and weak fragmentation above half-filling~\cite{weak2}. Although the dimensions of the largest fragment at half-filling and above half-filling have been previously 
investigated while analyzing the freezing transition~\cite{weak2}, the complete
fragmentation structure and other effects of fragmentation have not been exhaustively explored for this model; this motivates our present study. We 
note that a complete analytical 
understanding of a fragmented Hilbert space is often challenging since the fragments cannot be labeled by the quantum numbers of local or quasilocal symmetry 
operators. The significant steps in this direction include constructs like statistically localized integral of motions~\cite{stat_loc_2020}, commutant 
algebras~\cite{moudgalya_commu_2022}, and irreducible 
strings~\cite{menon_1997,barma_1994,menon_1995,aditya_2024,irr1,irr2}, and 
matrix-product operators ~\cite{pozsgay_2023}, which successfully capture the fragmentation structures of several kinetically constrained models.

To unravel the complete fragmentation structure of the Hilbert space of this
model, we first present a canonical representation of the root state by imposing some specific 
rules which uniquely label each fragment. This kind of procedure has been previously presented in the literature while studying the fragmentation 
structure of the Fredkin chain~\cite{fredkin1,fredkin2,fredkin3}. The identification of root states enables us to count the total number of fragments and frozen states using the 
transfer matrix method~\cite{gen,aditya_2024,menon_1997,menon_1995}. This further allows us to compute the growth of the Hilbert spaces with the system size for
each of the fragments, followed by a mapping of the transitions allowed by the East constraint to the widely-studied combinatorics sequence of Dyck words. The growth of 
each fragment can thus be exactly analytically computed with the help of the total number of Dyck sequences appearing in the Catalan triangle of order 
$l=1$~\cite{catalan}. We have verified that our numerically obtained results are in complete agreement with our analytical findings.

A numerical study of short-range spectral correlations of the energy eigenvalues 
in this model indicates that although the ETH is not satisfied within the full Hilbert which shows Poisson statistics, the largest fragments for all filling fractions follow the Krylov-restricted ETH~\cite{rev1,moudgalya_2021,HSF2_moudgalya_2020} in a weaker 
sense with the Gaussian orthogonal ensemble (GOE) statistics~\cite{huse,atas,BGS}. The weaker violation of the ETH 
is confirmed by the presence of subspace-restricted many-body scars~\cite{HSF2_moudgalya_2020,brighi_2023}, which are the consequence of the 
recursive quantum HSF~\cite{brighi_2023} within the classical fragments. Moreover, various static and dynamical measures of thermalization within the scar subspace 
reflect the signature of the ETH violation, which includes zero entanglement entropy of such eigenstates for certain entanglement cuts, the long-time revivals in the 
Loschmidt echo dynamics, and sudden jumps in the overlap amplitudes for the scar states which do not behave as smooth functions of the energy~\cite{rev1,rev2}. We also observe that the filling 
at which the model has the largest fragment shifts from half-filling ($L/2$) as $\sqrt{L}/2$, where $L$ is the system size; this can be readily confirmed using the asymptotic properties of 
the Catalan numbers~\cite{catalan}. This shift also impacts the ground state properties since the ground state in this system lies within the 
largest fragment. Accordingly, we observe that the filling where the ground state appears also moves with increasing system sizes. In particular, the ground state never occurs at half-filling; this is a combined outcome of the fragmentation structure and the lack of inversion and particle-hole symmetry in this model.

\begin{figure*}
\includegraphics[width=0.9\hsize]{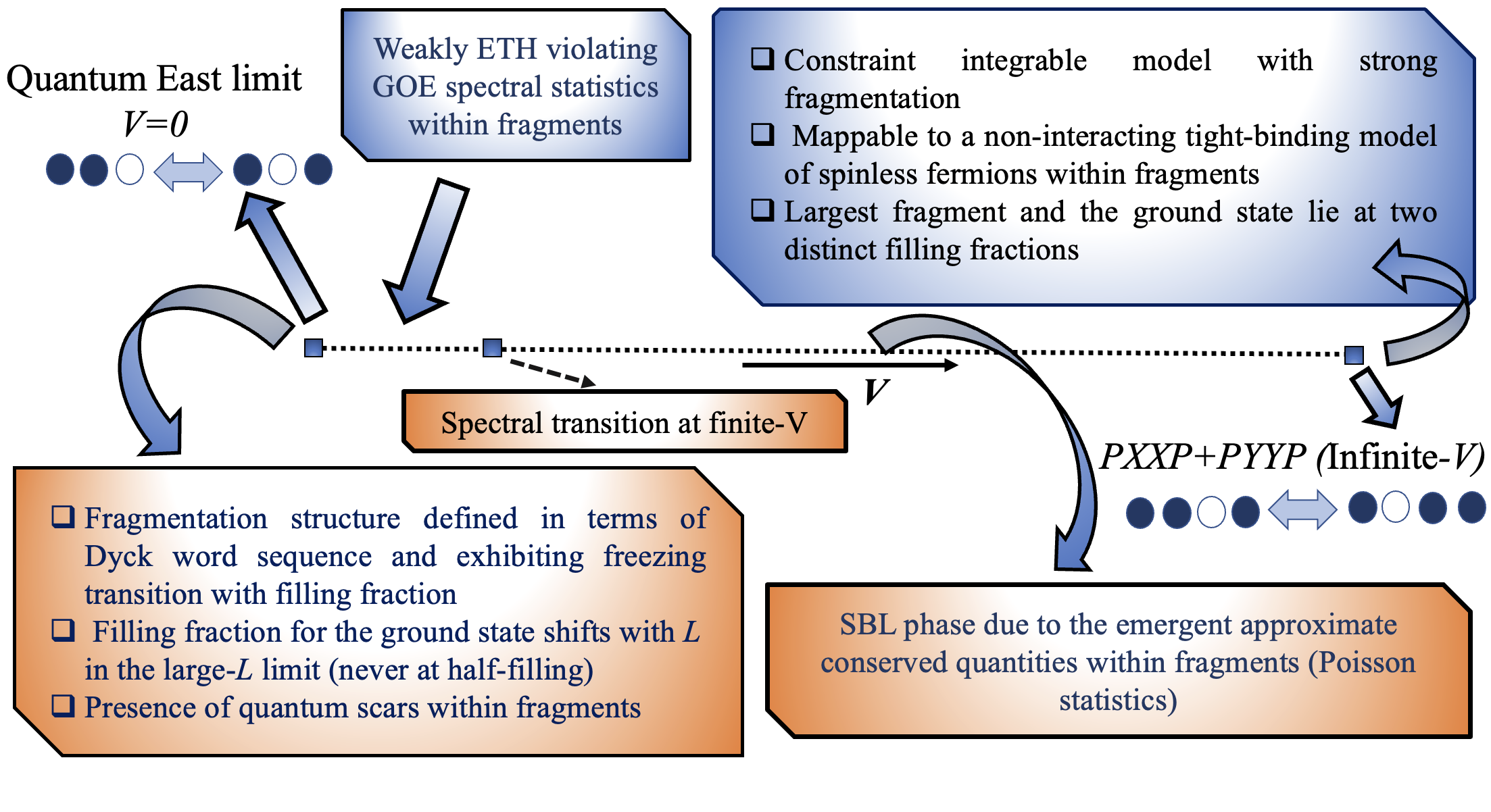}
\caption{Schematic of the main results obtained from our analysis. The PXXP$+$PYYP 
model will be discussed in Sec.~\ref{PXXP}, and the statistical bubble localized (SBL)
phase will be discussed in Sec.~\ref{sec12}.} \label{mainresults} \end{figure*}

To stabilize the ground state at a particular filling fraction, we investigate the effect of perturbations, namely, a nearest-neighbor 
density-density interaction with strength $V$, which is easily realizable in various experimental platforms~\cite{exp1,exp2,exp3,exp4}.
We note that a spectral transition occurs with increasing $V$ from the GOE statistics (weakly ETH-violating case with subspace-restricted 
scars~\cite{rev1,brighi_2023,subscar1,subscar2}) to the Poisson statistics (statistical bubble localization within fragments ~\cite{SBL1,SBL2}). This statistical bubble localized 
phase~\cite{SBL1,SBL2} is a manifestation of increasing $V$, which introduces large energy costs for certain configurations even within a fully connected fragment and 
eventually, for large $V$, produces secondary fragmentation within a single fragment due to the emergence of approximate conserved quantities. In particular, the $V\to\infty$ limit 
reduces to a constrained integrable model~\cite{pozsgay_2024, zadnik_2021}, which, interestingly, is invariant under inversion symmetry, unlike the East model with 
$V=0$. We further note that 
increasing the value of $V$ moves the ground state closer to half-filling while keeping the fragmentation structure of the East model intact. 
Nevertheless, this behavior abruptly changes exactly at the limit $V\to\infty$ 
due to the appearance of a different fragmentation structure. We analytically
unravel the structure of this new
fractured Hilbert space by mapping individual fragments to tight-binding models of non-interacting spinless fermions. This model displays some additional 
intriguing properties, such as the appearance of the largest fragment and of 
the ground state at two different filling fractions, which we are also able to
explain analytically. Finally, we conclude by proposing an experimentally 
achievable special kind of $t-V$ model with a specific kind of on-site 
potential $\mu$, which can give rise to this infinite-$V$ model~\cite{aditya_2024} in a particular limit. The results obtained from our analysis in various limits have been schematically shown in Fig. \ref{mainresults}.

The plan of this paper is as follows. In Sec. \ref{model}, we introduce our model and discuss the symmetries of the Hamiltonian. In Secs. \ref{HSFQE}, we describe the fragmentation structure of this model after identifying the canonical representation for each root state identifying each fragment. Further, we discuss fragmentation-induced anomalous ground state properties of this model. Subsequently, we describe the ETH properties within the full Hilbert space and within fragments; additionally, we examine the recursive quantum HSF-induced subspace-restricted scar states in Sec. \ref{ETH}. In Sec. \ref{DynQE}, we investigate the dynamical signatures of the ETH violation within scar subspace; further, we also inspect the long-time behavior of autocorrelators, which exhibit distinct bulk and boundary characteristics. We afterward discuss the effects of interactions in the East model in Sec. \ref{sec8}. In Sec. \ref{PXXP}, we examine the infinite-$V$ limit of this model, its fragmentation structure, and ground state properties. Further, we propose an experimental setup to realize this model in the same section. We then conclude this paper after summarizing our findings and offering the possible future avenues pertaining to this problem in Sec. \ref{Dis}.

\section{Model Hamiltonian }
\label{model}

In this section, we introduce our correlated-hopping model which involves spinless fermions in one dimension with terms including three consecutive sites. The model allows nearest-neighbor hoppings between sites $j$ and $j+1$ only if the site $j-1$ is occupied; this is called the strict East constraint. (This is known to be the simplest member of a more general family of facilitated correlated-hopping models with a similar kinetic constraint~\cite{brighi_2023}).  The Hamiltonian is given by 
\begin{equation}
H ~=~ J ~\sum_{j} ~n_{j-1} ~(c_{j+1}^\dagger c_j ~+~ c_{j}^\dagger 
c_{j+1}), \end{equation}
where $J$ denotes the strength of the hopping, and $L$ is the system size. We will set $J=1$ in the rest of this paper. While this model conserves the total particle number $N = \sum_j
c_j^\dagger c_j$, and is invariant under translation symmetry (for a system with periodic
boundary conditions (PBC)), it is not invariant under the inversion symmetry transformation given by $c_{j}\to c_{L+1-j}$. The energy eigenvalues of this
Hamiltonian appear in pairs with $\pm E$ as a consequence of the sublattice symmetry transformation, given by $c_j \to (-1)^j c_j$, which transforms $H\to-H$.

%We have studied several properties of this model, both analytically and numerically,
%which are relevant to HSF and thermalization.

\section{Fragmentation structure of the East model}
\label{HSFQE}

Due to the dynamical hindrance caused by the East constraint, this model reveals an intriguing HSF structure, which we will now unravel using some analytical methods,
namely, a transfer matrix method and enumerative combinatorics~\cite{gen,aditya_2024,menon_1995,menon_1997,barma_1994} as discussed below. 

\subsection{Characterization of fragments using root configurations}
\label{subsecA}

We will first investigate how one can represent each fragment by a root configuration, which can label individual fragments in a unique manner. This identification will also 
provide an efficient way to characterize the fragmentation structure of the Hilbert space within this model~\cite{rev1,aditya_2024}, such as the total number of fragments and frozen states and the growth with the system size of all the fragments at arbitrary fillings; these properties have not been comprehensively explored before. We will assume 
open boundary conditions (OBC) for most of our analysis since this is more convenient for using the different analytical methods.

We first need to specify certain rules, which will enable us to characterize each fragment by a distinctive representative state; we will call this the root state for the rest of our discussion. To proceed, we note that the East constraint in
our model only allows the transition 
\beq 110 ~\leftrightarrow ~101. \eeq
This can be represented in an equivalent form by using the notion of Dyck words~\cite{catalan,fredkin1,fredkin2,fredkin3} in terms of open and closed parenthesis as $(()\,\leftrightarrow\,()($, where $1$ and $0$ are replaced by an opening parenthesis, $($, and a closing parenthesis, $)$, respectively. Further we note that the number of opening and closing parenthesis along with the balanced pair of parenthesis, i.e., $()$ on both sides of the transition is the same. This
suggests that the growth of each fragment can be captured 
via the extensively studied combinatorics sequence of Dyck words/paths~\cite{catalan}. Next, it can be 
checked that for all simple and fully connected fragments, it is possible to move all occurrences of $10$'s to the leftmost end of the string by successively replacing $110$'s by $101$'s. This provides a canonical 
way of labeling each fragment by a distinct root state~\cite{fredkin1,fredkin2,fredkin3}. (We will discuss at the end of this
section why the opposite procedure of replacing $101$'s by $110$'s
does not guarantee a unique representation for root states). 
This method is identical to finding the faithful representations for all the equivalent classes as studied in representation theory. After successively removing all the $110$'s, the most general fully connected fragment can be represented by a unique root state with the following separable form $\psi_{L}\otimes\psi_{m}\otimes\psi_{R}$, where $\psi_{L}$
can be a null string ($\phi$) or a substring of $0$'s represented by $)\cdots)$ in the parenthesis notation, $\psi_{m}$ is given by substrings of $10$'s or in an alternative notation by $()\cdots()$, and $\psi_{R}$ can be either a null string ($\phi$) or a substring of $1$'s, i.e.,  $(\cdots($ or a substring of 0's followed by substring of 1's, i.e., $)\cdots)(\cdots($. Thereafter, the dimension of such a fragment can be computed with the help of the following 
rules.
\vspace{0.2cm}

\noi (i) One can always remove $\psi_{L}$ from the counting problem since 
the transition does not allow it to participate in the dynamics. This statement also holds trivially when $\psi_{L}$ is $\phi$.
\vspace{0.2cm}

\noi (ii) Subsequently, if $\psi_R$ is a substring made of only 0's, i.e., $)\cdots)$ or any number of 0's followed by any number of 1's, $)\cdots($, such a  $\psi_R$ also remains frozen under dynamical evolution. This statement again trivially holds when $\psi_R$ is $\phi$. These two scenarios comprise case I. But if $\psi_R$ is a substring of $1$'s, i.e., $(\cdots($, it has to be integrated into the counting problem; this gives rise to the second root category which we will call case II.

After implementing these rules, the growth of the dimension 
$D(L)$ of a fully connected fragment can be found from the Catalan triangle of order, $l=1$~\cite{catalan}, as follows.
\vspace{0.2cm}

\noi (i) For the first case when $\psi_{R}$ falls under case I, we obtain $D(L)=C(n,n)$, where $C(n,n)$ is the $n$-th Catalan number (diagonal elements of the Catalan triangle), given by $\frac{1}{n+1}\left(\begin{matrix}2n\\n\end{matrix}\right)$, and $n$ denotes the number of $10$'s.
\vspace{0.3cm}

\noi (ii) For the second case where $\psi_{R}$ is a member of case II, we have $D(L)=C(n+k,n)$, where $C(n+k,n)$ is the element of the Catalan triangle of order 
$l=1$~\cite{catalan} corresponding to $(n+k)$-th row and $n$-th column, where $n$ and $k$ denote the total number of $10$'s and $1$'s in $\psi_{m}$ and $\psi_{R}$, respectively. The element $C(p,q)$ associated to the
$p$-th row and $q$-th column of the Catalan triangle of order $l=1$ is given by $\left(\begin{matrix}p+q\\q\end{matrix}\right)\,-\,\left(\begin{matrix}p+q\\q-1\end{matrix}\right)$.
\vspace{0.3cm}

This procedure can be generalized further to an arbitrary string which contains blockaded regions that disconnect two active regions by frozen state configurations. All such 
strings containing blockades can thus be uniquely labeled by a root state with the following separable form $\psi_{L}\otimes\psi_{m_1}\otimes\psi_{I_1}\otimes\psi_{m_2}\otimes\psi_{I_2}\cdots\otimes\psi_R$ , where $\psi_L$ is a null string $(\phi)$ or a substring made of any number of 0's, $\psi_{m_i}$ is a substring made of 10's, $\psi_{I_i}$ is a blockade region comprising substring of 0's, and, finally, $\psi_{R}$ contains
only 0's or any number of 0's followed by any number of 1's or all 1's. The dimensions of such fragments with blockades can be computed with the help of similar rules as discussed above:

\noi (i) For the first case with $ \psi_R$ falling under case I as discussed before, $ D(L)~ =~ \prod_{i = 1}^{L-1} C(n_i, n_i) $ where $n_i$ is the number of pairs of $10$'s in $\psi_{m_i}$, and $C(n_{i},n_{i})$ is the $n_{i}$-th Catalan number.
\vspace{0.2cm}

\noi (ii) For $ \psi_R$ with the properties of case II, $ D(L)~ =~ \prod_{i = 1}^{L-1} C(n_i, n_i) \, C(n_L + k, 
n_L)$, where $n_i$ is the number of pairs of 10's in $\psi_{m_i}$ and $k$ is the number of $1$'s in $\psi_R$. Further, $C(n_{L}+k,n_{L})$ denotes the element of the Catalan triangle corresponding to the $(n_{L}+k)$-th row and $n_{L}$ column, and $C(n_{i},n_{i})$ is the $n_{i}$-th Catalan number.

An equivalent representation of the Dyck words is possible in terms of Dyck paths. A Dyck path is a path that starts at the origin (0,0) and ends at $(N,0)$ in the two-dimensional $x-y$ plane; further, the path contains only two moves, namely, diagonal up steps $(1,1)$ and diagonal down steps
$(1,-1)$. Moreover, all Dyck paths have to meet the condition that they should always have $y \geq 0$. Following such a prescription,  one can show the number of allowed Dyck paths with $N$ moves is given by the Catalan number $D_{N,0,0}=\frac{1}{N/2+1}\begin{pmatrix}N\\ N/2\end{pmatrix}$, where $N$ is even. Furthermore, if we consider a more general case, where the paths start at $(0,0)$ and end at $(N,m)$ with $m\geq0$, the number of allowed Dyck paths in the upper half plane is given by
\bea D_{N,0,m}=\frac{m+1}{N+1}\begin{pmatrix}N+1\\(N-m)/2\end{pmatrix}.
\eea
This is equal to the element of the Catalan triangle of order $l=1$ corresponding to the $n$-th row and $k$-th column, where $n=(N+m)/2$ and $k=(N-m)/2$. We use the Dyck path representation to illustrate all the fragments, allowed states, and the dimensions of the fragments for the East model with $L=5$ in Table \ref{dyckrep} using the rules mentioned above.

Before ending this section, we note that we cannot find an alternate canonical representation for root states by successively replacing strings of $101$'s by $110$'s in this model. To show 
this, we consider the configuration $11001$ as an example. It can checked that this string is a member of a classical fragment specified by the root state $10101$ due to our rule of successively replacing $110$'s by $101$'s. But if one replaces $101$'s by $110$'s instead, it may appear that the configuration $11001$ is already a canonical root state as it does not contain any $101$'s to be replaced with $110$'s. However, this is clearly not true due to the fact if we start from the root state $10101$ obtained using our earlier rule, and thereafter successively remove 
$101$'s, with the rule that the rightmost $101$'s are replaced first while replacing them with $110$'s, it will generate a different root state which is $11100$ (by the transitions $10101 \to 10110 \to 11010 \to11100$). This shows that replacing
$101$'s by $110$'s fails to generate a unique root state for this fragment. Similar problems arise for several other fragments as well.

%Finally we note that the identification of the root state by successively removing 101's fails to uniquely represent each fragment; further, implementing the 101 removal rule also gives an incorrect result for the total number of fragments.

\subsection{Counting of fragments}

To characterize the fragmentation structure of our model,
we will now count the total number of fragments as a function of $L$ using the transfer matrix method. To do this, we will consider OBC which allows us to efficiently implement the above method. Using the fact that a unique identification of the root states requires us to remove all occurrences of $110$'s, we find that the number of fragments, $N_{frag}$, grows asymptotically as $\tau^{L}$, where $\tau=(\sqrt{5}+1)/2$ is the golden ratio,~\cite{gen,aditya_2024,menon_1997,menon_1995,barma_1994} as shown in Appendix \ref{appA}. Further, our analytical results are found to be in agreement with the
values obtained by numerical enumeration.

\begin{comment}
One should further note that since our model comprises terms involving three consecutive sites, it is thus required to construct a $4\times 4$ transfer matrix to compute the total number of fragments, represented as $T(c_{i},c_{j})$, where $c_{i}$ denotes the configurations $\ket{11}$, $\ket{10}$, $\ket{01}$ and $\ket{00}$. In addition, we also observed, as mentioned earlier in the discussion, that the unique identification of the root state to label each fragment requires us to move all the pairs of $10$'s to the extreme left end of the strings by successively removing $110$'s. This, therefore, suggests that any resultant $110$ should be removed from the transfer matrix.
After incorporating this fact, the final transfer matrix thus admits the form as
\bea
 T =
\begin{bmatrix}
1 & 0 & 0 & 0 \\
0 & 0 & 1 & 1 \\
1 & 1 & 0 & 0 \\
0 & 0 & 1 & 1
\end{bmatrix}.
\label{trans1}
\eea
The eigenvalues of Eq. \eqref{trans1} are given by $\tau = (\sqrt{5}+1)/2 \simeq
1.618$, $-1/\tau \simeq -0.618$, 1 and 0. The total number of fragments, therefore, shows an exponential growth (in $L$) as $1.618^L$ in the asymptotic limit.
\end{comment}

The number of fragments $N_{frag}$ for a given $L$ can be analytically computed utilizing the fact that $N_{frag} = \sum_{i,j} 
M_{i,j}$, where $M = T_{1}^{L-2}$ for $L\geq 3$, and $T_{1}$ is the transfer matrix shown in Appendix \ref{appA}. This allows us to determine $N_{frag}$ for the first few values
of $L$, as shown in Table \ref{Table2}. In addition, we choose $N_{frag}(L)\,=\,1,2,4$ for $L=0,1,2$, respectively, for further convenience.
%Table 1 shows the number of fragments versus the system size $L$ for 
%$3 \le L \le 16$.

We further note from Table \ref{Table2} that $N_{frag}$ follows the relation
\begin{equation} 
N_{frag} (L) ~=~ F_{L + 3} ~-~ 1,
\end{equation}
where $F_{L}$ is the Fibonacci sequence with initial conditions $F_{0}=0$ and $F_{1}$=1.

\begin{center}
\begin{table*}[htb]
\begin{tabular}{|c|c|c|c|} 
\hline
Root state & Allowed transitions within fragments & Dyck path representation & $D_{frag}$ \\
& & (after implementing our rules) & \\
\hline
 $00000$ & $00000$ & No representation (frozen state) & $1$ \\

\hline
 $ 10000$ & $ 10000$ &
 \begin{tikzpicture}[scale=0.5]
\draw[step=1cm,gray,very thin] (0,0) grid (2,1);
\draw[black, thick, ->, >=latex, line width=1.5pt,  midway] (0,0) -- (1,1);
\draw[black, thick, ->, >=latex, line width=1.5pt,  midway] (1,1) -- (2,0);  
\end{tikzpicture} (frozen state)& 
$C(1,0) = 1$\\

\hline
 $ 01000$ & $ 01000$ & 
 \begin{tikzpicture}[scale=0.5]
\draw[step=1cm,gray,very thin] (0,0) grid (2,1);
\draw[black, thick, ->, >=latex, line width=1.5pt,  midway] (0,0) -- (1,1);
\draw[black, thick, ->, >=latex, line width=1.5pt,  midway] (1,1) -- (2,0); 
\end{tikzpicture} (frozen state)&
$C(1,0) = 1$\\

\hline

$ 00100$ & $ 00100$ & 
\begin{tikzpicture}[scale=0.5]
\draw[step=1cm,gray,very thin] (0,0) grid (2,1);
\draw[black, thick, ->, >=latex, line width=1.5pt,  midway] (0,0) -- (1,1);
\draw[black, thick, ->, >=latex, line width=1.5pt,  midway] (1,1) -- (2,0);
\end{tikzpicture} (frozen state)&
$C(1,0) = 1$\\

\hline
 $ 00010$ & $ 00010$ & 
 \begin{tikzpicture}[scale=0.5]
\draw[step=1cm,gray,very thin] (0,0) grid (2,1);
\draw[black, thick, ->, >=latex, line width=1.5pt,  midway] (0,0) -- (1,1);
\draw[black, thick, ->, >=latex, line width=1.5pt,  midway] (1,1) -- (2,0);
\end{tikzpicture} (frozen state)& 
$C(1,0) = 1$\\

\hline
 $ 00001$ & $ 00001$ & No representation (frozen state)  & $ C(1,0) = 1$\\

\hline
 $ 11111$ & $ 11111$ & No representation (frozen state) & $C(5,0) = 1$\\

\hline
 $ 10010$ & $ 10010$ & 
 \begin{tikzpicture}[scale=0.5]
\draw[step=1cm,gray,very thin] (0,0) grid (2,1);
\draw[black, thick, ->, >=latex, line width=1.5pt,  midway] (0,0) -- (1,1);
\draw[black, thick, ->, >=latex, line width=1.5pt,  midway] (1,1) -- (2,0);
\end{tikzpicture}
$\otimes$ \fbox{\text{blockade}} $\otimes$
\begin{tikzpicture}[scale=0.5]
\draw[step=1cm,gray,very thin] (0,0) grid (2,1);
\draw[black, thick, ->, >=latex, line width=1.5pt,  midway] (0,0) -- (1,1);
\draw[black, thick, ->, >=latex, line width=1.5pt,  midway] (1,1) -- (2,0);
\end{tikzpicture} & 
$C(1,1) \times C(1,1) = 1$\\

\hline
 $ 10001$ & $ 10001$ & 
 \begin{tikzpicture}[scale=0.5]
\draw[step=1cm,gray,very thin] (0,0) grid (2,1);
\draw[black, thick, ->, >=latex, line width=1.5pt,  midway] (0,0) -- (1,1);
\draw[black, thick, ->, >=latex, line width=1.5pt,  midway] (1,1) -- (2,0);
\end{tikzpicture} (frozen state)
& $C(1,0) = 1$\\

\hline
 $ 00011$ & $ 00011$ & No representation (frozen state) & $C(2,0) = 1$\\

\hline
 $ 01001$ & $ 01001$ & 
 \begin{tikzpicture}[scale=0.5]
\draw[step=1cm,gray,very thin] (0,0) grid (2,1);
\draw[black, thick, ->, >=latex, line width=1.5pt,  midway] (0,0) -- (1,1);
\draw[black, thick, ->, >=latex, line width=1.5pt,  midway] (1,1) -- (2,0);
\end{tikzpicture} (frozen state)
& $C(1,0) = 1$\\

\hline
 $ 10011$ & $ 10011$ & \begin{tikzpicture}[scale=0.5]
\draw[step=1cm,gray,very thin] (0,0) grid (2,1);
\draw[black, thick, ->, >=latex, line width=1.5pt,  midway] (0,0) -- (1,1);
\draw[black, thick, ->, >=latex, line width=1.5pt,  midway] (1,1) -- (2,0);
\end{tikzpicture} (frozen state)
& $C(1,0) = 1$\\

\hline
$ 00111$ & $ 00111$ & No representation (frozen state) & $C(3,0) = 1$\\

\hline
 $ 01111$ & $ 01111$ & No representation (frozen state) & $C(4,0) = 1$\\

\hline
 $10100$ & $10100$, $11000$ & 
 
 \begin{tikzpicture}[scale=0.5]
\draw[step=1cm,gray,very thin] (0,0) grid (4,1);
\draw[black, thick, ->, >=latex, line width=1.5pt,  midway] (0,0) -- (1,1);
\draw[black, thick, ->, >=latex, line width=1.5pt,  midway] (1,1) -- (2,0);

\draw[black, thick, ->, >=latex, line width=1.5pt,  midway] (2,0) -- (3,1);
\draw[black, thick, ->, >=latex, line width=1.5pt,  midway] (3,1) -- (4,0);
\end{tikzpicture},
\begin{tikzpicture}[scale=0.5]
\draw[step=1cm,gray,very thin] (0,0) grid (4,2);
\draw[black, thick, ->, >=latex, line width=1.5pt,  midway] (0,0) -- (1,1);
\draw[black, thick, ->, >=latex, line width=1.5pt,  midway] (1,1) -- (2,2);

\draw[black, thick, ->, >=latex, line width=1.5pt,  midway] (2,2) -- (3,1);
\draw[black, thick, ->, >=latex, line width=1.5pt,  midway] (3,1) -- (4,0);
\end{tikzpicture}
& $C(2,2) = 2 $ \\

\hline
 $00101$ & $00101$, $00110$ & 
\begin{tikzpicture}[scale=0.5]
\draw[step=1cm,gray,very thin] (0,0) grid (3,1);
\draw[black, thick, ->, >=latex, line width=1.5pt,  midway] (0,0) -- (1,1);
\draw[black, thick, ->, >=latex, line width=1.5pt,  midway] (1,1) -- (2,0);

\draw[black, thick, ->, >=latex, line width=1.5pt,  midway] (2,0) -- (3,1);

\end{tikzpicture},
\begin{tikzpicture}[scale=0.5]
\draw[step=1cm,gray,very thin] (0,0) grid (3,2);
\draw[black, thick, ->, >=latex, line width=1.5pt,  midway] (0,0) -- (1,1);
\draw[black, thick, ->, >=latex, line width=1.5pt,  midway] (1,1) -- (2,2);

\draw[black, thick, ->, >=latex, line width=1.5pt,  midway] (2,2) -- (3,1);

\end{tikzpicture}
& $C(2,1) = 2$ \\

\hline
$01010$ & $01010$, $01100$ & 
\begin{tikzpicture}[scale=0.4]
\draw[step=1cm,gray,very thin] (0,0) grid (4,1);
\draw[black, thick, ->, >=latex, line width=1.5pt,  midway] (0,0) -- (1,1);
\draw[black, thick, ->, >=latex, line width=1.5pt,  midway] (1,1) -- (2,0);

\draw[black, thick, ->, >=latex, line width=1.5pt,  midway] (2,0) -- (3,1);
\draw[black, thick, ->, >=latex, line width=1.5pt,  midway] (3,1) -- (4,0);
\end{tikzpicture},
\begin{tikzpicture}[scale=0.5]
\draw[step=1cm,gray,very thin] (0,0) grid (4,2);
\draw[black, thick, ->, >=latex, line width=1.5pt,  midway] (0,0) -- (1,1);
\draw[black, thick, ->, >=latex, line width=1.5pt,  midway] (1,1) -- (2,2);

\draw[black, thick, ->, >=latex, line width=1.5pt,  midway] (2,2) -- (3,1);
\draw[black, thick, ->, >=latex, line width=1.5pt,  midway] (3,1) -- (4,0);
\end{tikzpicture}
& $C(2,2) = 2$\\

\hline
 $01011$ & $01011$, $01101$, $01110$ & 
 
 \begin{tikzpicture}[scale=0.3]
\draw[step=1cm,gray,very thin] (0,0) grid (4,2);
\draw[black, thick, ->, >=latex, line width=1.5pt,  midway] (0,0) -- (1,1);
\draw[black, thick, ->, >=latex, line width=1.5pt,  midway] (1,1) -- (2,0);

\draw[black, thick, ->, >=latex, line width=1.5pt,  midway] (2,0) -- (3,1);
\draw[black, thick, ->, >=latex, line width=1.5pt,  midway] (3,1) -- (4,2);
\end{tikzpicture},

\begin{tikzpicture}[scale=0.3]
\draw[step=1cm,gray,very thin] (0,0) grid (4,2);
\draw[black, thick, ->, >=latex, line width=1.5pt,  midway] (0,0) -- (1,1);
\draw[black, thick, ->, >=latex, line width=1.5pt,  midway] (1,1) -- (2,2);

\draw[black, thick, ->, >=latex, line width=1.5pt,  midway] (2,2) -- (3,1);
\draw[black, thick, ->, >=latex, line width=1.5pt,  midway] (3,1) -- (4,2);
\end{tikzpicture},

\begin{tikzpicture}[scale=0.3]
\draw[step=1cm,gray,very thin] (0,0) grid (4,3);
\draw[black, thick, ->, >=latex, line width=1.5pt,  midway] (0,0) -- (1,1);
\draw[black, thick, ->, >=latex, line width=1.5pt,  midway] (1,1) -- (2,2);

\draw[black, thick, ->, >=latex, line width=1.5pt,  midway] (2,2) -- (3,3);
\draw[black, thick, ->, >=latex, line width=1.5pt,  midway] (3,3) -- (4,2);
\end{tikzpicture}

& $C(3,1) = 3$\\

\hline

 $10111$ & $10111$, $ 11110$, $11101$, $11011$ & 
 \begin{tikzpicture}[scale=0.3]
\draw[step=1cm,gray,very thin] (0,0) grid (5,3);
\draw[black, thick, ->, >=latex, line width=1.5pt,  midway] (0,0) -- (1,1);
\draw[black, thick, ->, >=latex, line width=1.5pt,  midway] (1,1) -- (2,0);

\draw[black, thick, ->, >=latex, line width=1.5pt,  midway] (2,0) -- (3,1);
\draw[black, thick, ->, >=latex, line width=1.5pt,  midway] (3,1) -- (4,2);
\draw[black, thick, ->, >=latex, line width=1.5pt,  midway] (4,2) -- (5,3);
\end{tikzpicture},
\begin{tikzpicture}[scale=0.3]
\draw[step=1cm,gray,very thin] (0,0) grid (5,4);
\draw[black, thick, ->, >=latex, line width=1.5pt,  midway] (0,0) -- (1,1);
\draw[black, thick, ->, >=latex, line width=1.5pt,  midway] (1,1) -- (2,2);

\draw[black, thick, ->, >=latex, line width=1.5pt,  midway] (2,2) -- (3,3);
\draw[black, thick, ->, >=latex, line width=1.5pt,  midway] (3,3) -- (4,4);
\draw[black, thick, ->, >=latex, line width=1.5pt,  midway] (4,4) -- (5,3);
\end{tikzpicture},

\begin{tikzpicture}[scale=0.3]
\draw[step=1cm,gray,very thin] (0,0) grid (5,3);
\draw[black, thick, ->, >=latex, line width=1.5pt,  midway] (0,0) -- (1,1);
\draw[black, thick, ->, >=latex, line width=1.5pt,  midway] (1,1) -- (2,2);

\draw[black, thick, ->, >=latex, line width=1.5pt,  midway] (2,2) -- (3,3);
\draw[black, thick, ->, >=latex, line width=1.5pt,  midway] (3,3) -- (4,2);
\draw[black, thick, ->, >=latex, line width=1.5pt,  midway] (4,2) -- (5,3);
\end{tikzpicture},

& $C(4,1) = 4$ \\

& & \begin{tikzpicture}[scale=0.3]
\draw[step=1cm,gray,very thin] (0,0) grid (5,3);
\draw[black, thick, ->, >=latex, line width=1.5pt,  midway] (0,0) -- (1,1);
\draw[black, thick, ->, >=latex, line width=1.5pt,  midway] (1,1) -- (2,2);

\draw[black, thick, ->, >=latex, line width=1.5pt,  midway] (2,2) -- (3,1);
\draw[black, thick, ->, >=latex, line width=1.5pt,  midway] (3,1) -- (4,2);
\draw[black, thick, ->, >=latex, line width=1.5pt,  midway] (4,2) -- (5,3);
\end{tikzpicture}&
\\
\hline
 $10101$ & $11100$, $11010$, $11001$, $10110$, $10101$ & 

 \begin{tikzpicture}[scale=0.3]
\draw[step=1cm,gray,very thin] (0,0) grid (5,3);
\draw[black, thick, ->, >=latex, line width=1.5pt,  midway] (0,0) -- (1,1);
\draw[black, thick, ->, >=latex, line width=1.5pt,  midway] (1,1) -- (2,2);

\draw[black, thick, ->, >=latex, line width=1.5pt,  midway] (2,2) -- (3,3);
\draw[black, thick, ->, >=latex, line width=1.5pt,  midway] (3,3) -- (4,2);
\draw[black, thick, ->, >=latex, line width=1.5pt,  midway] (4,2) -- (5,1);
\end{tikzpicture}

\begin{tikzpicture}[scale=0.3]
\draw[step=1cm,gray,very thin] (0,0) grid (5,2);
\draw[black, thick, ->, >=latex, line width=1.5pt,  midway] (0,0) -- (1,1);
\draw[black, thick, ->, >=latex, line width=1.5pt,  midway] (1,1) -- (2,2);

\draw[black, thick, ->, >=latex, line width=1.5pt,  midway] (2,2) -- (3,1);
\draw[black, thick, ->, >=latex, line width=1.5pt,  midway] (3,1) -- (4,2);
\draw[black, thick, ->, >=latex, line width=1.5pt,  midway] (4,2) -- (5,1);
\end{tikzpicture},

\begin{tikzpicture}[scale=0.3]
\draw[step=1cm,gray,very thin] (0,0) grid (5,2);
\draw[black, thick, ->, >=latex, line width=1.5pt,  midway] (0,0) -- (1,1);
\draw[black, thick, ->, >=latex, line width=1.5pt,  midway] (1,1) -- (2,2);

\draw[black, thick, ->, >=latex, line width=1.5pt,  midway] (2,2) -- (3,1);
\draw[black, thick, ->, >=latex, line width=1.5pt,  midway] (3,1) -- (4,0);
\draw[black, thick, ->, >=latex, line width=1.5pt,  midway] (4,0) -- (5,1);
\end{tikzpicture},

\begin{tikzpicture}[scale=0.3]
\draw[step=1cm,gray,very thin] (0,0) grid (5,2);
\draw[black, thick, ->, >=latex, line width=1.5pt,  midway] (0,0) -- (1,1);
\draw[black, thick, ->, >=latex, line width=1.5pt,  midway] (1,1) -- (2,0);

\draw[black, thick, ->, >=latex, line width=1.5pt,  midway] (2,0) -- (3,1);
\draw[black, thick, ->, >=latex, line width=1.5pt,  midway] (3,1) -- (4,2);
\draw[black, thick, ->, >=latex, line width=1.5pt,  midway] (4,2) -- (5,1);
\end{tikzpicture},

& $C(3,2) = 5 $\\

 &&\begin{tikzpicture}[scale=0.3]
\draw[step=1cm,gray,very thin] (0,0) grid (5,1);
\draw[black, thick, ->, >=latex, line width=1.5pt,  midway] (0,0) -- (1,1);
\draw[black, thick, ->, >=latex, line width=1.5pt,  midway] (1,1) -- (2,0);

\draw[black, thick, ->, >=latex, line width=1.5pt,  midway] (2,0) -- (3,1);
\draw[black, thick, ->, >=latex, line width=1.5pt,  midway] (3,1) -- (4,0);
\draw[black, thick, ->, >=latex, line width=1.5pt,  midway] (4,0) -- (5,1);
\end{tikzpicture}&\\
\hline
        
\end{tabular}
\caption{Complete characterization of the fragmentation structure for $L=5$ in the East model, showing the root state labeling each fragment, the allowed transitions within them, their Dyck path representations, and the dimensions of the fragments calculated using our rules.}
\label{dyckrep}
\end{table*}
\end{center}

\begin{table}[h]
\centering
\begin{tabular}{|c|c|c|c|c|c|c|c|c|c|c|c|c|c|c|c|c|}
\hline
$L$ & 3 & 4 & 5 & 6 & 7 & 8 & 9 & 10 & 11 & 12 & 13 & 14 & 15 \\
\hline
\(N_{frag}\) & 7 & 12 & 20 & 33 & 54 & 88 & 143 & 232 & 376 & 609 & 986 & 1596 & 2583 \\
\hline
\end{tabular}
\caption{$N_{frag}$ for the first few values of $L$ as obtained analytically. These agree with our numerical results.} 
\label{Table2}
\end{table}

\subsection{Frozen fragments}

We will now discuss the growth of the total number of frozen fragments, which
consist of a single state that does not participate in the dynamics. These states are eigenstates of our Hamiltonian with zero energy. The simplest examples of such states are $11\cdots 11$ and $00\cdots00$.
%which typically show exponential growth in systems with fractured Hilbert space. 
We can find the growth with the system size of these states
using the transfer matrix method~\cite{gen,aditya_2024} in the same way as the computation of the total number of fragments. Since the frozen states do not participate in dynamics, the transfer matrix for this computation should not have any $110$ and $101$'s. Using this fact, a straightforward analytical calculation using the transfer matrix method~\cite{gen,aditya_2024,menon_1997,menon_1995,barma_1994} demonstrates that $N_{froz}(L)$ grows as $1.466^{L}$; this again agrees with our numerically obtained results. The details of this computation, both with OBC and PBC, are shown in Appendix \ref{appB}.

\subsection{Description of the largest fragment}

\begin{comment}
We saw in the previous sections that by writing the root configuration of a particular fragment we can get the total number of states within that fragment from the Catalan triangle. But now we are interested to know the filling at which the quantum East model has the largest fragment in the large-$L$ limit. Let the filling at which we get the largest fragment be \(L/2 + a\). In the language of the Catalan triangle, $ n = L/2
+ a $ and $ k = L/2 - a $ so that  $n + k = L$.
\end{comment}

After discussing the total number of fragments and frozen states, we will now find the filling fraction at which our model has the largest fragment. This question is important
to address since it is already known that this model exhibits a freezing transition~\cite{weak2} as a function of filling. We will use our canonical representation method to answer this question. Before proceeding further, we note
that the root identification method mentioned earlier already shows that the states in a fully connected fragment follow a specific sequence of the Dyck words~\cite{catalan}, as shown in Sec. \ref{subsecA}. 
Hence the growth of any fragment can be readily computed with the help of the Catalan triangle of order $l=1$ after incorporating the rules discussed earlier. It can also be shown that the fully connected fragment with the largest dimension at a filling $N_{f}=L/2+a$ with $a>0$ is represented by the following root state $10\cdots..1011\cdots1$. Following the rules discussed in Sec. \ref{subsecA}, the dimension of such a fragment is found to be
\bea
    D(L)=C(n+k,n) = \frac{(2 a + 1)}{ (\frac{L}{2} + a + 1)} \frac{L!}{(\frac{L}{2} - a )! (\frac{L}{2} + a )!},\non\\
    \label{offCat}
\eea
where $n=(L/2-a)$ and $k=2a$ are the total number of $10$'s and $1$'s in $\psi_{m}$ and $\psi_{R}$ in the given root state, respectively.
Since we are interested in the thermodynamic limit, we utilize Stirling's formula, which gives
\begin{comment}
\bea
D(L) & = \frac{(2 a + 1)}{ (\frac{L}{2} + a + 1)} \frac{1}{\sqrt{2 \pi}} \frac{L^{L + 1/2}}{(\frac{L}{2} - a )^{\frac{L}{2} - a} (\frac{L}{2} + a )^{\frac{L}{2} + a} {(\frac{L^2}{4} - a^2)^{1/2}}} \\
    & = \frac{(2 a + 1)}{ (\frac{L}{2} + a + 1)} \frac{1}{\sqrt{2 \pi}} \frac{L^{ L + \frac{1}{2}}}{(\frac{L^2}{4}- a^2)^{(\frac{L+1}{2})} (\frac{1 - \frac{2a}{L}}{1 + \frac{2a}{L}})^{-a}} \\
    & = 2^L \frac{(2 a + 1 )}{\frac{L}{2} + a + 1} \frac{2}{\sqrt{2 \pi L}}(1 - \frac{4 a^2}{
    L^2})^{-\frac{1}{2}} (1 - \frac{4 a^2}{
    L^2})^{-\frac{L}{2}}(1 + \frac{4 a}{L})^{-a}.
\eea
\end{comment}
\bea
D(L)&\,=\,&2^{L}\, \frac{(2 a + 1 )}{\left(\frac{L}{2} + a + 1\right)}P,\non\\
{\rm where}~~~~~~ P&\,=\,& \sqrt{\frac{2}{2 \pi L}}\left(1 - \frac{4 a^2}{
    L^2}\right)^{-\frac{L+1}{2}}\left(1 + \frac{4 a}{L}\right)^{-a}.\non\\
    \label{largf}
\eea

\begin{figure}[h!]
    \centering
    \subfigure[]{
    \includegraphics[width=0.75\hsize]{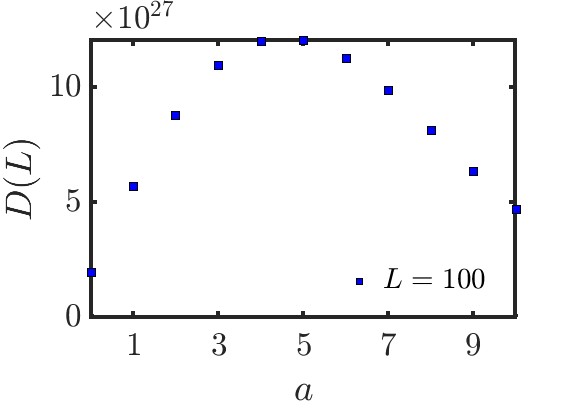}
    }\\
    \subfigure[]{
    \includegraphics[width=0.75\hsize]{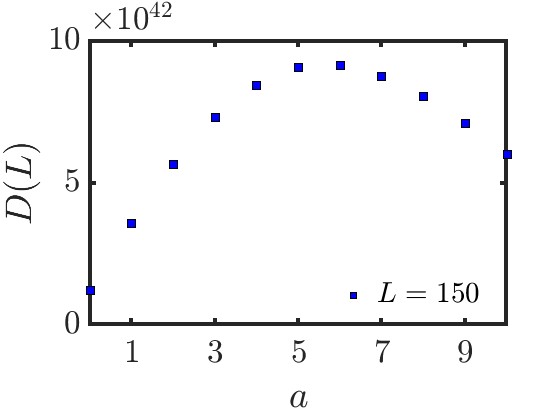}
    }
  \caption{(a-b) Plots showing the dimension of the largest fragment given by $D(L)$ (obtained using the Eq. \eqref{offCat} versus $a$, where the filling, $N_{f}=L/2+a$ for $L=100$ and $L=150$, respectively. These two plots confirm that $D(L)$ appears to be largest for $L/2+a$  where $a\simeq\sqrt{L}/2$ in the large-$L$ limit, which is $5$ and $6$ for $L=100$ and $150$, respectively.} \label{fig1100}
\end{figure}

The above expression for $P$ can be further simplified by taking a logarithm on both sides, expanding the expression up to $O(1/L)$, and then exponentiating the truncated series expansion; this thus yields $P\,\simeq\,\sqrt{\frac{2}{\pi L}}\,e^{-2a^2/L}$. Thereafter, the dimension of the largest fragment at a filling $N_{f}=L/2+a$ reduces to 
\bea
D(L)\,\simeq\, 2^{L+1} \frac{(2 a + 1)}{L} \sqrt{\frac{2}{\pi L}} e^{-\frac{2 a^2}{L}}.\label{largeLdim}
\eea
\begin{comment}
\begin{equation}
   \begin{aligned}
   & \frac{d}{d a} \{(2 a + 1) \exp(\frac{-2a^2}{L})\} = 0, \\
   & a = \frac{-\frac{1}{2} \pm \sqrt{\frac{1}{4} + L}}{2}.
   \end{aligned}
\end{equation}    
\end{comment}
Extremizing the expression in Eq. \eqref{largeLdim}, we obtain the filling fraction at which this model has the largest fragment,
\bea
N_{f}&~=~&L/2+a,\non\\
{\rm where}~~~a&~=~&\frac{-\frac{1}{2} + \sqrt{\frac{1}{4} + L}}{2}\,\simeq\,\sqrt{L}/2.\label{shift}
\eea
Eq. \eqref{shift} implies that in the thermodynamic limit,
the filling at which the largest fragment lies shifts away from half-filling ($L/2$) as $\sqrt{L}/2$ as $L$ increases. 
%due to the distinct fragmentation structure in terms of the Dyck word sequence~\cite{catalan}. 
We further numerically validate this result using Eq. \eqref{offCat}, as shown in Fig. \ref{fig1100}. In Fig. \ref{fig1100} (a-b), we plot $D(L)$ versus $a$ for two large system sizes given by $L=100$ and $L=150$, respectively. Both the plots confirm that the largest fragment shifts as $L/2+a$, where $a=\sqrt{L}/2$ 
%with increasing system sizes in the large-$L$ limit,
is equal to $5$ and $6$ (approximately) for $L=100$ and $150$, respectively. We will later discuss how this shift impacts the ground state properties of this model in the thermodynamic limit.

\begin{comment}
In the large-$L$ limit we see that $ a \simeq \frac{\sqrt{L}}{2} $. Hence the filling at which the quantum East model has the largest fragment is given by $ (\frac{L}{2} + a ) $, where $ a $ is $ \frac{\sqrt{L}}{2} $.
\end{comment} 
\begin{comment}
In the literature on HSF there is a notion of strong and weak HSF. Here in this East model we find a very interesting feature of the fragmentation. It shows either strong or weak HSF depending on the filling. We can see it very clearly by examining the ratio $ D_{max} /D $, where $D_{max}$ is the dimension of largest fragment in that filling and D is the total Hilbert space dimension. We find that the model shows strong HSF below half-filling whereas above half-filling it shows weak HSF.
\end{comment}

We will now demonstrate that this model exhibits a freezing transition, i.e., a transition from strong fragmentation (below half-filling) to weak fragmentation (above half-filling) as a function of the filling fraction. The transition occurs exactly at half-filling.
%which refers to the critical filling fraction for such a transition. 
This has been shown in the literature~\cite{weak2} earlier while obtaining the growth of the largest fragment at half-filling using the Catalan numbers (the 
diagonal elements of the Catalan triangle) in a recursive manner~\cite{weak2}. In this paper, we will validate this result utilizing our root identification method, which can capture this transition in a much easier manner. To do so, we
need to first identify the appropriate root state representing the largest fragment at an arbitrary filling, and then find the growth of that fragment with the help of the Catalan triangle. To elucidate this 
below half-filling, let us consider the root state representing the largest fragment at a filling $N_{f}=L/2-\alpha L$, where we assume that
$0 < \alpha \ll 1$. We note that such a root state has the following form $0\cdots010\cdots10$ or $10\cdots100\cdots0$, where the numbers of $10$'s in the $\psi_{m}$ part of the root state is $(L/2-\alpha {L})$, and $2\alpha L$ is the number of $0$'s in $\psi_{L}$ or $\psi_{R}$, respectively.
Now, given such a root state, the dimension, $D(L)$ of such a fragment can be readily obtained using the rules given in Sec. \ref{subsecA}, as

\begin{comment}
From the fragmentation structure we can write the root state for the largest fragment below half-filling. The general structure of this root state for the largest fragment below half-filling is $ 00000 ...101010...10 $. So the total number of states within this fragment can be found from the Catalan triangle and it is equal to the total number of states in the fragment for which the root state is $ 101010...10 $ since in the counting process we can remove all the zeros from the left side after writing it in the form with maximum spreading. Let us take the filling of the system $ n = (\frac{L}{2} -\alpha L).$ Then the total number of states within that fragment will be $ C(n, n) $ and it is given by 

\end{comment}
\bea
    D(L)& = &C\left(\frac{L}{2} - \alpha L , \frac{L}{2} - \alpha L\right)\non\\
    & = &\frac{1}{L/2 -\alpha L + 1} ~\frac{(L - 2 \alpha L )!}{((L/2 - \alpha L)!)^2}.\label{larbhalf}
    \eea
Using Stirling's formula, the above expression can be recast as follows in the large-$L$ limit,
\begin{comment}
    & \frac{(L - 2 \alpha L)!}{((\frac{L}{2} - \alpha L)!)^2} = \frac{(\frac{L- 2 \alpha L}{e})^{(L - 2 \alpha L)}}{(\frac{L- 2 \alpha L}{2 e})^{(L - 2 \alpha L)}}, \\
    & \frac{(L - 2 \alpha L)!}{((\frac{L}{2} - \alpha L)!)^2} = 2^{L( 1 - 2 \alpha)}.
    \eea
\end{comment}   
\bea
D_{max}(L)&\,\simeq\,& \frac{\mu^{L}}{L^{3/2}},\non\\
{\rm with}~~~\mu &~\simeq~& 2^{1 - 2 \alpha}, 
\label{largbhalfasym} \eea
to lowest order in $\alpha$.

\begin{comment}

If we write $  2^{L( 1 - 2 \alpha)} = \mu ^L $, where $  \mu ^L $ is the dimension of the largest fragment then $ \mu = 2^{( 1 - 2 \alpha)}$. So in the thermodynamic limit, $ \frac{D_{max}}{D} = \frac{\mu ^L}{2^L} = ( 2^{- 2 \alpha})^L \rightarrow 1 $ for $ \alpha \rightarrow 0 $. Thus we can say that near half-filling the dimension of the largest fragment is comparable with the total Hilbert space dimension which shows there is weak HSF. But for large values of $ \alpha $, i.e., far below 
half-filling $ \frac{D_{max}}{D} = \frac{\mu ^L}{2^L} = (2^{-2 \alpha})^L \rightarrow 0 $ which indicates strong HSF. Thus we can say we have a transition of the fragmentation structure from strong to weak depending on the value of the filling of the system.
\end{comment}

On the other hand, the dimension of the full Hilbert space for $N_{f}=L/2-\alpha L$, $D_{sum}$ is given by
\bea
D_{sum}(L)&\,\simeq\,& \frac{L!}{(L/2-\alpha  L)! (L/2+\alpha L)!}, \label{sym}
\eea
which in the large-$L$ limit becomes $2^{L} e^{-2 \al^2 L} /\sqrt{L}$ 
to lowest order in $\al$, utilizing Stirling's formula.
Eqs. \eqref{largbhalfasym} and \eqref{sym} thus indicate that $D_{max}(L)/D_{sum}(L)\to 0$ exponentially as $L\to\infty$ for $\alpha>0$. This implies strong fragmentation below half-filling. However, as one approaches half-filling, we find that $D_{max}(L)/D_{sum}(L)\simeq 1/L$, which means this ratio again goes to zero in the thermodynamic limit, but only polynomially, which is much slower than an exponential fall. This, therefore, signifies a change in the fragmentation structure, and therefore, $N_{f}=L/2$ is called the critical filling fraction for the strong-to-weak fragmentation transition. One can perform a similar analysis for the Hilbert space above half-filling, where the ratio $D_{max}(L)/D_{sum}$ for $N_{f}=L/2+a$ with $a>0$ reduces to
\bea
\frac{D_{max}(L)}{D_{sum}(L)}=\frac{2a+1}{L/2+a+1},\label{abovehalf}
\eea
where $D_{max}(L)$ is given in Eq. \eqref{offCat} and $D_{sum}$ is the dimension of the full Hilbert space for $N_{f}=L/2+a$. If we write $a=cL$ (where $c$ is a number of order 1) and $L\rightarrow\infty$, Eq. \eqref{abovehalf} reduces to $\simeq 4c/(2c+1)$, which is a constant in the thermodynamic limit. This thus validates the fact that this model manifests a weakly fractured Hilbert space above half-filling.

\section{Ground state properties of the East model}
\label{sec5}

\begin{comment}
For any quantum mechanical system the analysis of the ground state is very important. It tells us about the low-temperature behaviour of the model where the quantum effects are usually the  strongest. In this quantum East model we found that the ground state lies in the largest fragment of the system. We observe a shift of the ground state energy with increasing system size. We also discussed in the previous section about the dependence of largest fragment on the filling of the system. So, as the ground state lies in the largest fragment, we can say that the shift of the ground state energy with $L$ is due to the shift of the largest fragment with the filling in the large-$L$ limit.

So the quantum East model shows has the remarkable feature that depends on the filling of the system. We checked this with both PBC
and OBC as follows.
\end{comment}

\begin{figure}[h!]
    \centering
    \subfigure[]{
    \includegraphics[scale=0.5]{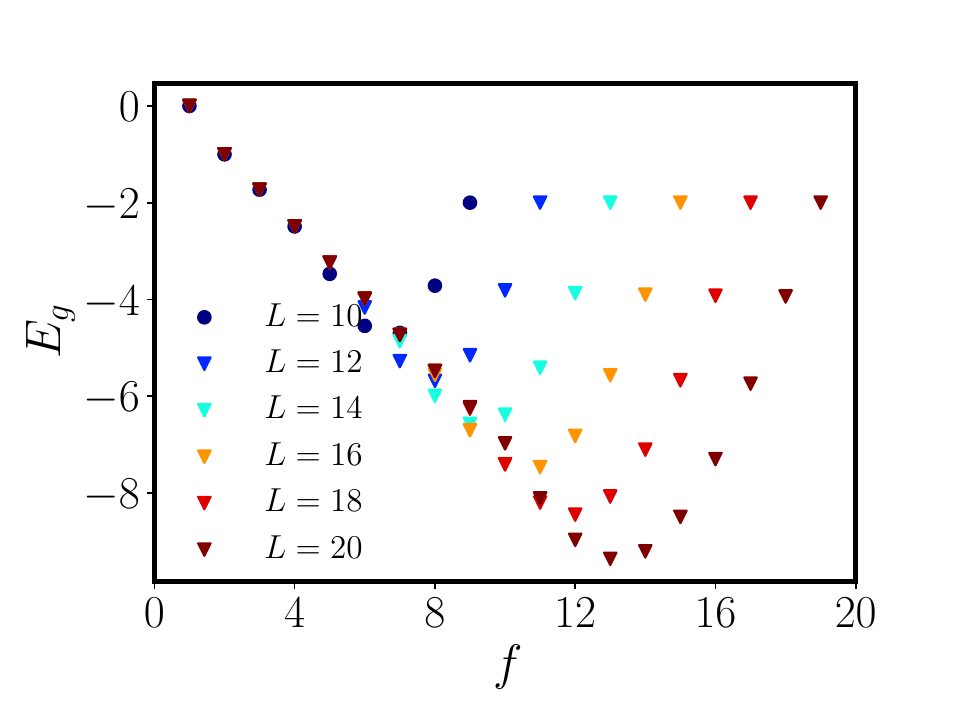}
    }
    \hfill
    \subfigure[]{
    \includegraphics[scale=0.5]{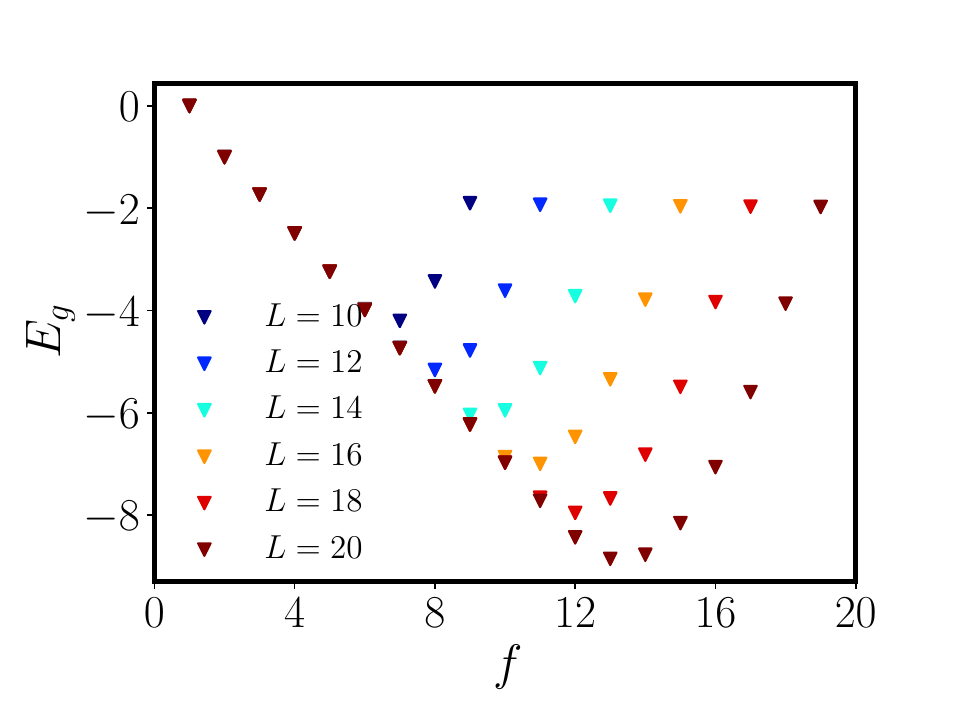}
    }
   \caption{(a-b) Plots showing the ground state energy $E_{g}$ as a function of filling $f$ for the East model with PBC and OBC, respectively, for different system sizes. Both plots confirm a shift in the filling away from $L/2$ for the ground state. }
    \label{fig4}
\end{figure}

We will now discuss some of the anomalous ground state properties of the East model; these are consequences of the unusual fragmentation structure of the Hilbert space. It is well-known that understanding the ground state is crucial 
for comprehending the low-energy properties of many-body quantum systems. However, it has recently been observed that a fractured Hilbert space can give rise to anomalous ground state behavior~\cite{aditya_2024,fredkin1,fredkin2,ground1}, which can significantly impact the low-energy properties of such systems, as
we will show in the discussion below. We have already established earlier that the filling at which this model has the largest fragment shifts from half-filling
as $L/2+ \sqrt{L}/2$.
In addition, a numerical exact diagonalization analysis of the full Hilbert space and of specific fragments confirms that the ground state lies within the largest fragment, regardless of the boundary conditions (OBC or PBC). Qualitatively, this
occurs since the larger dimension of a fragment usually favors the reduction of 
the minimum energy eigenvalue cost (although there are some exceptions). Accordingly, the filling fraction for the ground state also shifts from 
half-filling as $\sqrt{L}/2$ in the large-$L$ limit; this is a consequence of the distinct fragmentation structure of this model. In Figs. \ref{fig4} (a-b), we demonstrate the shift of the ground state filling as a function of $L$ with PBC and OBC, respectively, for small system sizes. In both cases, we observe that the ground state energy $E_{GS}$ lies at $N_{f}=L/2+2$ for $L=10,12, 14$, and $N_{f}=L/2+3$ for $L=16,18$ and $20$, respectively. One should note that the shift in filling for the ground state as $\sqrt{L}/2$ is an asymptotic result; however, this 
finite-size behaviors shown in Fig. \ref{fig4} does indicate that there is a 
shift in the filling for the ground state as one increases the system size.

\section{ETH properties of the Hilbert space} \label{ETH}

\begin{comment}
In the theoretical understanding of thermalisation for closed quantum systems, the eigenstate thermalisation hypothesis (ETH) represents a cornerstone. While analysing the fragmentation structure of a model, the existence of atypical fragments like integrable fragments and frozen fragments imply that within the full Hilbert space the model does not obey the strong ETH, which represents the 
non-thermalising behaviour of the model with respect to the full Hilbert space. Suppose that a Hamiltonian written in a basis given by the product of local states has some fragmentation structure, and the energy eigenstates have non-zero overlap only within a single fragment. Then if we calculate the reduce density matrix from the eigenvectors of a local observable it will tend to a value corresponding to a restricted microcanonical ensemble where all the eigenvectors corresponding to a given energy density are equally probable. Thus though the full Hilbert space does not obey ETH, a subsector can obey the ETH in this way, and this can be called subsector restricted ETH.
\end{comment}

In this section, we will discuss the thermalization properties of the Hilbert space of this model. In fragmented systems, the presence of exponential numbers of frozen eigenstates, integrable fragments, and blockades typically causes a complete breakdown of the ETH within the full Hilbert space. At the same time, fragmentation can also yield novel thermalization properties within the different subspaces, which is called Krylov- or 
subspace-restricted thermalization~\cite{rev1,HSF2_moudgalya_2020, aditya_2024}. Furthermore, there are also some models known where quantum many-body
scars are found within some of the fragments. We will now elucidate the complete thermalization properties of our model within the full Hilbert space and within individual fragments in the following sections.

\subsection{ETH properties within the full Hilbert space at different fillings}

\begin{comment}
Here we will investigate the validity of ETH in our model for the full Hilbert space as well as within different sectors for different fillings. We will study this within the largest fragment both below half-filling and above half-filling. To investigate this we examine the bipartite entanglement entropy $ S_{L/2} $ as a function of the energy $E$ without resolving the fragmentation structure within the full Hilbert space, both below half-filling and above half-filling. 
\end{comment}

\begin{figure*}
\centering
\subfigure[]{
\includegraphics[width=0.325\hsize]{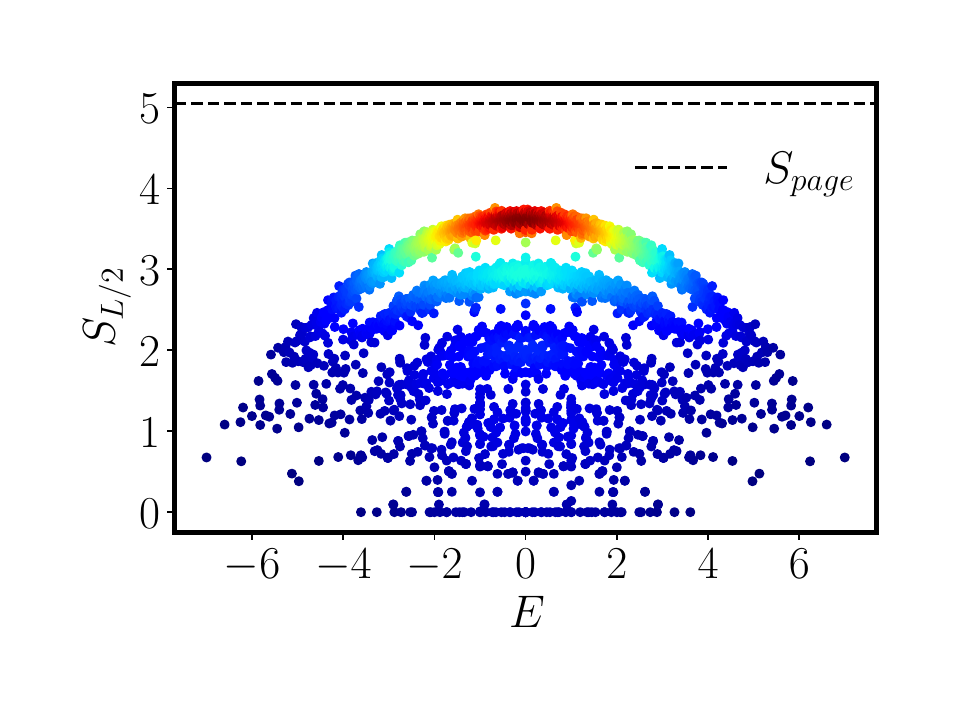}
}%
\subfigure[]{\includegraphics[width=0.325\hsize]{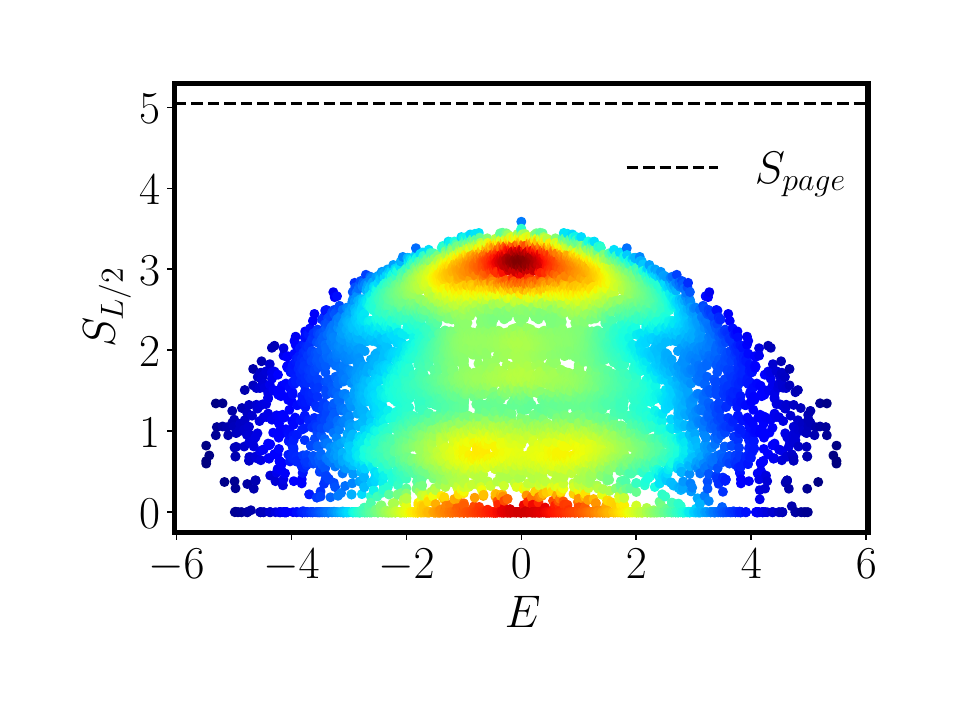}
}%
\subfigure[]{
    \includegraphics[width=0.325\hsize]{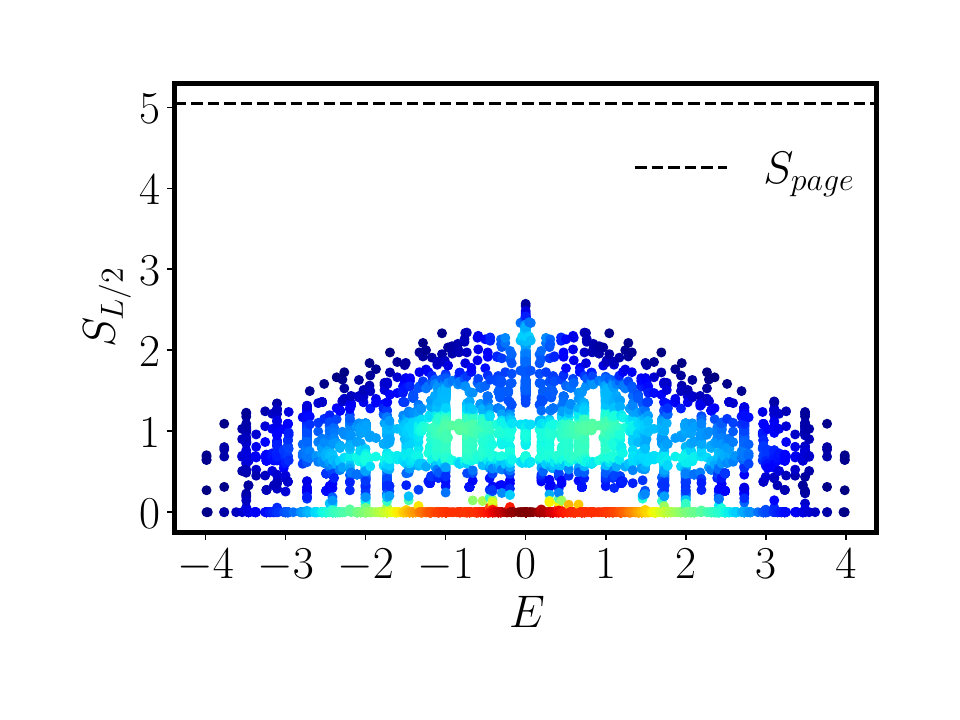}
}%
\caption{Plots showing the half-chain entanglement entropy $ S_{L/2}$ versus $E$ for the full Hilbert space for $ L = 16 $ for three different fillings, i.e.,  $L/2 + 3$ (above half-filling), $L/2$ (at half-filling) and $L/2 - 2 $ (below half-filling). In all cases, the maximum value of $S_{L/2}$ is much below the Page value, $S_{page}$, as depicted by the red dashed line. (a) 
For $N_f = L/2 + 3$, the entanglement spectrum shows several well-defined rainbow patterns, with the majority of states lying within the uppermost rainbow pattern. (b) For $N_f = L/2$, the entanglement spectrum shows that 
several rainbow patterns are beginning to separate to from one another, and
the major fraction of states lie in either the uppermost rainbow or in the
states with zero entanglement. (c) For $N_f = L/2 - 2$, the entanglement spectrum looks quite different from the other two cases. The rainbow patterns have coalesced to form an almost continuous entanglement spectrum, and 
there is a large number of states with zero entanglement.}
\label{fig5}
\end{figure*}

We will first inspect the bipartite entanglement spectrum for the full Hilbert space for different fillings, i.e., at half-filling, above half-filling, and below half-filling. As this model manifests weak and strong fragmentation above and below half-filling~\cite{weak2}, respectively, one might expect that the entanglement spectrum for the different cases should be distinct from one another. In Figs. \ref{fig5} (a-c), we show the half-chain entanglement $S_{L/2}$ versus the energy $E$ for $L=16$ for three different fillings given by $N_{f}=L/2+3$, $L/2$ and $L/2-2$. In all cases, we see that the value of $S_{L/2}$ is much lower than the Page value given by $S_{{\rm page}}=\frac{1}{2} (L\ln(2)-1)$~\cite{page}; this implies that the ETH is not satisfied within the full Hilbert space. In addition, the entanglement spectrum also carries the signature of the freezing transition~\cite{weak1,weak2}; this can be argued as follows. In Fig. \ref{fig5} (a), we see that there are several well-defined rainbow patterns separated from one another; moreover, the majority of states lie in the uppermost rainbow pattern, which signifies that the largest fragment at this filling grows faster than all the other fragments.
%to follow the growth of the full Hilbert space. 
However, Fig \ref{fig5} (c) reveals that there are no separated well-defined rainbow patterns for $N_f=L/2-2$; rather, there is a smooth spectrum of the entanglement entropy where the largest value of $S_{L/2}$ is much lower than the $S_{{\rm page}}$ due to the strong fragmentation below half-filling. Also, the maximum density of states lies near the states with $S_{L/2}\simeq 0$, which is primarily governed by the frozen eigenstates and by fragments consisting of active regions separated by blockaded regions.
In addition, exactly at the freezing transition point~\cite{weak1,weak2}, i.e., at $N_{f}=L/2$, we note an intermediate behavior in the entanglement spectrum. To
be specific, we see that several well-defined rainbow patterns are beginning to form, and at the same time, the majority of states divide themselves into two regions, in the uppermost rainbow pattern (dominant region for above half-filling) as well as in the region comprising the zero-entanglement eigenstates (dominant region for below half-filling). This intermediate characteristic occurring at half-filling also points towards a transition occurring at this filling fraction. 

\begin{figure}[h!]
\centering
\includegraphics[width=0.85\columnwidth]{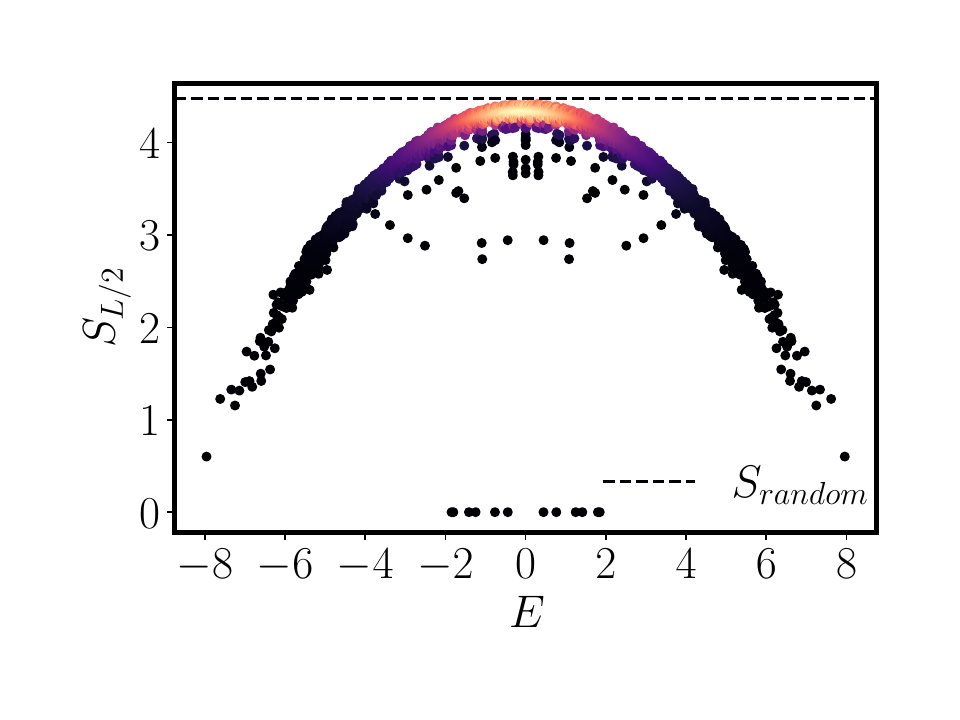}
\caption{Plot exhibiting $S_{L/2}$ versus $E$ in the largest fragment for $L = 18$  and filling $N_{f}= L/2 + 3$; the fragment dimension= is $9996$. We observe a sharp, well-defined rainbow pattern, where the maximum value of $S_{L/2}$ is much smaller than $S_{page}$ for the full Hilbert space; nevertheless, it almost reaches the value of $S_{L/2}$ for a typical thermal random state within the fragment.} 
%This implies that the eigenstates within this fragment follow a weaker version of subspace-restricted thermalization. The weak violation of the ETH arises from the presence of many states with exactly zero entanglement near the middle of the spectrum.}
\label{fig6}
\end{figure}

To understand the ETH properties within very large fragments, we consider the largest fragment of the full Hilbert space for $L=18$; this occurs at $N_{f}=L/2+3$ with the fragment dimension equal to $9996$. As shown in Fig. \ref{fig6}, we note a well-defined rainbow pattern, characteristic of a thermalizing system, with a number of states with exactly zero entanglement lying
near the middle of the spectrum. The zero-entanglement states can be potential candidates for quantum many-body scars. The rainbow pattern implies that sufficiently large fragments thermalize within their own Hilbert spaces, but in 
a weak sense. This weak violation of Krylov-restricted thermalization is supported by the presence of subspace-restricted quantum many-body scars~\cite{brighi_2023,HSF2_moudgalya_2020,subscar1,subscar2} as discussed below.
In addition, we note that although all sufficiently large fragments follow the weak form of Krylov-restricted thermalization for all fillings, the number of outlying states that violate the ETH~\cite{ETH1,ETH2,ETH3} increases as one moves from weak to strong fragmentation side, i.e., from above to below half-filling.

\begin{comment}

To understand the ETH properties within different sectors we have taken the largest fragment below and above half-filling and studies the bipartite entanglement entropy $ S_{L/2}$ as a function of the energy $ E$.
From the figures we see for the full Hilbert space that the maximum value of the entanglement spectrum does not reach the thermal value or the Page value both below half-filling and above half-filling. In the plots we also see that
the spectra do not show a single rainbow-like curve; instead they show a large spreading, which also indicates a non-thermalising behaviour. Thus for the full Hilbert space the quantum East model does not shows a thermalising behaviour for all fillings. 
\end{comment}

Another standard measure for checking the integrability~\cite{berry} or non-integrability~\cite{BGS} of a model is the nearest-neighbor energy eigenvalue correlation~\cite{levelstat1,Vikram}, namely, the level spacing statistics between consecutive energy eigenvalues in the sorted energy spectrum. However, the level spacing analysis of consecutive energy levels is often difficult in many-body quantum systems since it needs
an appropriate unfolding procedure, which requires us to know the exact form of the density of states~\cite{huse}. However, the extraction of the density of states is prohibitively hard in many cases. To circumvent this problem, it is preferable to analyze the spectral statistics in terms of the ratio of two consecutive level spacings of the sorted energy spectrum~\cite{huse,serbyn}. This is defined as $ r_n = \delta_{n+1} / \delta_n $, where $ \delta_n = E_{n+1} -E_n $ with $ E_n$ being the $n$-th energy eigenvalue. This ratio analysis does not require any information on the profile of the density of states in such systems, which makes it more convenient to study
than the level spacing statistics. Now, it is well-established that integrable systems follow the the Poisson distribution~\cite{huse,serbyn,berry} due to the lack of any level repulsion in the energy spectrum with the form
\begin{equation}
P(r) =\frac{1}{(1+r)^2},
\end{equation}
\begin{comment}
While if a model is non-integrable and is defined by a real Hermitian Hamiltonian, the distribution obeyed by $ r $ has the form of a
GOE given by
\end{comment}
while a non-integrable model in the presence of time-reversal symmetry obeys the GOE statistics~\cite{levelstat1,BGS,huse,serbyn} given by the random matrix 
theory of real Hermitian Hamiltonians as
\begin{equation}
P(r) = \frac{27}{8} \frac{(r + r^2)}{(1 + r + r^2)^{5/2}}.
\end{equation}

It is further convenient to introduce a variable $ \tilde{r}$ rather than 
$r$, which is given by 
\begin{equation}
\tilde{r}_n = \frac{{\rm min}(\delta_{n+1} , \delta_{n})}{{\rm max}(\delta_{n+1} , \delta_{n})} = {\rm min}( r_n , 1/r_n ).
\end{equation}
We note that  $ \tilde{r} $ follows the distribution $ P( \tilde{r} ) = 2 P(r)\,\theta(1-r)$, where $ \la\tilde{r}\ra= 0.536$ and $0.386 $ for the GOE and the Poisson distributions, respectively.

\begin{comment}
Here in our model we have analysed this level spacing distribution numerically. We have checked the level spacing distribution for both the full Hilbert space and within a particular fragment. For the analysis of the full Hilbert space we have chosen system size $ L = 16 $ and half-filling with OBC. To understand the level spacing distribution within a particular fragment we have chosen system size $ L = 18 $ and taken the largest fragment at the filling $\frac{L}{2} + 3$ with OBC. To numerically calculate the level spacing statistics of the consecutive energy levels for both cases, we have computed the energy spectrum and sorted the eigenvalues where, to resolve all the discrete symmetries of the system and avoid any accidental degeneracies, we have added some uniform on-site potential with disorder $ w = 0.01 $. We note that the presence of on-site disorder preserves the fragmentation structure of the model.
\end{comment}

\begin{figure}[h!]
    \centering
    \subfigure[]{
    \includegraphics[scale=0.5]{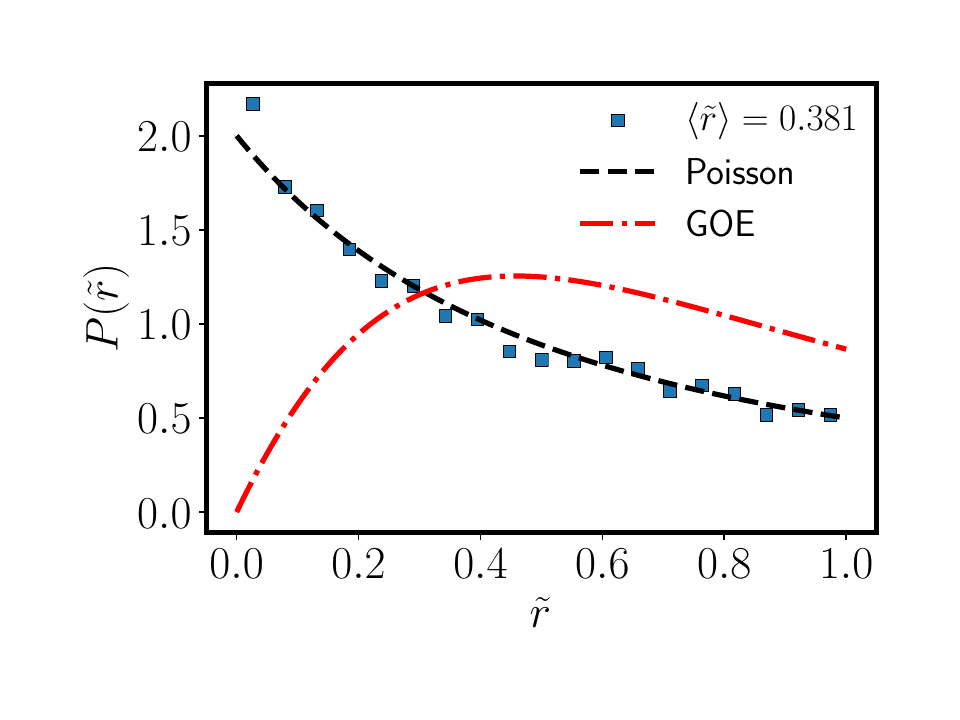}
    }
    \hfill
    \subfigure[]{
    \includegraphics[scale=0.5]{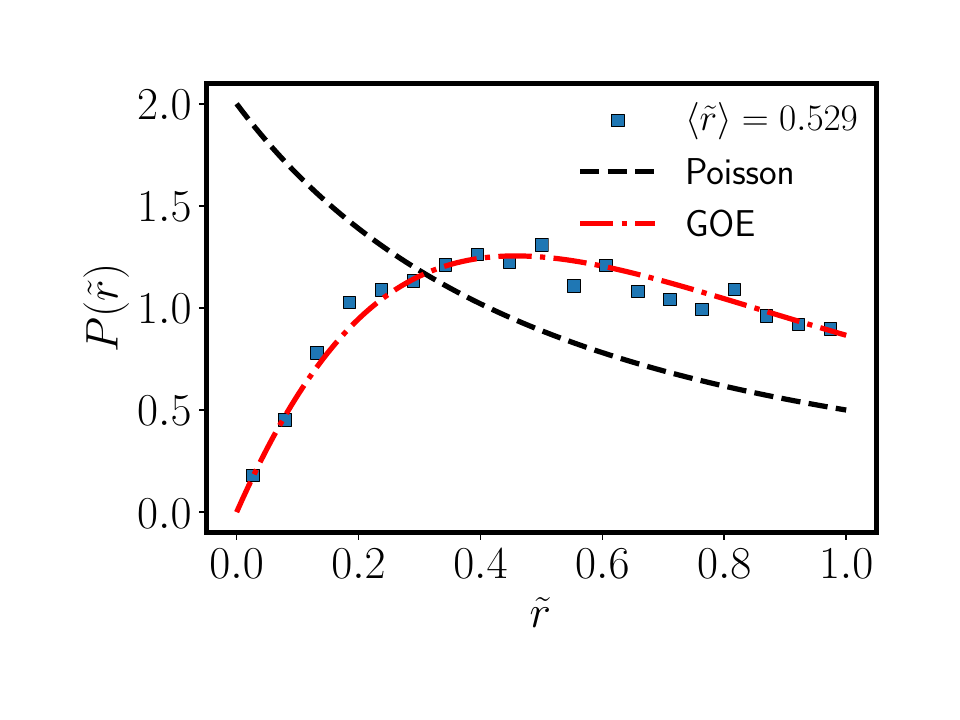}
    }
   \caption{(a-b) The distribution of consecutive level spacing ratios for the full Hilbert space at $N_{f}=L/2$ and for the largest fragment for $N_{f}=L/2+3$, for system sizes $L=16$ and $L=18$, respectively. We include a small random on-site potential with $w=0.01$ in order to break any accidental degeneracy and undesired discrete symmetries.
%; however, this addition preserves the fragmentation structure of our model. 
(a) We observe that the level spacing ratios follow the Poisson distribution for the full Hilbert space.
%as the exponentially many invariant subspaces prevent any level repulsion within the full Hilbert space. 
(b) The level spacing ratio within the largest fragment obeys the GOE distribution.} %which suggests non-integrability within the largest fragment.}
    \label{fig7}
\end{figure}

In Figs. \ref{fig7} (a-b), we have shown the distribution of consecutive level spacing ratios within the full Hilbert space for half-filling and within the largest fragment for filling $N_{f}=L/2+3$ for $L=16$ and $18$, respectively, with OBC. In both cases, we introduce a small diagonal disorder of strength $w=0.01$ in order to break any undesirable discrete symmetries as well as to circumvent any accidental degeneracies~\cite{Herviou,aditya_2024,HSF2_moudgalya_2020}, which can potentially create ambiguities in the level spacing analysis. We note that the presence of a diagonal disorder keeps the fragmentation structure of the East model intact. In Fig. \ref{fig7} (a), we observe that the level statistics of the full Hilbert space is indistinguishable from the Poisson statistics~\cite{Herviou,Kwan,aditya_2024} due to the presence of exponentially many invariant subspaces, which act as conserved quantities and thus prevent any level repulsion in the spectrum. On the other hand, Fig. \ref{fig7} (b) depicts that the largest fragment at $N_{f}=L/2+3$ demonstrates the GOE distribution~\cite{HSF2_moudgalya_2020,Kwan,Herviou,aditya_2024}, which implies that the model within the largest fragment is non-integrable, further reflecting the signature of a weaker version of the Krylov- or subspace-restricted ETH~\cite{aditya_2024,rev1,HSF2_moudgalya_2020} within this fragment. One should note that this weak violation of the ETH within sufficiently large fragments is supported by several exactly zero-entanglement eigenstates for specific entanglement cuts in the middle of the spectrum, as pointed out in Fig. \ref{fig6}. These zero-entanglement eigenstates within the sufficiently large fragments are candidates for quantum many-body scars~\cite{HSF2_moudgalya_2020,brighi_2023,subscar1,subscar2}, which will be
the central object for the next part of our discussion.

\begin{comment}
From Fig. \ref{fig7} (a) we can clearly see that for the full Hilbert space, the level spacing distribution of the consecutive energy eigenvalues matches with the Poisson distribution of $ \tilde{r} $. This indicates the behaviour of an integrable system. We can understand this since the exponentially large number of fragments are dynamically disconnected; hence they behave as a large number of conserved quantum numbers which forbid the level repulsion of the energy spectrum as we generally see in integrable models.

For the study of level repulsion within the largest fragment, we have followed the same steps as before keeping the filling $\frac{L}{2} + 3$ for system size $ L = 18$ with uniform on-site disorder potential $ w = 0.01$. In Fig. \ref{fig7} (b) we see that the consecutive energy level spacing distribution within the largest fragment matches with the GOE distribution of $ \tilde{r}$ which is generally found in non-integrable systems. So, finally we see that although the full Hilbert space of the quantum East model behaves like integrable systems, the behaviour of the largest fragment is like non-integrable systems.
\end{comment}

\section{Recursive quantum Hilbert space fragmentation and many-body scars}

\begin{comment}
In the literature of thermalization, quantum many-body scars (QMBS) are another interesting finding. If a system thermalises in the thermodynamic limit then the dimension of the largest fragment is generally comparable to the dimension of the full Hilbert space. But, we can have an exponentially smaller number of states which show persistent oscillation or revivals in the long-time dynamics of the system. These states are dynamically disconnected from the thermal blocks.
\end{comment}

As mentioned earlier, there are many exact zero-entanglement states 
%(the zero entanglement might vary from fragment to fragment) 
near the middle of the spectrum within almost all fully connected and sufficiently large fragments in the East model; this is also illustrated in Fig. \ref{fig6}. These zero-entanglement states demand a more thorough investigation since they show features 
similar to quantum many-body scars~\cite{QMBS1,QMBS2,QMBS3,udupa}. Scars have been extensively studied in models with kinetic constraints, for instance, the PXP model~\cite{QMBS1,rev2}. However, in our case, similar states emerge within HSF-induced fragments; hence we call them subspace-restricted scars. Such scars have 
been found before in other fragmented systems~\cite{subscar1,subscar2,brighi_2023,HSF2_moudgalya_2020}.
Before proceeding further, we note that quantum many-body scars are highly-excited eigenstates of non-integrable quantum many-body systems which violate the ETH in a weak
sense. This is a weak violation~\cite{ETH1,ETH2,ETH3} in the sense that almost all the eigenstates of such systems satisfy the ETH, except for the scar states~\cite{QMBS1,QMBS2,QMBS3} which form a set of measure zero in the thermodynamic limit. This partial violation of the ETH can significantly affects the non-equilibrium dynamics of such many-body systems~\cite{rev1,rev2}. For instance, quantities such as the Loschmidt echo and the occupation number starting from an initial product state which has a high overlap with such scar-like states show long-time persistent revivals under unitary time evolution. This is which is a prominent signature of the system retaining the memory of the initial state for a long period of time.

\begin{figure*}
\centering
\subfigure[]{\includegraphics[width=0.33\hsize]{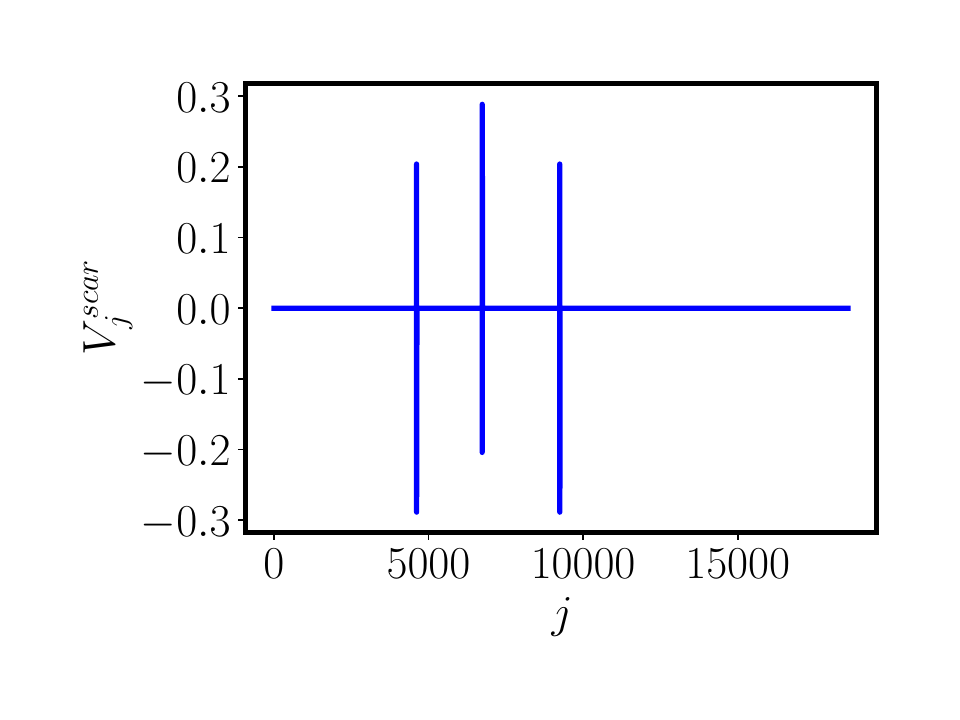}}%
\subfigure[]{\includegraphics[width=0.33\hsize]{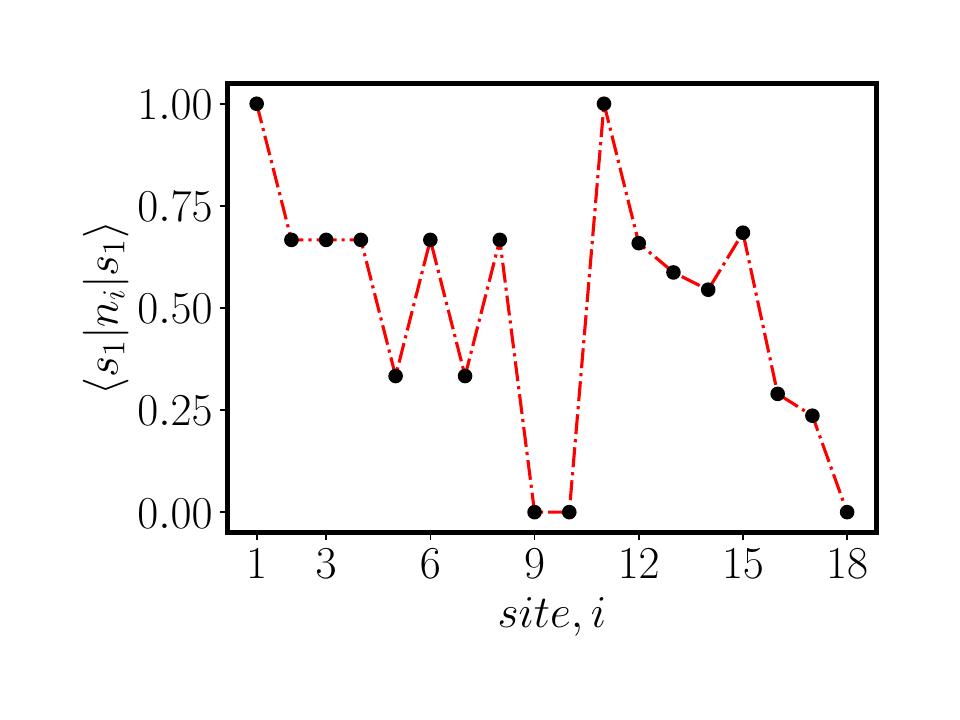}}%
\subfigure[]{\includegraphics[width=0.33\hsize]{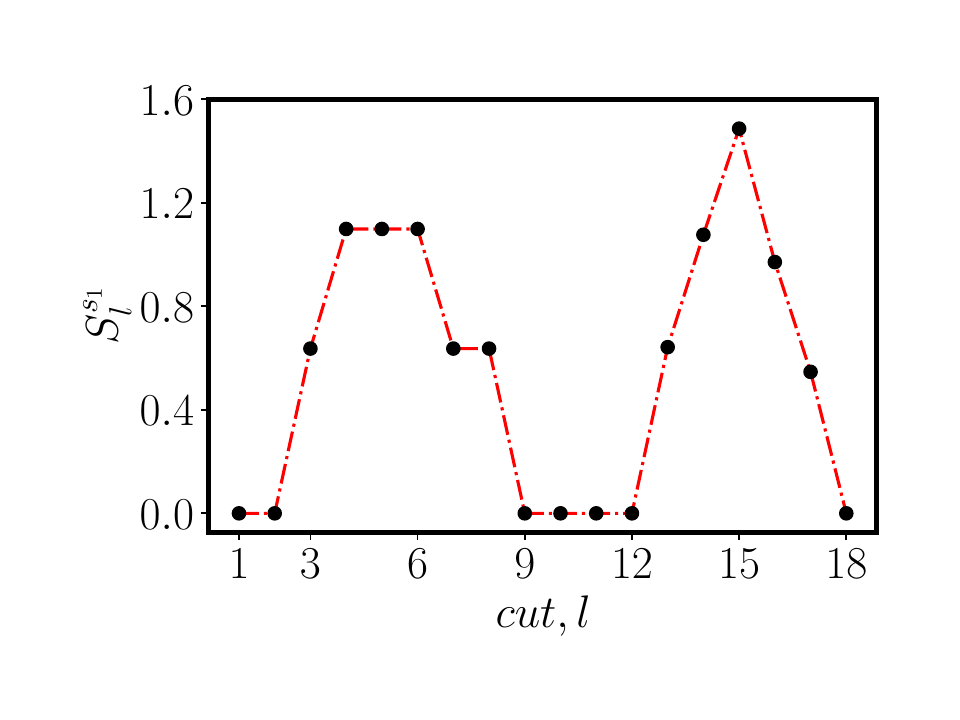}}%   
\caption{(a) Plot showing the structure of the wave function for a scar eigenstate within the largest fragment of $L=18$ and $N_{f}=L/2+3$ (the fragment contains
9996 states). This scar has a non-zero large overlap with exactly 21 Fock basis states, which is much smaller than the size of the full Hilbert space size. (b) The density profile of a scar eigenstate $\ket{s_{1}}$ versus the site index $i$ for a specific 
scar state within this fragment, which appears to have a state $\ket{00}\otimes\ket{1}$ at sites $i=9,10$, and $11$. (c) The entanglement entropy $S_{l}$ versus the position $l$ of the cut for the same scar state shows zero entanglement at $l=9,10, 11$, and $12$; this is the consequence of the specific product form of these scar states.}
\label{fig8}
\end{figure*}

To begin our analysis of scars, we observe that there are twelve exact zero-entanglement states in the middle of the spectrum for a bipartite entanglement cut within the largest fragment for $L=18$ and $N_{f}=L/2+3$ (this fragment has dimension $9996$); this is
shown in Fig. \ref{fig6}. This number is surprisingly small compared to the Hilbert space dimension of this fragment. 
% To further analyze the characteristics of these scar states, 
Next, we plot the wave functions of these states to investigate the number of Fock basis states with which these states have significant overlaps. In Fig. \ref{fig8} (a), we demonstrate the wave function of a scar eigenstate. We find, surprisingly, that it has
a non-zero overlap with only 21 basis states in the Fock space, showing that it has
the following separable form, $\ket{\phi_{L}}\otimes\ket{00}\otimes\ket{\phi_{R}}$, where $\ket{\phi_{L}}$ is either $\ket{11100110}$ (for the first 7 basis states), $\ket{11010101}$ (for the second 7 basis states) or $\ket{10111001}$ (the third 7 basis states). Furthermore, $\ket{\phi_{R}}$ are all the possible basis states for a subsystem with 8 sites and 7 particles having the constraint that the leftmost particle is strictly frozen in all the basis states. %Further, $\ket{\phi_{R}}$ can also be thought of as a non-interacting tight-binding model of a system with $L=7$, where a hole is hopping in the background of 6 particles.
In fact, it can further be shown that the left-restricted part of these scar eigenstates is $\ket{\phi_{L}'}=\frac{1}{\sqrt{3}}\left(\ket{11100110}-\ket{11010101}+\ket{10111001}\right)$; this is an eigenstate with zero eigenvalue of the left-restricted Hamiltonian with $L=8$ and $N_{f}=5$. Further, we note that this left-restricted eigenstate is a zero-energy eigenstate of a classical fragment labeled by the root state $10101111$. This left-restricted part cannot interact with the right part, $\ket{\phi_{R}}$ due to the intermediate blockaded region given by $\ket{00}$. Consequently, the dynamics of such a scar subspace essentially comes from the $\ket{\phi_{R}}$ part, which can be described by an effective Hamiltonian within the basis of seven product states, i.e., $\ket{11111110}$, $\ket{11111101}$, $\ket{11111011}$, $\ket{11110111}$, $\ket{11101111}$, $\ket{11011111}$, and $\ket{10111111}$. The effective Hamiltonian within this right-restricted part reduces to a single hole hopping in a background of $7$ particles (the leftmost particle is frozen) with OBC. The single-particle dispersion of this effective problem is given by $E_{p}=-2\cos (\pi p/8)$, where $p=1,2,\cdots,7$; this has the eigenvalues
$E=\pm 1.8476, \pm 1.414, \pm 0.7654,$ and $0$. We have then numerically verified that 6 out of the total 12 scar states within this fragment appear to have the same analytically obtained eigenvalues, except for the state with $E=0$. The zero-energy scar state does not
disappear, but it becomes numerically difficult to isolate it due to the presence of a large number of zero-energy thermal (i.e., non-scar) states which can hybridize with the scar state.
%, which can potentially lead to a 
%hybridization of this state with a thermal zero-energy eigenstate.
%as a consequence of numerical artifact. 
We note that it is absolutely crucial to consider appropriate entangled combinations of left-restricted basis states to understand the many-body scars described above. This bears a close analogy to quantum fragmentation~\cite{moudgalya_commu_2022}; however, 
the crucial difference with the usual quantum fragmentation lies in the fact that our scars occur within the classical fragments in a recursive manner~\cite{brighi_2023}. This feature was first demonstrated in the longer-range variant of the East model discussed in Ref. \onlinecite{brighi_2023}. 

In the fragment described above, we have identified another six scar states, whose left-restricted part is again a zero-energy eigenstate of the East model for a smaller subsystem with $L=9$ and $N_{f}=6$, and has the form $\ket{\phi_{L}'}=\frac{1}{2} (\ket{101111100}-\ket{110111010}-\ket{111100101}$ $+\ket{111010110} )$. This is a member of the classical fragment which has the root state, $101010111$. The left-restricted part is once again separated from the right-restricted part by a blockades part given by
$\ket{00}$. Finally, the right-restricted part only participates in the dynamics as in
the previous case, and the dispersion can be captured by a tight-binding model with $L=7$ and OBC, where a hole hops in the background of $6$ particles, with the leftmost particle being strictly frozen. The dispersion of the rightmost part is described by $E_{p}=-2\cos ( \pi p /7)$, where $p=1,2,\cdots,6$. The energy eigenvalues are 
therefore $\pm 1.2469, \pm 0.445$, and $\pm 1.8019$.

In Fig. \ref{fig8} (b), we show the density profile for a scar state as a function of the site index $i$. This illustrates that the scar state has a product part given by 
$\ket{00}$ exactly at $i=9$ and $10$. This makes the entanglement $S_{l}$ exactly
equal to zero if the bipartite cut is made at $l=9$ and $10$, as shown in Fig. \ref{fig8} (c). Fig. \ref{fig8} (b) also shows that the leftmost site of $\ket{\phi_{R}}$ is frozen at $n_{i}=11$; this again gives zero entanglement if a 
cut is made at $l=11$, as we see in Fig. \ref{fig8} (c).

\begin{comment}
In our quantum East model some more interesting things happen. We find some scar-like states within the largest fragment. Though the largest fragment shows a weakly thermalizing behaviour, from the entanglement spectrum we can see that there are some very low-entanglement states. To analyse whether they are scar-like or not, we have taken the eigenvectors corresponding to the low-entanglement states. Then we have chosen one of those eigenvectors and plotted its overlaps with all the states in the Fock space basis. This shows that the eigenvector has a large overlap with only a few basis states, while the overlap with the other basis states are close to zero. This behaviour is found to be the same for all the 
low-entanglement eigenvectors which we studied. For the system size $L = 18$ at filling $ \frac{L}{2} + 3$ we have the largest fragment where we have a total of 14 zero-entanglement states which has large overlap with 21 basis states (the numbers of these basis states are 4616, 4614, 9244, 6736, 6738, 9246, 9247, 6735, 9243, 6739, 4617, 4613, 4612, 4618, 6740, 9242, 6734, 9248, 9245, 6737, 4615), whereas the size of this largest fragment is $9996$. Further we have calculated the Loschmidt echo for these states which shows persistent oscillation even at large times. So we can say that these states retain their initial memories even in thermodynamic limit; this is a property of scar-like states.
\end{comment}

\begin{figure*}
\centering
\subfigure[]{\includegraphics[width=0.33\hsize]{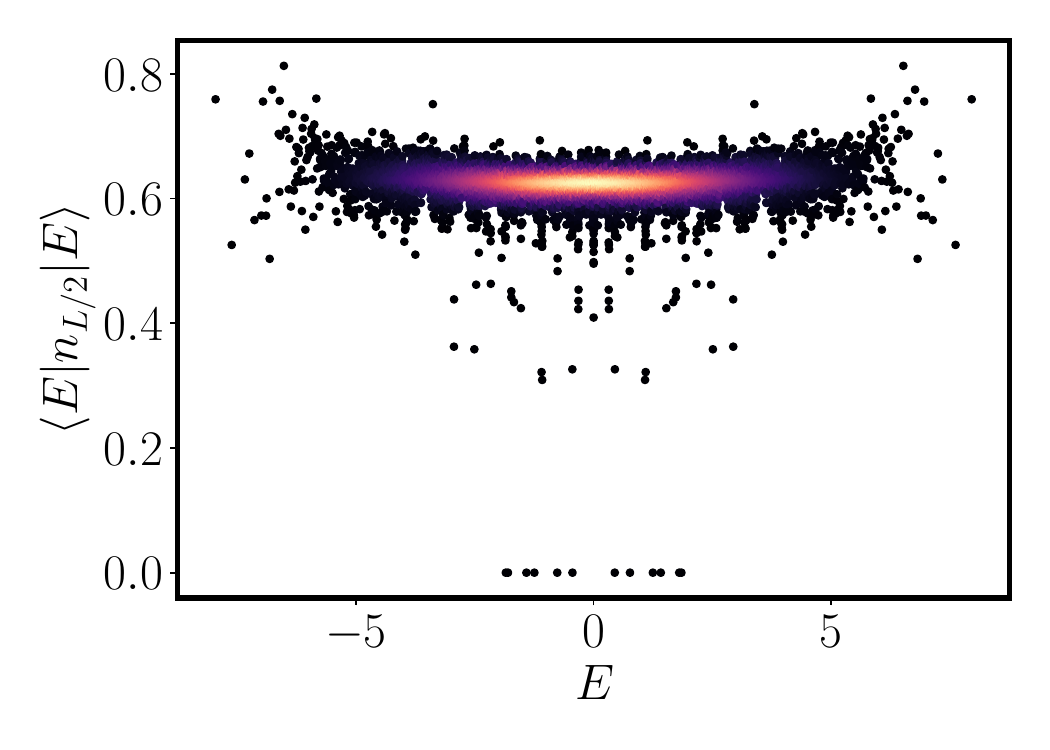}}%
\subfigure[]{\includegraphics[width=0.33\hsize]{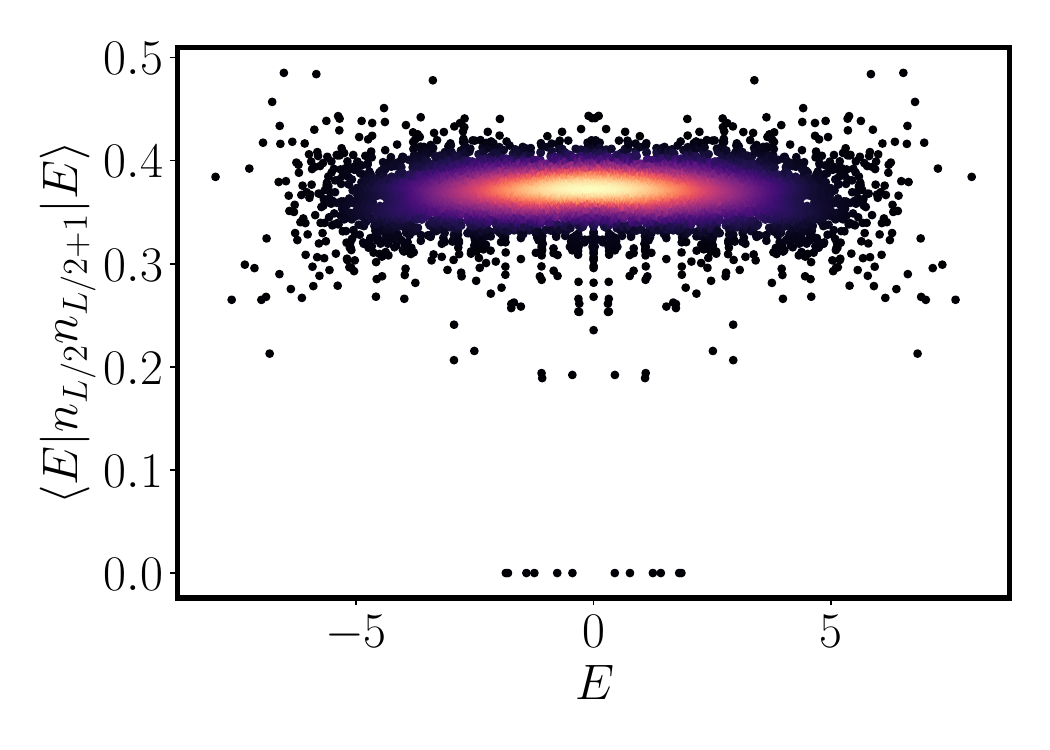}}%
\subfigure[]{\includegraphics[width=0.33\hsize]{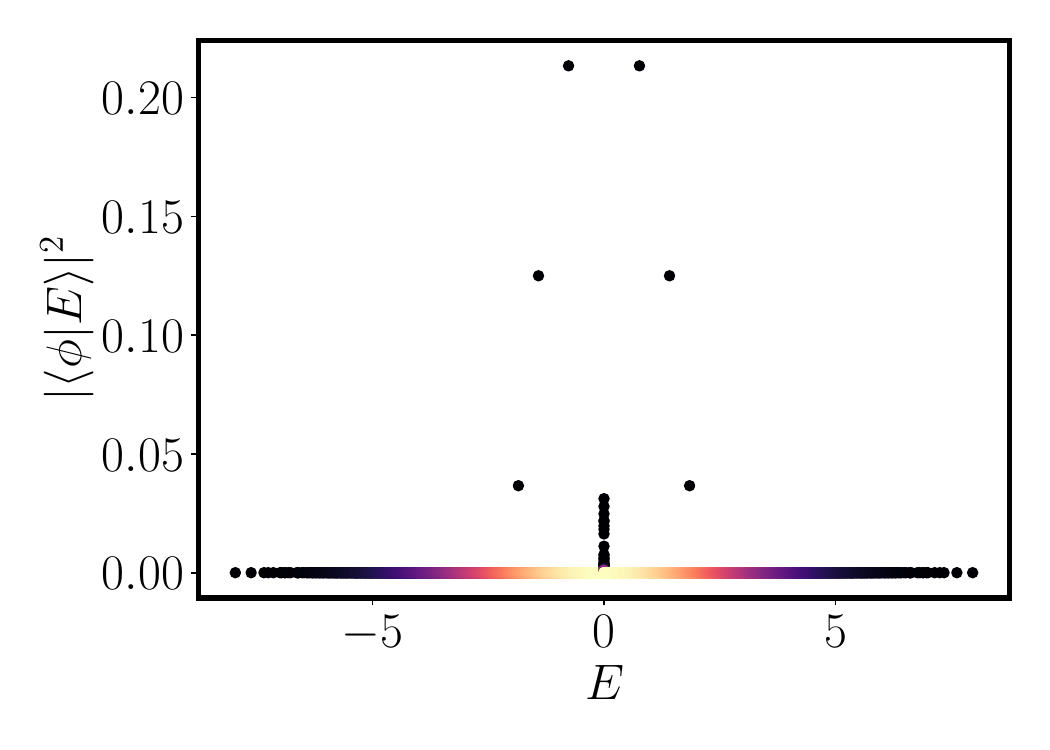}}%
\caption{(a-b) Plots showing the expectation values of $n_{L/2}$ and $n_{L/2}n_{L/2+1}$ in the eigenstates lying in the largest fragment for $L=18$ and $N_{f}=L/2+3$. (c) The overlap of a specific state $\ket{\phi}$ (whose left-restricted part is a zero-energy eigenstate of the Hamiltonian for the East model with 8 sites and 5 particles) 
with all other eigenstates within this fragment. 
Plots (a-b) demonstrate that both expectation values are smooth functions of $E$ except for a few eigenstates in the middle of the spectrum.
%; the latter states lead to a violation of the ETH. 
(c) We see that the state $\ket{\phi}$ has large overlaps with only six eigenstates which have $E\neq 0$. In addition, this state has  
small overlaps of the order of $10^{-2}$ with a number of states at $E=0$.}
\label{fig10}
\end{figure*}

In Figs. \ref{fig10} (a-b), we examine the expectation values of the few-body observables $n_{L/2}$ and $n_{L/2}n_{L/2+1}$ in all the eigenstates within the largest fragment for $L=18$ and $N_{f}=L/2+3$ (this fragment has 9996 states). The expectation values of these observables appear to be a smooth function of the energy for almost all the eigenstates, except for a few outlying states near the middle of the spectrum which are the scar eigenstates within this fragment; these states lead to a weak violation of
ETH. Furthermore, $\la s_{j}| n_{L/2}|s_{j}\ra $ and $\la s_{j}| n_{L/2}n_{L/2+1}|s_{j}\ra $ turn out to be zero for the scar eigenstates, where $s_{j}$ denotes 
the $j$-th scar state. The zero expectation values can be explained from the structures of these states, which has the form $\ket{00}$ on the sites $j=L/2=9$ and $j=L/2+1=10$; this leads to the zero expectation values for these states. In Fig. \ref{fig10} (c), we consider the state 
\bea \ket{\phi} &=& \frac{1}{\sqrt{3}}\left(\ket{11100110}-\ket{11010101}+\ket{10111001}\right) \non\\
&& \otimes \ket{00} \otimes \ket{11111110} \eea 
with which the scar states have a large non-zero overlap, and we show the overlap amplitudes of this state with all the eigenstates as a function of $E$ within the fragment; the figure is consistent
with what we see in Figs. \ref{fig10} (a-b). We then note that there are exactly six eigenstates with $E=\pm 1.8478, \pm 1.414, \pm 0.7654$, which have a large overlap of 
order 1, as expected from our previous discussion. These sudden jumps in the overlap amplitudes suggest that this quantity is not a smooth function of $E$. This is not a behavior expected from the ETH~\cite{rev1,rev2,ETH1,ETH2,ETH3}, and these six eigenstates therefore violate the ETH~\cite{rev1,rev2,ETH1,ETH2,ETH3} within the fragment. At the same time, we observe another bunch of few eigenstates with $E=0$ showing a small overlap of the order of $10^{-2}$ with the given state. %This observation also implies that these scars are not perfect scars due to the fact that this eigensubspace is fully protected from the thermal subspace. This further indicates that the subspace will eventually thermalize in the thermodynamic limit; also, the persistent oscillation in the dynamics will eventually decay due to the mixing with the thermal subspace, which we will examine in the next part of our discussion. 
This is probably due to the zero-energy eigenstate of the scar subspace, which gets hybridized with other thermal zero-energy states during numerical diagonalization. Accordingly, these zero-energy states also demonstrate a small overlap with this given state, as can be seen from Fig. \ref{fig10} (c). Furthermore, we also note that the recursive quantum-HSF-induced scars~\cite{brighi_2023} might not always show zero entanglements for cut at the center of the system; we have shown this in Appendix \ref{appC} within a relatively smaller classical fragment with $L=14$ and $N_{f}=L/2+1$. In this fragment, the scar states exhibit zero entanglement for a cut at
the site $L/2+2$. 

As we have mentioned earlier, a recursive quantum HSF has been introduced in Ref. \onlinecite{brighi_2023} in the East model for a longer-range variant with 
$r=2$ (our model corresponds to $r=1$). While recursively building the left-restricted part of the scar states, Ref. \onlinecite{brighi_2023} presented a necessary condition that the expectation value of $n_{l}$ has to be zero where $l$ is the rightmost site of the left-restricted part for the model with $r=2$. However, our analysis for the $r=1$ model demonstrates that the only necessary condition for this mechanism to occur is that the left-restricted part $\ket{\phi_{L}'}$ and the right-restricted part $\ket{\phi_{R}}$ should not interact with each other due to a blockaded region consisting of $0$'s residing in between the two parts whose length is $\geq (r+1)$. In addition, the other necessary condition is the left-restricted entangled state should be the 
zero-energy eigenstate of the left-restricted classical fragment of the East model, as was mentioned in Ref. \onlinecite{brighi_2023}. We have explicitly verified this fact using our analytical construction. We also discuss the robustness of these scar states for different perturbations in Appendix \ref{appG}. In that analysis, we consider three different perturbations, namely, a random on-site potential $\mu_{j}$, a random hopping $J_{j}$, and a nearest-neighbor density-density interaction $V$; each of these keeps the fragmentation structure of the Hilbert space intact. We show in
Appendix \ref{appG} that
while these scar states get perturbed only gradually by the effects of $\mu_{j}$ and $V$, they vanish completely when a small random hopping is introduced.

\section{Dynamical signatures of the lack of thermalization in the East model}
\label{DynQE}

We will now discuss the dynamical signatures of the lack of thermalization due to various mechanisms, namely, the structure imposed by HSF and the presence of subspace-restricted quantum many-body scars.

\label{sec7}
\subsection{Dynamics due to subspace-restricted quantum many-body scars}

\begin{figure}[h!]
    \centering
    \subfigure[]{
    \includegraphics[scale=0.45]{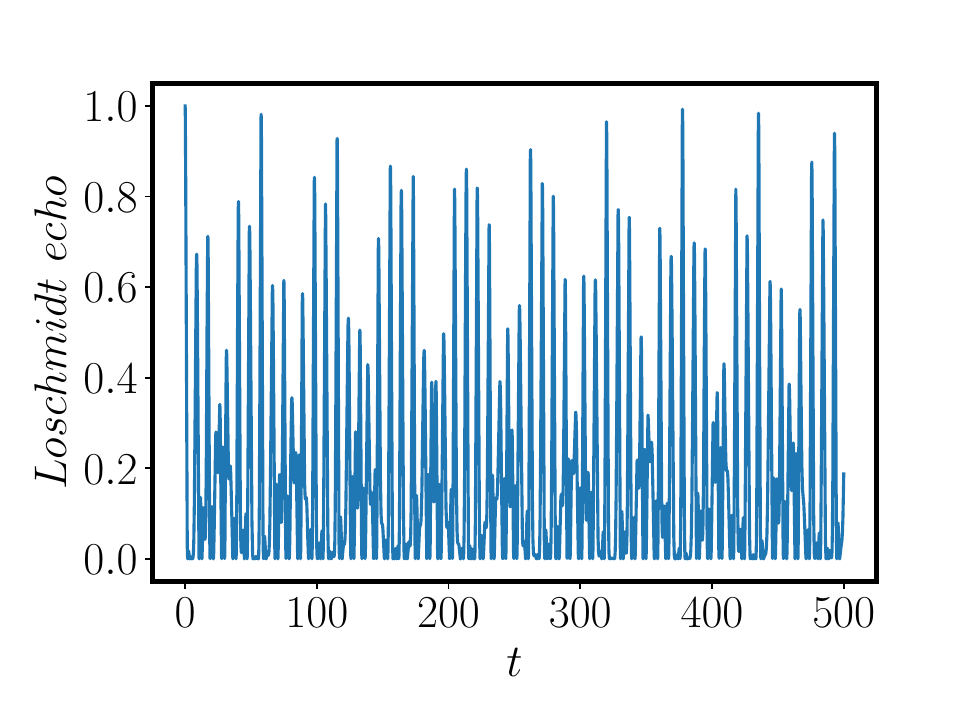}
    }\\
    \subfigure[]{
    \includegraphics[scale=0.45]{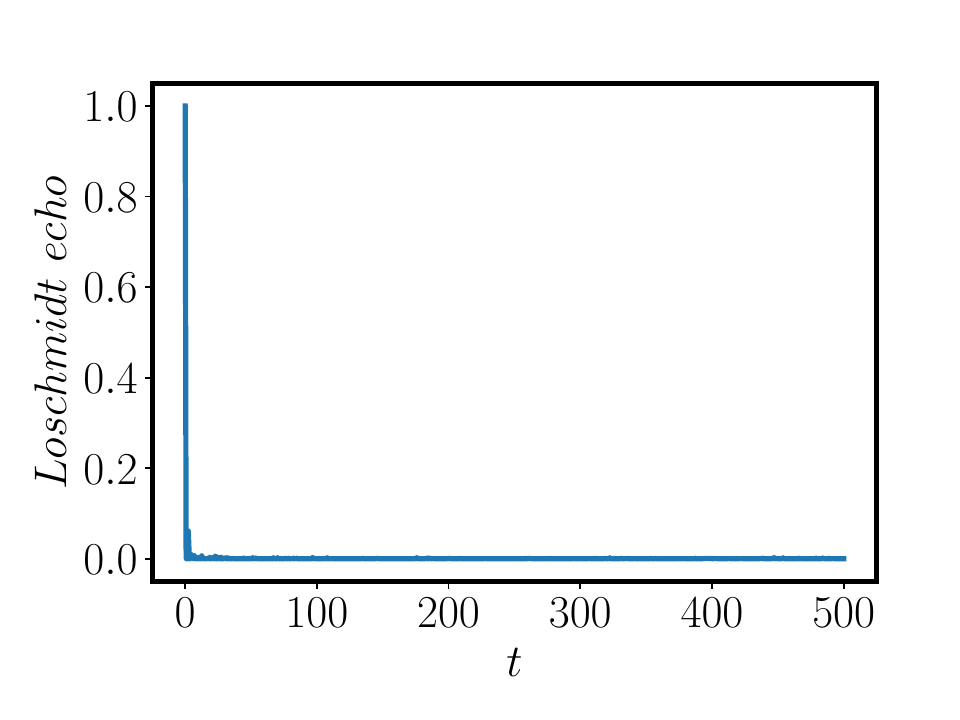}
    }
\caption{(a) Dynamics of the Loschmidt amplitude with time, starting from a basis state with which the scar eigenstates have large non-zero overlaps. We choose the initial state to have the form $\ket{\phi_{L}}\otimes\ket{00}\otimes\ket{10111111}$, where $\ket{\phi_{L}}=\frac{1}{\sqrt{3}}(\ket{11100110}-\ket{11010101}$ $+\ket{10111001})$. (b) Dynamics of the Loschmidt amplitude, starting from the root state of the fragment which does not have a significant overlap with the scar states. We consider the largest fragment for $L=18$ and $N_{f}=L/2+3$ with OBC for both cases. (a) In the first case, the Loschmidt amplitude exhibits persistent oscillations for a long time; however, the amplitude never reaches unity, suggesting that these subspace-restricted scars are not perfect scars. (b) In the second case, the Loschmidt amplitude quickly decays to zero, which is a signature of thermalization within this fragment.}
\label{fig9}
\end{figure}

We will start by examining the dynamical signatures of quantum
many-body scars~\cite{QMBS1,QMBS2,QMBS3,rev2,rev1} within sufficiently large non-integrable fragments. In Fig. \ref{fig9} (a), we show the dynamics of the Loschmidt amplitude, given by $|\la\psi|\psi(t)\ra|^2$, where the initial state is taken to be 
\bea \ket{\psi} &=& \frac{1}{\sqrt{3}}\left(\ket{11100110}-\ket{11010101}+\ket{10111001}\right) \non \\
&& \otimes\ket{00}\otimes\ket{11111110} \eea
with which the scar eigenstates have large non-zero amplitudes. As shown in Fig. \ref{fig9} (a), the Loschmidt amplitude exhibits long-time revivals; this is similar to what is seen in 
the case of scar states. However, it is difficult to compute the period of revival for this case since there are multiple energy scales arising from to the diagonalization of this scar subspace as discussed before.
%; this makes the extraction of the period of oscillation not so straightforward. 
As a comparison, we show in Fig. \ref{fig9} (b) the dynamics starting from the root state for the same fragment, i.e., $\ket{10\cdots101\cdots1}$ which is not a
part of the scar subspace; this shows a strikingly distinct behavior from Fig. \ref{fig9} (a). In this case, we see that the Loschmidt echo quickly goes to zero, clearly demonstrating thermalization.

\begin{figure}[h!]
\includegraphics[width=0.85\columnwidth]{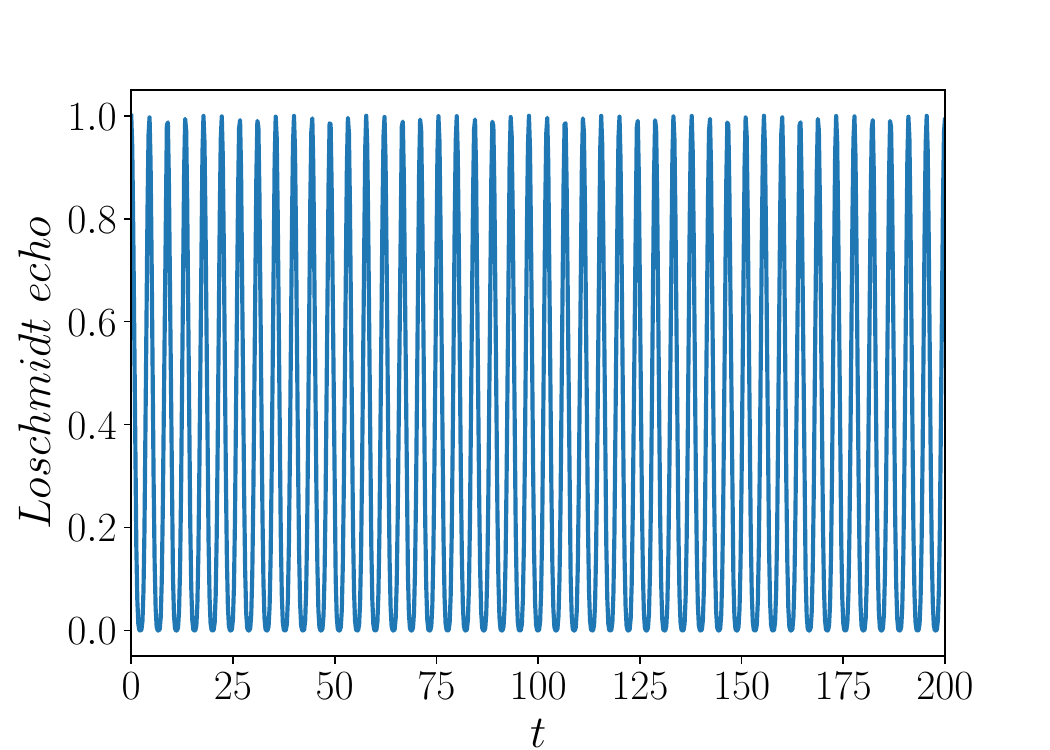}
    \caption{Plot of the Loschmidt echo versus $t$ for the same fragment as shown in Fig. \ref{QHSFobs}. We start from an initial state with a form $\ket{\phi_{L}}\otimes\ket{00}\otimes \ket{1110}$, where the left-restricted part is chosen the same as in Fig. \ref{fig9}. This initial state has large overlaps with the scar states, which results in persistent oscillation in the dynamics with a period of oscillation $T=\sqrt{2}\pi$.}
\label{QHSFdynamics}
\end{figure}

We also examine the Loschmidt echo dynamics within another comparatively smaller fragment for $L=14$ and $N_{f}=8$, starting with an initial state 
\bea | \psi \ra &=& \frac{1}{\sqrt{3}}\left(\ket{11100110}-\ket{11010101}+\ket{10111001}\right) \non \\
&& \otimes\ket{00}\otimes\ket{1110}. \eea
This state has a strong overlap with the scar states within the invariant subspace discussed in Appendix \ref{appC}. 
We see perfect revivals with a period of oscillation $T=\sqrt{2}\pi$, as shown in Fig. \ref{QHSFdynamics}. In this case, one can readily understand the period of oscillation analytically as follows. The effective Hamiltonian within this scar 
subspace is given by
\bea
H=\begin{pmatrix}0& 1 &0 \\1& 0&1 \\ 0&1 &0\end{pmatrix}, \label{effH}
\eea
where the basis states in the right-restricted part are given by $\ket{1110}$, $\ket{1101}$, and $\ket{1011}$. The eigenvalues of Eq. \eqref{effH} are $\sqrt{2}$, $0$, and $-\sqrt{2}$, and the corresponding eigenvectors are \bea
\psi_{1}=\frac{1}{2}\begin{pmatrix}
    1 \\ \sqrt{2} \\1
\end{pmatrix},~~ \psi_{2}=\frac{1}{\sqrt{2}}\begin{pmatrix} 1 \\0 \\-1\end{pmatrix},~~ \psi_{3}=\frac{1}{2}\begin{pmatrix} 1 \\ -\sqrt{2} \\1\end{pmatrix},
\non\\
\eea
respectively. In terms of these three eigenstates, the time-evolved state can be recast as 
\beq | \psi(t) \ra =\frac{1}{2}\left(\psi_{1}e^{-i\sqrt{2}t}+\psi_{3}e^{i\sqrt{2}t}+\sqrt{2}\psi_{2}\right). \eeq
Hence, the Loschmidt echo as a function of $t$ can be simplified to $\frac{1}{4}\left(\frac{3}{2}+2\cos(\sqrt{2}t)+\frac{1}{2}\cos(2\sqrt{2}t)\right)$ after some analytical manipulations; this expression demonstrates a perfect revival pattern with 
a period of oscillation given by $T=\sqrt{2}\pi$. This is exactly what we
see in Fig. \ref{QHSFdynamics}.

\subsection{Anomalous behaviors of bulk and boundary autocorrelation functions due to HSF}

\begin{figure*}
    \centering
    \subfigure[]{
    \includegraphics[width=0.33\hsize]{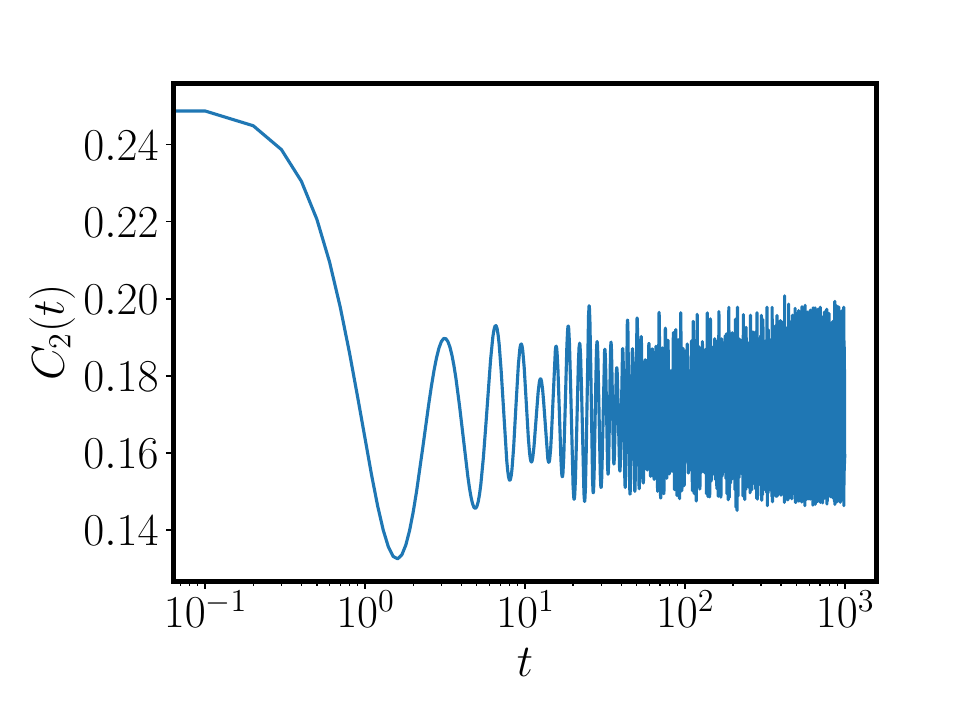}
    }%
    \subfigure[]{
    \includegraphics[width=0.33\hsize]{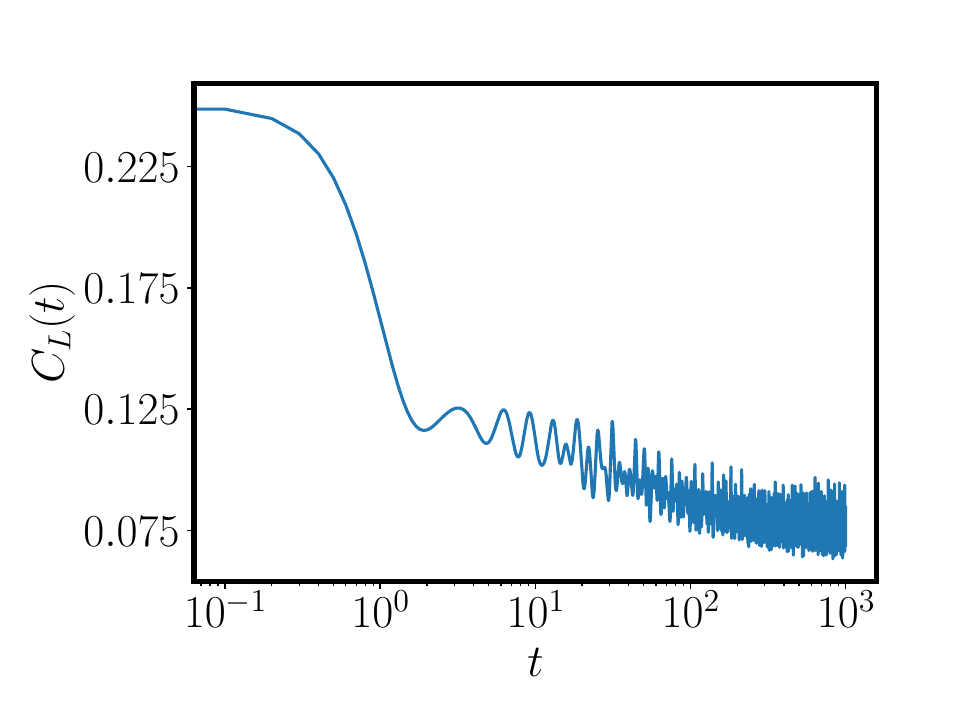}}%
\subfigure[]{\includegraphics[width=0.33\hsize]{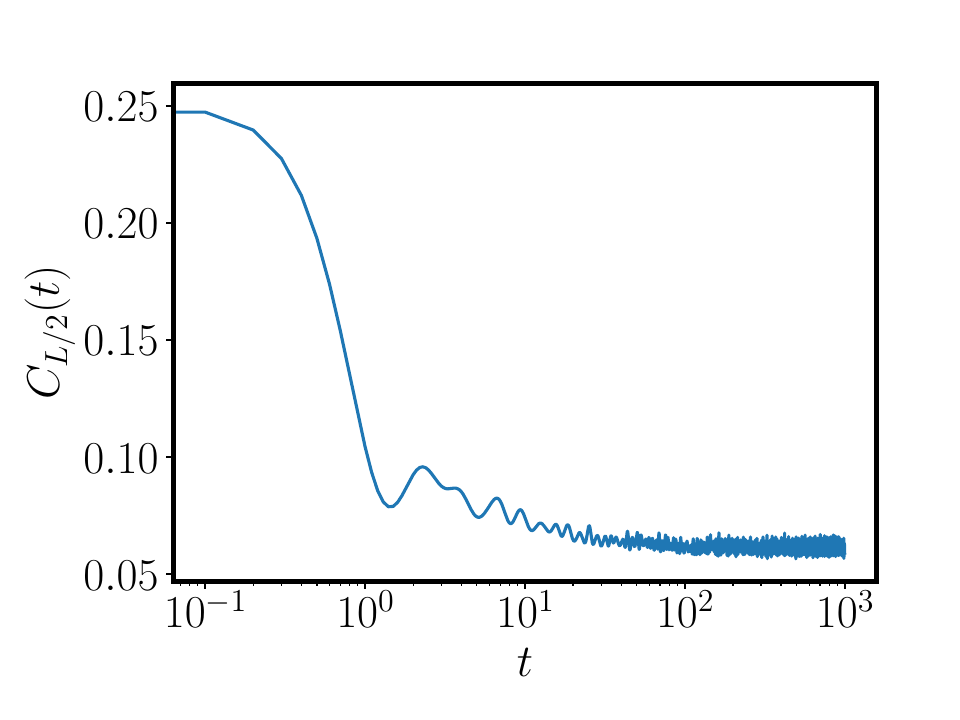}}
\caption{(a-c) Plots depicting the behaviors of $C_{2}$, $C_{L}$ and $C_{L/2}$ with $t$ starting from a random initial state. (a) $C_{2}(t)$ oscillates around a finite saturation value ($\sim 0.17$) at long times. (b) The right boundary site, $C_{L}(t)$, demonstrates a distinct behavior from $C_{2}$. We see a gradual decay to a saturation value ($\sim 0.09$) which is much smaller compared to the left boundary site. 
This is a signature of broken inversion symmetry. (c) The autocorrelation in the
bulk, $C_{L/2} (t)$, depicts a similar behavior as $C_{L}(t)$. However, the saturation value ($\sim 0.05$) is much smaller compared to the two boundary sites.}\label{fig11}
\end{figure*}

In this section, we will demonstrate how the fragmentation structure in this model 
gives rise to anomalous behaviors of the bulk and boundary autocorrelation functions.
This includes a non-uniform profile of the long-time saturation values of 
autocorrelators near the edges of the chain; this has also been observed earlier in the literature~\cite{stat_loc_2020,moudgalya_commu_2022,HSF2_moudgalya_2020,Li_2023,hart_2024}. This behavior is more prominent in this model due to two reasons. The first one is the lack of inversion symmetry, which creates an asymmetry between the behaviors at the
two boundaries (left and right), and the second is the strong-to-weak fragmentation transition~\cite{weak2}.

To address the first question, we first examine the behavior of the unequal-time autocorrelation functions
\bea
C_j(t)=\bra{\psi_0} (n_j(t) - 1/2 ) (n_j(0) - 1/2 )\ket{\psi_0},
\eea
starting from a random typical initial state, given by $ \ket{\psi_0} = \sum_j r_j \ket{\phi_j}$, where $r_{j}$ 
are numbers chosen from a uniform distribution in $[-1,1]$ with zero mean (satisfying the 
normalization condition $\sum_{j}|r_{j}|^2=1$), and 
$\ket{\phi_{j}}$'s are the Fock basis states. In Figs. \ref{fig11} (a-c), we plot the long-time behavior of autocorrelation functions for site $l=2$, $l=L$, and $l=L/2$, respectively, for $L=20$ at half-filling with OBC. In Fig. \ref{fig11} (a), we observe that $C_{2}(t)$ oscillates around a finite value of $\sim 0.17$, unlike a thermalizing system, where we expect a saturation value of $C_{2}(\infty) = 0$ at half-filling. (We do not study the autocorrelation $C_1 (t)$ at the leftmost site since the number at that site remains frozen for all times).
To analytically understand the finite saturation value of the autocorrelation function at the site $j=2$, we consider the first three sites of an infinitely long chain undergoing transitions allowed by the East constraint; these three sites can include eight states in the number basis, $111$, $110$, $011$, $101$, $100$, $010$, $001$, and $000$. Therefore, the probability for the occurrence of each of these basis states in a typical thermal state is $1/8$ in this three-site approximation.
Furthermore, the finite saturation value of the autocorrelator at $j=2$ can be supported by those states out of these eight states, which remain dynamically frozen at the site $j=2$, regardless of the occupation numbers on any other sites of the chain. Taking this fact into consideration, we see that only five states, i.e., $000$, $001$, $100$, $010$, and $011$ host dynamical frozen configuration at $j=2$. This gives a saturation value 
equal to $1/8\times1/4\times5 \simeq 0.156$, which is quite close to our numerically obtained mean saturation value. However, 
the other boundary in the East model behaves in a completely distinct manner, as shown in Fig. \ref{fig11} (b), due to the lack of inversion symmetry. In this case, we see that $C_{L}(t)$ initially shows a sharp transient decay; following this, it gradually decays to a finite saturation value 
of about $0.09$ around which it oscillates at long times. The saturation value at the rightmost site, $C_{L}(\infty)$, is much smaller than at the leftmost site. In Fig. \ref{fig11} (c), we demonstrate the behavior of the autocorrelation function inside the
bulk, $C_{L/2}(t)$, which again shows a decay profile similar to the rightmost site. However, the long-time saturation value of $\sim 0.05$ for this case is smaller compared to the two boundary sites, but it is clearly positive  unlike a thermalizing system. This thus suggests that apart from distinct bulk and boundary profiles of autocorrelators, which are typically observed in fragmented systems, the two boundaries also illustrate quite different features due to the lack of inversion symmetry in this model; to the best of our knowledge, this feature has not observed in any other models in the context of HSF.

\begin{figure}[h!]
    \centering
    \subfigure[]{
    \includegraphics[width=0.75\hsize]{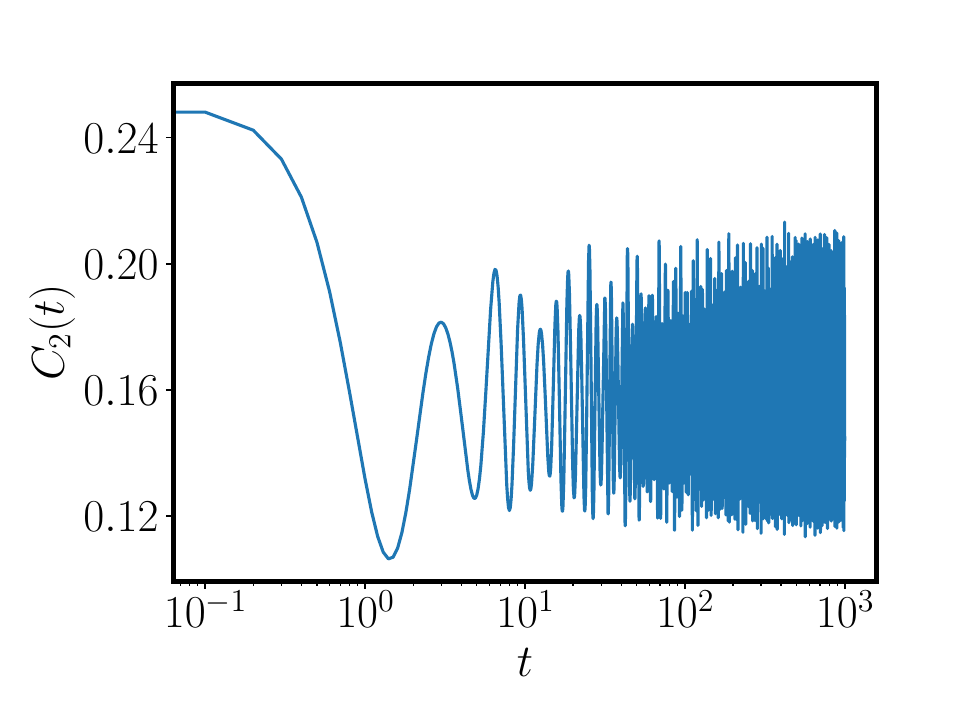}
    }\\
    \subfigure[]{
    \includegraphics[width=0.75\hsize]{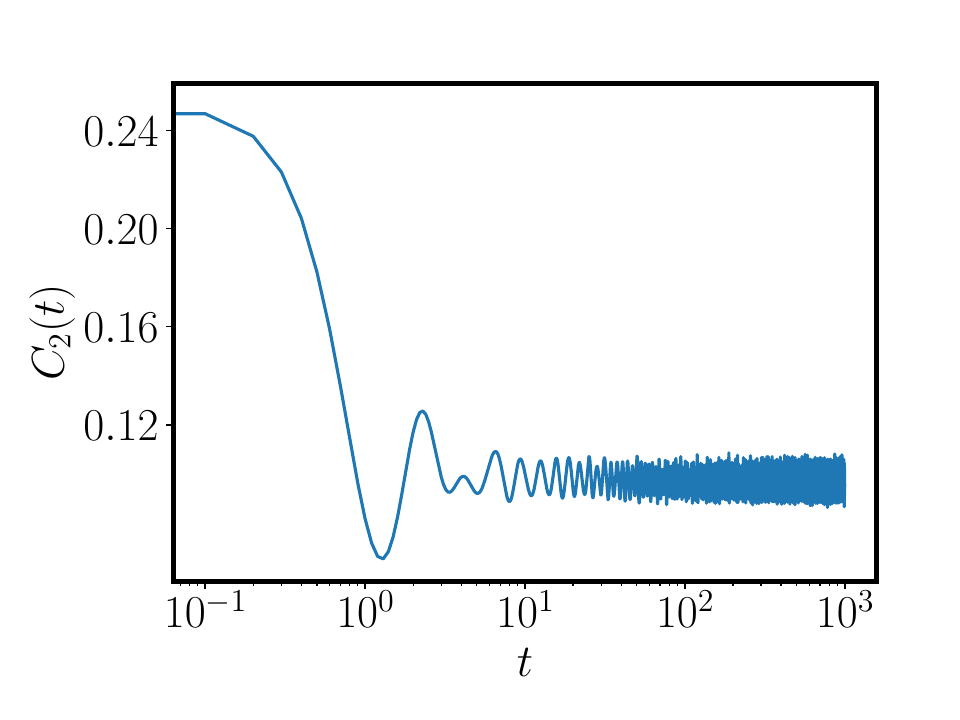}}\\

\caption{(a-b) The behavior of $C_{2}(t)$ with $t$ starting from a random initial state for $L=20$ for $N_{f}=L/2-2$ and $L/2+3$, respectively, with OBC. $C_{2}(t)$ for the first case oscillates around a saturation value ($\sim 0.16$), which is much larger than the the value found in the latter case ($\sim 0.1)$. These two distinct behaviors of long-time saturation value of left boundary autocorrelators again illustrate that there is a transition from the strong-to-weak fragmentation as one moves from below to above half-filling in this model. }\label{fig111}
\end{figure}

To address the second question related to the strong-to-weak fragmentation as the
filling is varied, we examine the boundary autocorrelation function $C_{2}(t)$ below and above half-filling for a system with $L=20$, for $N_{f}=L/2-2$ and $N_{f}=L/2+3$, respectively. This is shown in Figs. \ref{fig111} (a-b). In both cases, the initial state is chosen to be a typical random initial state, similar to the one
considered in Fig. \ref{fig11}. Furthermore, the filling above half-filling,
$N_f = L/2 +3$, is chosen such that the largest fragment within the full Hilbert space for $L=20$ lies at this filling. As demonstrated in Figs. \ref{fig111} (a-b), the different fragmentation structures have a profound effect on the long-time saturation values of the boundary autocorrelators below and above half-filling. In Fig. \ref{fig111} (a), we see that the saturation value ($\sim 0.16$) about which $C_{2}(t)$ oscillates around is much larger than the case above half-filling ($\sim 0.1$) which
is shown in Fig. \ref{fig111} (b). The larger saturation value of $C_{2}$ is purely a manifestation of the strong fragmentation structure, where the largest fragment growth is exponentially smaller than the full Hilbert space size, while in the latter case, the growth of the largest fragment is the same as that of the full Hilbert space
(up to powers of $L$). Another remarkable thing to note is that the Hilbert space size for the first case is $125970$, whereas in the second case, the Hilbert space size is $38760$. It is generally observed that the amplitude of late-time oscillations becomes more sharply well-defined with increasing system sizes; however, we see exactly the
opposite feature in our model. This suggests that there are exponentially many disconnected fragments which govern the long-time behavior of $C_{2}(t)$. In addition, the amplitude of oscillations in the first case being much larger than in 
the second case again confirms that the dimension of the largest fragment is exponentially smaller than the full Hilbert space size.

%\subsection{Dynamical behavior of the scar-like states}
%We have mentioned previously that for this quantum East model, we have found some scar-like states within the largest fragment. We have tried to understand their behavior and dynamical properties. To do so we have chosen system size $ L = 18 $ with OBC we got 14 scar-like states with zero entanglement entropy which has large overlap on 21 basis states ( these basis states numbers are 4616, 4614, 9244, 6736, 6738, 9246, 9247, 6735, 9243, 6739, 4617, 4613, 4612, 4618, 6740, 9242, 6734, 9248, 9245, 6737, 4615). Then, we studied the Loschmidt echo of a particular basis state among these states after choosing a scar state.

\section{Effects of nearest-neighbor density-density interactions with finite strength}
\label{sec8}

\begin{comment}
We will now discuss the effects of interactions in the quantum East model. The motivation behind this is that while studying the ground state properties of the model, we saw that the filling shifts away from half-filling as $\sqrt{L}$ for system size $L$. We would like to stabilize the ground state at a particular filling by adding interactions. We can add many kinds of interactions but we have chosen to have nearest-neighbour density-density interaction since this is the simplest and most easily realizable in experiments involving cold atom systems with tunable interactions. For example, the large $V$ limit of Rydberg atom chains show van der Waals interactions which create a Rydberg blockade in certain limiting cases.
\end{comment}

The robustness of the fragmentation structure under perturbations~\cite{robust1,robust2,robust3} is, in general, an interesting question to address and has been studied extensively in recent times. The rich behavior of the fragmentation structure in the East model, namely, weak and strong fragmentation for different fillings~\cite{weak2}, and several fragmentation-induced anomalous properties~\cite{brighi_2023} (for instance, the filling fraction for the ground state shifting with the system size, and the presence of quantum many-body scars within 
different fragments) motivates us to examine the effects of perturbations in this model. 
In this direction, a commonly studied perturbation, which can be realized in various experimental platforms~\cite{exp1,exp2,exp3,exp4} (for instance, the Rydberg blockade limit of a Rydberg simulator) is a uniform nearest-neighbor density-density interaction
with strength $V$. As we will discuss below, this perturbation can give rise to some intriguing features in this fragmented system.
% albeit the fact that this perturbation has a very simple structure.

\subsection{Effects of finite-$V$ on the ground state properties}

\begin{figure}[h!]
\centering
\includegraphics[width=0.95\columnwidth]{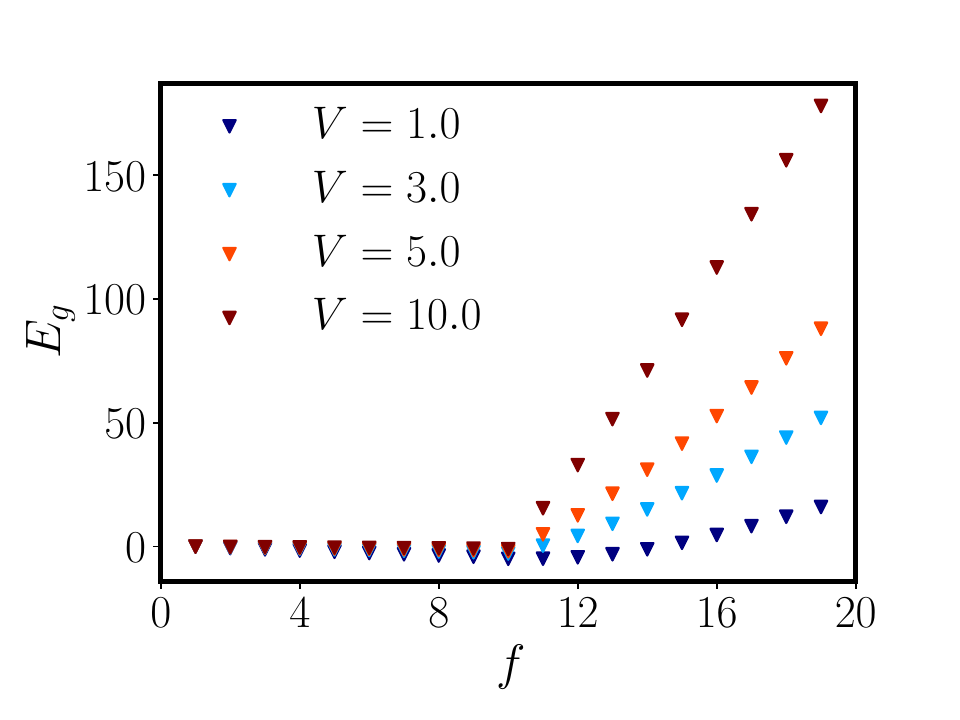}
\caption{Plot showing the ground state energy as a function of filling for the 
East model with a density-density interaction for several values of $V$, for $L=20$ with PBC. Contrary to Fig. \ref{fig4} which corresponds to the East model with $V=0$, the ground state for the interacting system stabilizes at $N_{f}=L/2$ 
%when interactions are introduced this case due to the introduction of a finite-$V$ term 
beyond a threshold value of $V$.} \label{fig17}
\end{figure}

The model under consideration has the Hamiltonian
\bea
H~=~\sum_{j} n_{j-1}\left(c_{j}^{\dagger}c_{j+1}+{\rm H.c.}\right)\,+\,V\sum_{j} n_{j}n_{j+1},\non\\
\label{vmode}\eea
where 
%the nearest-neighbor hopping is only allowed between sites $j$ and $j+1$ when the site $j-1$ is occupied, which is the so-called East constraint, 
$V$ is the strength of the nearest-neighbor density-density interaction. We note that an interaction with a finite strength does not change the fragmentation structure of the East model because it is diagonal in the number basis. However, we will see later that taking the limit $V \to \infty$ drastically changes the fragmentation structure of the model.
We will first study the effects of a finite $V$ on the ground state properties of the model.
%to draw comparisons with those of the East model in the absence of this term. 
We have seen earlier that the East model without interactions has some unusual properties for the ground state, namely, the filling for the ground state shifts with increasing system sizes, as shown in Fig. \ref{fig4}.
%, which we also capture analytically using the asymptotic properties of the Catalan number~\cite{catalan}. 
However, we find that shift in the ground state filling becomes stabilized at half-filling after a threshold value of $V$. In Fig. \ref{fig17}, we show the ground state energy versus filling for $L=20$ with PBC for several values of $V$. We note that the ground state energy shifts towards half-filling as $V$ increases; eventually it stabilizes at half-filling. The reason for the stabilization of the ground state at half-filling can be qualitatively understood as follows. Although interactions with finite $V$ do not change the fragmentation structure, it introduces a large energy cost for those configurations of states where two or more fermions sit next to each other. Hence, the configurations that are energetically favorable typically lie at half-filling or below half-filling. On the other hand, the natural tendency of the fragmentation structure due to the East constraint is to give rise to the largest fragment away from half-filling, and such a fragment typically favors the ground state due to its larger size compared to others; this pushes the ground state away from half-filling. The energy cost due to finite-$V$ therefore competes with the fragmentation structure imposed by the East constraint.
%whose inherent inclination is the exact opposite of the former. 
These opposing proclivities eventually brings the ground state towards half-filling with increasing $V$, and it finally stabilizes at half-filling after a threshold strength of $V$.

\begin{comment}
%We have seen the effect of interaction in the ground state of the East model. 
The fragmentation structure of the Hilbert space does not change with the strength of a finite interaction $V$. However, due to the energy cost coming from the nearest-neighbour 
density-density interaction, the ground state shifts towards half-filling. Moreover, at half-filling, a simple second-order perturbative argument tells us that the ground state energy changes by $- \frac{V^2}{J}$, which favours the ground state to be at half-filling. But exactly at $V \rightarrow \infty$ we have already seen there are very interesting ground state properties. Hence the special point in parameter space given by $V \rightarrow \infty$ has a rather singular behaviour. To numerically check the effects of 
nearest-neighbour 
density-density interactions on the ground state of the East model, we have taken system size $L = 20$ with PBC and considered different interaction strengths. Then we have plotted the ground state energies versus filling for different interaction strengths. The plots shown in Fig. \ref{fig17} clearly indicate that the ground state gradually shifts towards 
half-filling with increasing interaction strength.
\end{comment}

\subsection{Effects of finite-$V$ on the thermalization properties}
\label{sec12}

\begin{figure*}
    \centering
    \subfigure[]{
    \includegraphics[width=0.33\hsize]{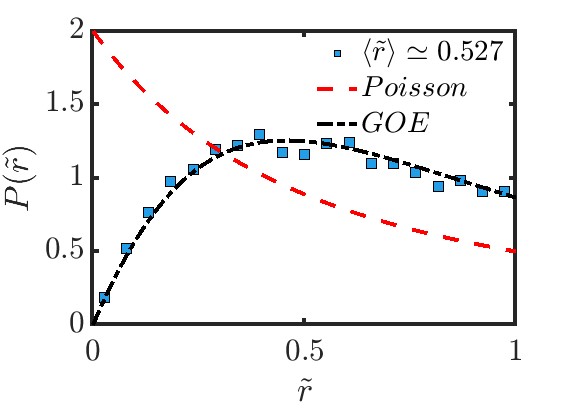}%
    }%
    \subfigure[]{
    \includegraphics[width=0.324\hsize]{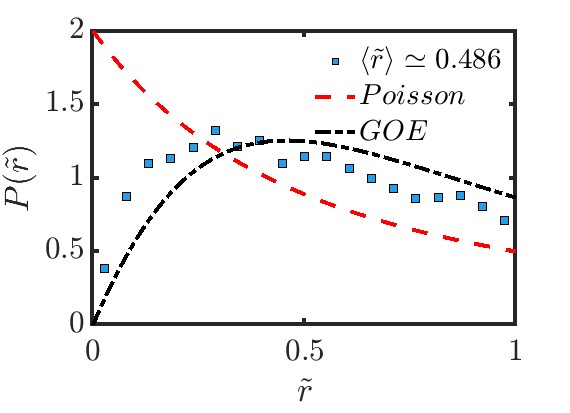}}%
    \subfigure[]{
    \includegraphics[width=0.332\hsize]{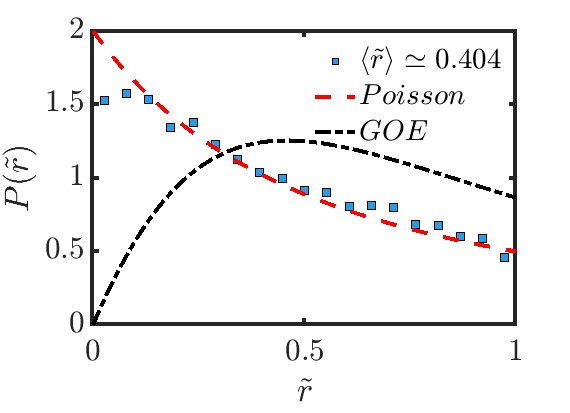}
    }%
\caption{ (a-c) Plots showing the level spacing statistics between consecutive level ratios for $V=0$, $3$, and $5$, respectively, within the largest fragment for $L=18$ and $N_{f}=L/2+3$ (the fragment size is 9996). The value of $\la\tilde{r}\ra$ shows a transition from the GOE to the Poisson distribution with increasing $V$, namely, $\la\tilde{r}\ra\sim0.527$ (the GOE value), $0.486$ and $0.404$ (close to the Poisson value) for $V=0$, $3$ and $5$, respectively, as illustrated in plots
(a-c). The distribution of level spacing ratios indicates that there is a transition occurring from non-integrability (subspace-restricted thermalization) to integrability (statistical bubble localized phase) within the largest fragment as $V$ increases.}
\label{fig18}
\end{figure*}

\begin{figure}
    \centering
    \subfigure[]{
    \includegraphics[width=0.78\hsize]{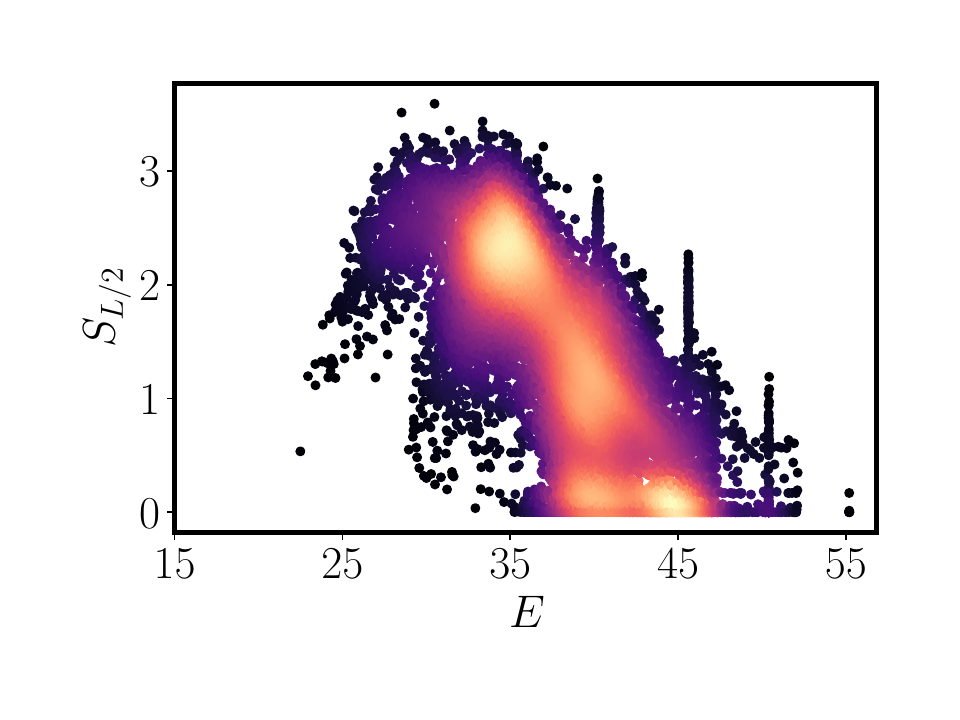}}\\
     \subfigure[]{
    \includegraphics[width=0.75\hsize]{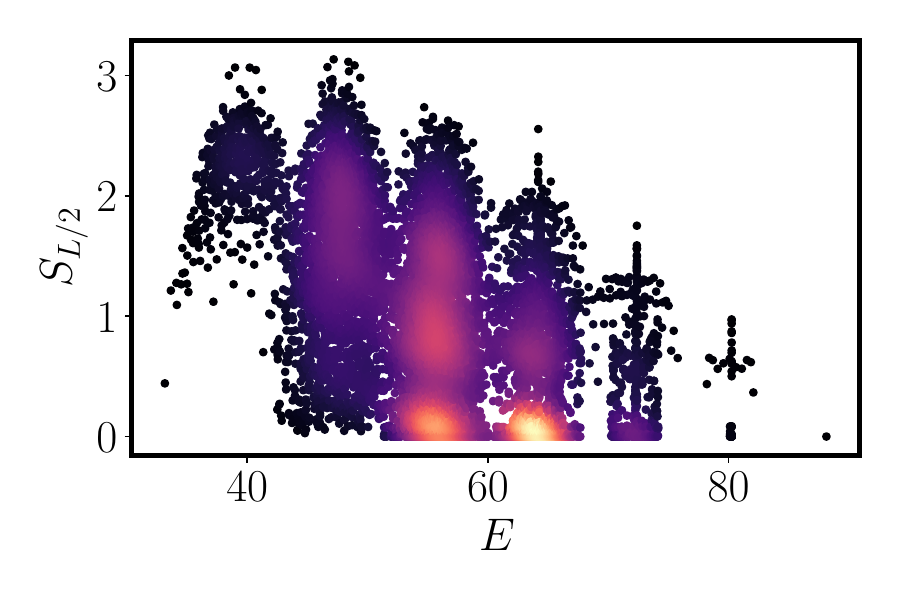}}\\
\caption{(a-b) Plots showing the spectrum of $S_{L/2}$ versus $E$ for $V=5$ and $V=8$ in the largest fragment for $L=18$ and $N=L/2+3$. Both these plots demonstrate that $S_{L/2}$ is not a smooth function of $E$ due to an emergent statistical bubble localization. This emergent bubble localization further causes secondary fragmentation within a primary fragment; this is distinctly visible in the entanglement spectrum as we see in Fig. (b) for $V=8$.}
\label{fig1111}
\end{figure}

We will now describe how the thermalization properties of the East model change as the interaction strength is
increased. To do this analysis, we restrict ourselves
to the largest fragment of the East model, which remains unchanged in the presence of a finite $V$. In Figs. \ref{fig18} (a-c), we show the level spacing distributions of consecutive energy levels within the largest fragments for several values of $V=0$ (the 
original East model), $3$ and $5$, for $L=18$ and $N_{f}=L/2+3$ (this fragment has 9996 states).
As we have discussed before, the largest fragment of the East model follows the GOE statistics~\cite{huse,atas,BGS} with $\la\tilde{r}\ra=0.5298$ (close to the GOE value), as shown in Fig. \ref{fig18} (a).
%since the subspace-restricted ETH gets satisfied within this fragment in a weaker sense, where most of the eigenstates obey the subspace-restricted ETH~\cite{rev1,HSF2_moudgalya_2020,aditya_2024} expect for measure zero eigenstates displaying quantum scarring property. 
However, the level spacing distribution shows a transition from the GOE~\cite{huse,atas,BGS} to the Poisson distribution~\cite{huse,atas,berry} as $V$ increases. In Figs. \ref{fig18} (b-c), we show a similar analysis for $V=3$ and $V=5$, which have $\la\tilde{r}\ra\simeq 0.503$ and $\la\tilde{r}\ra\simeq 0.417$ (close to the Poisson value), respectively. This emergent Poisson statistics with increasing $V$ is due to the statistical bubble localization~\cite{SBL1,SBL2} within the largest fragment, which causes a breakdown of the subspace-restricted ETH. The reason for this is that increasing $V$ introduces a large energy cost for certain configurations where two or more nearest-neighbor sites are occupied; this gives rise to a secondary fragmentation within a primary fully connected fragment by generating some emergent approximate conserved quantities. The ``bubble-neck" structures of the eigenstates in the bubble localized phase~\cite{SBL1,SBL2}, therefore, split up the active region into many subsectors, and the lack of level repulsion between these subsectors finally produces the Poisson level statistics~\cite{huse,atas,berry}. Further, the entanglement entropy is no longer a smooth function of the energy as shown in Fig. \ref{fig1111} (a). 

In the next section, we will study the limit $V \to \infty$ limit and show that this leads to a constraint-integrable model~\cite{pozsgay_2024, zadnik_2021}.

\section{East model with density-density interactions in the infinite-$V$ limit}
\label{PXXP}

We will now examine the model given in Eq. \eqref{vmode} in the limit $V\to\infty$. The effective Hamiltonian in this limit can be obtained by comparing the energy cost on the two sides of the allowed transitions. To find the effective Hamiltonian, we note that the East transitions can be recast involving four consecutive sites, as (i) $1100\leftrightarrow1010$ and (ii) $1101\leftrightarrow1011$. Although the original East transitions involve only three consecutive sites, the computation of the energy cost due to the $V$ term demands the consideration of these four-site configurations. The energy costs on the two sides of the above transitions are as follows: (i) $V$ (left side) and $0$ (right side), and (ii) $V$ (left side) and $V$ (right side). Hence only the second transition (in which the two sides have the same energy cost) is allowed in the limit $V\to\infty$. This gives rise to the following Hamiltonian involving four consecutive sites,
\bea
H~=~\sum_{j}n_{j-1}\left(c_{j}^{\dagger}c_{j+1}+{\rm H.c.}\right)n_{j+2}. \label{infiniteV}
\eea
Eq. \eqref{infiniteV} describes a correlated-hopping model, where a nearest-neighbor hopping is allowed only when the sites to their left and right are both occupied. Following a Jordan-Wigner transformation from spinless fermions to spin-1/2's, the model is described by a spin Hamiltonian of the form
\beq
H\,=\,\sum_{j} P_{j-1}\left(\sigma_{j}^{x}\sigma_{j+1}^{x}+\sigma_{j}^{y}\sigma_{j+1}^{y}\right)P_{j+2},
\eeq
where $P_{j}= (1+\sigma_{j}^{z})/2$ projects on to a state with $\uparrow$ spin on the $j$-th site, and $\sigma_{j}^{x}$, $\sigma_{j}^{y}$, $\sigma_{j}^{z}$ 
denote the Pauli matrices. This model can thus be called the PXXP$+$PYYP model~\cite{trans1}. Since this model only allows the transitions $1101 \leftrightarrow 1011$, it can be considered to be one-half of the folded-XXZ model~\cite{zadnik_2021,trans1} which allows both $1101 \leftrightarrow 1011$ and
$0100 \leftrightarrow 0010$. Surprisingly, the PXXP$+$PYYP model is invariant under spatial inversion, unlike the East model with $V=0$. 
%Further, the energy eigenvalues in this model appear in $E$ and $-E$ pairs due to the sublattice symmetry transformation, given by $c_{j}\to (-1)^j c_{j}$, that transforms $H\to -H$. 
We will now discuss the fragmentation structure and other features of this model in detail in the next section.

\begin{comment}
We now add a nearest-neighbour interaction term to the East model Hamiltonian. The Hamiltonian then looks as follows,
\begin{equation}
    \mathcal{H} = \sum_{j = 0}^{L-1} n_{j-1} (c_j^\dagger c_{j+1} + c_{j+1}^\dagger c_j ) + V \sum_{j=0}^{L-1} n_j n_{j+1}.
\end{equation}
The East form means that for four consecutive sites the only allowed processes are (a) $ 1\,1\,0\,0 \Leftrightarrow 1\,0\,1\,0$ and (b) $ 1\,1\,0\,1 \Leftrightarrow 1\,0\,1\,1$. In case (a) the energy cost of the state on the left side is $V$ while on the right side it is zero. For case (b) the energy cost on both sides is $V$. So in the $V \rightarrow \infty$ limit, for case (a) we need to change the energy by an amount of $V$ which is very costly and energetically not favourable. So, for the process $ 1\,1\,0\,1 \Leftrightarrow 1\,0\,1\,1$ we can write an effective four-site model for large interaction strength. The effective model in the infinite-$V$ limit is given by
\begin{equation}
    \mathcal{H}_{eff} = \sum_{j = 0}^{L-1} n_{j-1} (c_j^\dagger c_{j+1} + c_{j+1}^\dagger c_j) n_{j+2}.
\end{equation}
Now, we are interested in studying the fragmentation structure and different properties of this effective model. Unlike the East model this model has the inversion symmetry $c_j \rightarrow c_{-j}$. So, the inversion symmetry returns for the East model with 
nearest-neighbour density-density interactions in the large $V$ limit. Further, the energy spectrum has a $E \to - E$ symmetry due to the sublattice transformation $c_j \rightarrow (-1)^j c_j$ which flips $H \to - H$.
\end{comment}

\subsection{Characterization of root states}
\label{characPXXP}

We will first elucidate how to label each fragment in 
this model by a unique root state, similar to the East model discussed Sec. \ref{subsecA}. In doing so, we
utilize a similar procedure as discussed before, i.e., we will bring all the configurations appearing within a particular fragment to a uniquely defined root configuration having a simple form.
%which can represent each fragment. 
This identification will enable us to determine all the root states for the different fragments. We now note that there are two alternative ways to define the root states here since this model is inversion symmetric, unlike the East model where there is only one way to specify the root states due to the lack of inversion symmetry. Out of these two ways, we will use the root identification procedure which is similar to the East model.
%to put forward the whole analysis in a more reader-friendly manner.
To begin with, it can be checked that all the states within a simple and fully connected fragment can be reduced to a unique root state by moving all pairs of $10$'s to the leftmost side of the string while successively removing all $1101$'s. This generates the form $\psi_{L}\otimes\psi_{m}\otimes\psi_{R}$, where $\psi_{L}$ is either $\phi$ or has only $0$'s, $\psi_{m}$ has only $10$'s, and $\psi_{R}$ is either
$\phi$, all $0$'s or $0$'s followed by $1$'s, or all $1$'s. We will now discuss the rules to determine the sizes of the fragments.

\noi (i) We can check that $\psi_{L}$ never participates in the dynamics, and therefore it should be removed from the computation of the dimensions of the different fragments.

\noi (ii) In addition, if $\psi_{R}$ is $\phi$, $0$'s or $0$'s followed by $1$'s, then $\psi_{R}$ should also be removed from the calculation of the dimensions of fragments; such root states constitute class I. But if
when $\psi_{R}$ comprise $1$'s, such $1$'s should be integrated into the counting process, and consequently, such root states comprise class II.

After specifying these rules, we further note that the pairs of $10$'s always hop as a unit within a specific fragment. In addition, the rightmost $1$ for all the configurations within fragments falling under class II, i.e., $\psi_{R}$ comprising $1$'s, always remains dynamically inactive. However, its presence is essential for the facilitated hopping processes allowed by this model. Accordingly, this $1$ should also be removed from the calculation of the dimensions of such fragments.
Next, we denote the number of $10$'s and $1$'s in $\psi_{m}$ and $\psi_{R}$ as $N_{A}$ and $N_{B}$, respectively. With these rules, the dimension of a fully connected fragments is found to be
\bea
D(L)~=~\frac{(N_{A}+N_{B}-1)!}{N_{A}! (N_{B}-1)!},
\eea
where $L=2N_{A}+N_{B}$, and the filling fraction is $f=\frac{N_{A}+N_{B}}{L}$ provided that $\psi_{L}$ is $\phi$.

We can generalize these rules to compute the dimensions of more complicated fragments where there are multiple dynamically active regions separated from each other
by inactive blockade configurations. For such fragments,
the most general root state takes the separable form $\psi_{L}\otimes\psi_{B_1}\otimes\psi_{I_1}\otimes\psi_{B_2}\otimes\psi_{I_2}$\,$\cdots\otimes\psi_{I_n}\otimes\psi_{B_n}$, where $\psi_{L}$ is $\phi$ or all $0$'s, and $\psi_{I_i}$ has all $0$'s ($2$ or more in number). In addition, $\psi_{B_{i}}$ (except the rightmost one) can be further divided into two blocks $\psi_{m_{i}}\otimes\psi_{R_i}$, where $\psi_{m_i}$ consists of pairs of $10$'s, and $\psi_{R_i}$ contains $\phi$ or $0$'s followed by $1$'s or all $1$'s. Further, the rightmost $\psi_{B_{n}}$ again consists of two parts, $\psi_{m_{n}}$ and $\psi_{R_{n}}$, where $\psi_{m_{n}}$ is $10$'s, and $\psi_{R_{n}}$ contains either $\phi$, all $0$'s, $0$'s followed by $1$'s or all $1$'s. We should again note that $\psi_{L}$ never participates in the dynamics, and the role of $\psi_{I_i}$ is to keep the active regions given by $\psi_{B_i}$ dynamically separated from each other. These blocks, thus, do not play any active role in counting the dimensions of fragments. The $\psi_{B_i}$'s are therefore independent and fully connected dynamically active fragments which remain isolated from each other due to the presence of blockaded regions. The dimension of such generalized fragments can, therefore, be obtained by taking a product of the dimensions of each active region given by the rules discussed earlier. We thus obtain the general expression 
\bea
D_{{\rm gen}}(L)\,=\,\prod_{i=1}^{n}\frac{(N_{A_{i}}+N_{B_{i}}-1)!}{N_{A_i}! (N_{B_i}-1)!},
\eea
where $N_{A_i}$ and $N_{B_i}$ are the number of $10$'s and $1$'s in $\psi_{m_i}$ and $\psi_{R_i}$, which form a block $\psi_{B_i}$.

\subsection{Number of fragments in the PXXP$+$PYYP model}

\begin{comment}
To find the number of fragments, it is convenient to rewrite the state $ 1\, \fbox{1\,0} 1 $ for the four-site effective model as $\fbox{1\,0}\, 1\,1$. As some examples, this rule allows us to write some five-site states as follows:\\

a) 1\,1\,1\,1\,1\, (frozen state)\\

b) 1\,1\,1\,0\,1\, $\Leftrightarrow$ \fbox{1 0} \,1\,1\,1 \\

c) 1\,0\,1\,1\,1\, $\Leftrightarrow$ \fbox{1 0} \,1\,1\,1 \\

d) 1\,1\,0\,0\,1\, (frozen state)\\

e) 1\,0\,1\,0\,1\, $\Leftrightarrow$ \fbox{1 0} \fbox{1 0}\,1 \\

f) 1\,0\,0\,1\,1\, $\Leftrightarrow$ \fbox{1 0} \,0\,1\,1 \\

g) 1\,0\,0\,0\,1\, $\Leftrightarrow$ \fbox{1 0} \,0\,0\,1 \\

From the above examples we can clearly see that the $'1'$ in the last site always remains frozen. Next, just as in the three-site East model among the configurations $1\,1\,0$ and $1\,0\,1$ we chose $1\,0\,1$ and removed $ 1\,1\,0$, here for this effective four-site model we again have two allowed configurations $ 1\,1\,0\,1$ and $ 1\,0\,1\,1$. It is possible to remove either one of them. Either choices will reduce all the states to a particular canonical configurations. We will choose to remove  $1\,1\,0\,1$ analogous to the East model.
\end{comment}

We will now count the number of fragments in this model with OBC using the transfer matrix method, similar to the procedure discussed for the East model~\cite{brighi_2023,weak2}. We note that identifying the root configuration for each fragment requires us to remove all occurrence of $1101$'s; hence the resultant transfer matrix~\cite{gen,aditya_2024,menon_1995,menon_1997,barma_1994}, $T_3$ should not allow any occurrences of $1101$. Using this fact, a transfer matrix computation shows that $N_{frag}$ in this model grows asymptotically as $1.867^{L}$; we have also confirmed this
numerically. The details of the transfer matrix computation are shown in Appendix \ref{appD}.

We note that the number of fragments in this model grows much faster with $L$ compared to the East model. This is understandable since the infinite-$V$ limit forbids more
transitions than the East model. Hence the degree of fragmentation in the infinite-$V$
model is higher than in the East model.

\begin{comment}
Hence, as the system size $L$ becomes large,
the total number of fragments for this model grows exponentially as $1.8667^L$. A very important observation following is that this model is more constrained than the East model; hence the degree of fragmentation is larger, and the number of fragments grows faster here. 
We have also checked this growth of the total number of fragments numerically and
we find that the result matches the one found analytically (see Fig. \ref{fig13}).

\begin{figure}[h!]
\centering
\includegraphics[width=0.85\columnwidth]{N_frag_vs_L_infinite_V_model_obc.pdf}
\caption{Plot showing $N_{frag}$ versus $L$ with OBC.  The numerical fitting shows that the number of fragments grows as $1.867^L$, which is in agreement with our analytically obtained result.} \label{fig13}
\end{figure}
\end{comment}

We can again count the total number of fragments $N_{frag}$ for a given system size
$L$ with OBC by using the fact that $N_{frag} = \sum_{i,j} 
M_{i,j}$ where $M = T_3^{L-3}$. The number of fragments for the first few values of
$L$ is shown in Table \ref{tablefragPXXP}; we find that these agree with our numerical results. In addition, we have $N_{frag}(L)=1,2$ and $4$ for $L=0$, $1$ and $2$, respectively.
\begin{comment}Table 4 shows the number of fragments $N_{frag}$ of this infinite-$V$ model with system size $L$. $N_{frag}$ for $L = 0, 1$ are given by respectively $1, 2$ respectively. Interestingly, we have also found a recursion relation for the total number of fragments for a system size $L$,
\end{comment}

\begin{table}[h]
\centering
\begin{tabular}{|c|c|c|c|c|c|c|c|c|c|c|c|c|c|c|c|c|}
\hline
$L$ & 3 & 4 & 5 & 6 & 7 & 8 & 9 & 10 \\
\hline
\(N_{frag}\) & 8 & 15 & 28 & 52 & 97 & 181 & 338 & 631  \\
\hline
\end{tabular}
\caption{$N_{frag}$ versus $L$.} 
\label{tablefragPXXP}
\end{table}
We also note from Table \ref{tablefragPXXP} that $N_{frag}$ follows recursion relation
\bea N_{frag} (L) &= &N_{frag}( L-4 )\, - \,N_{frag}( L-3 )\, \non\\
&&+ \,2 N_{frag}( L-1 ), \eea
for $L\geq 4$.

\subsection{Frozen fragments in the PXXP$+$PYYP model}

\begin{comment}
The infinite - $V$ model turns out to have a huge number of frozen states. These are single states (i.e., not superpositions of states) in the particle number basis having zero eigenvalue  of the Hamiltonian. The number of frozen state is larger for this model compared to the previous East model as the dynamics is more constrained. To calculate the growth of the number of frozen state with system size we have found another transfer matrix where we have used the basis $\ket{111}, \ket{110}, \ket{101}, \ket{011}, \ket{100}, \ket{010}, \ket{001}, \ket{000}$ to define the rows and columns of the  matrix. An important point to keep in mind while writing the transfer matrix for the frozen states is that we have to disallow both the configurations $1101$ and $1010$ anywhere in the system. The $8 \times 8$ transfer matrix for the frozen states of this four-site model is given by
\end{comment}

We will now find the growth with system size of the number of frozen eigenstates in the PXXP$+$PYYP model using a similar transfer matrix method. Since these eigenstates are completely inactive dynamically, the counting of such states requires us to remove 
both $1101$ and $1011$ from the transfer matrix construction. Then one finds that the number of fragments $N_{froz}$ grows as $1.8124^{L}$ for a system with OBC;
we have verified this numerically. The details of the transfer matrix calculation~\cite{gen,aditya_2024,menon_1995,menon_1997,barma_1994} are shown in Appendix \ref{appE}. We again see that $N_{froz}$ in this model grows faster than in the East model; this is due to the fact that the infinite-$V$ model forbids some processes which are allowed in the East model with $V=0$.

\begin{comment}
    So, in the thermodynamic limit the number of frozen states grows with the system size $L$ as $1.812^L$ which is much larger compared to the East model. We have also calculated the number of frozen states numerically and verified that it agrees with our analytical results. The analytical expression for counting this frozen state for a given system size $L$ is  where $Y = (T_{froz})^{L-3}$. The number of frozen states for different system sizes with OBC is given in the table below.
\end{comment}

\begin{comment}
\begin{figure}[H]
\centering
\includegraphics[width=0.6\textwidth]{number_frozen_state_with_system_size_L_infinite_V_model_with_obc.pdf}
\caption{Plot of the total number of frozen state as a function of the system size $L$
for OBC.  The fit shows that the number of fragments grows as $1.8122^L$.} \label{fig15}
\end{figure}
\end{comment}

\subsection{Description of the largest fragment}

We will now compute the dimension of the largest fragment to get more insight into the fragmentation structure of this model. We have shown in Sec. \ref{characPXXP} that the dimension of a simple and fully connected fragment specified by a root state 
$\psi_{L}\otimes\psi_{m}\otimes\psi_{R}$ is given by
\bea
D^{PXXP}(L)=\frac{(N_{A}+N_{B}-1)!}{N_{A}! (N_{B}-1)!},\label{Dpxxp}
\eea
where $N_{A}$ and $N_{B}$ are the total number of $10$'s and $1$'s in $\psi_{m}$ and $\psi_{R}$, respectively. For simplicity, let us now assume that $\psi_{L}$ is $\phi$;
this allows us to use the relations $L=2N_{A}+N_{B}$ and filling $f=N_{A}+N_{B}$. 
To study the thermodynamic limit $L \to \infty$, we parameterize $N_{A}=\alpha L$ and $N_{B}=\gamma L$, where $\alpha$, $\gamma \geq 0$. These parameters satisfy 
$2 \al + \ga = 1$ and $f=(1- \al) L$. After using Stirling's approximation, Eq. \eqref{Dpxxp} simplifies to
\bea
D (L) ~=~ \left(\frac{(1-\alpha)^{1-\alpha}}{\alpha^{\alpha}\,(1-2\alpha)^{1-2\alpha}}\right)^{L},\label{DinfPXXP}
\eea
where we set $\gamma=1-2\alpha$. Extremizing the Eq. \eqref{DinfPXXP} 
with respect to $\alpha$, we find that the dimension of the largest fragment in the thermodynamic limit occurs at $\alpha=\frac{1}{2}\left(1-\frac{1}{\sqrt{5}}\right)\simeq 0.276$. The filling $f$ for the largest fragment is then
given by $(1-\alpha)L \simeq 0.724 ~L$. Strikingly, the dimension of the largest fragment then turns out to be $1.618^{L}$, which is exponentially smaller than the full Hilbert space size of $2^{L}$. Hence this model exhibits strong fragmentation~\cite{rev1,aditya_2024}, unlike the East model, where we observe weak fragmentation at this filling which lies above half-filling.

\subsection{Integrability of the PXXP$+$PYYP model and ground state analysis}\label{sec10}

\begin{figure}[htb]
\centering
\includegraphics[width=0.85\columnwidth]{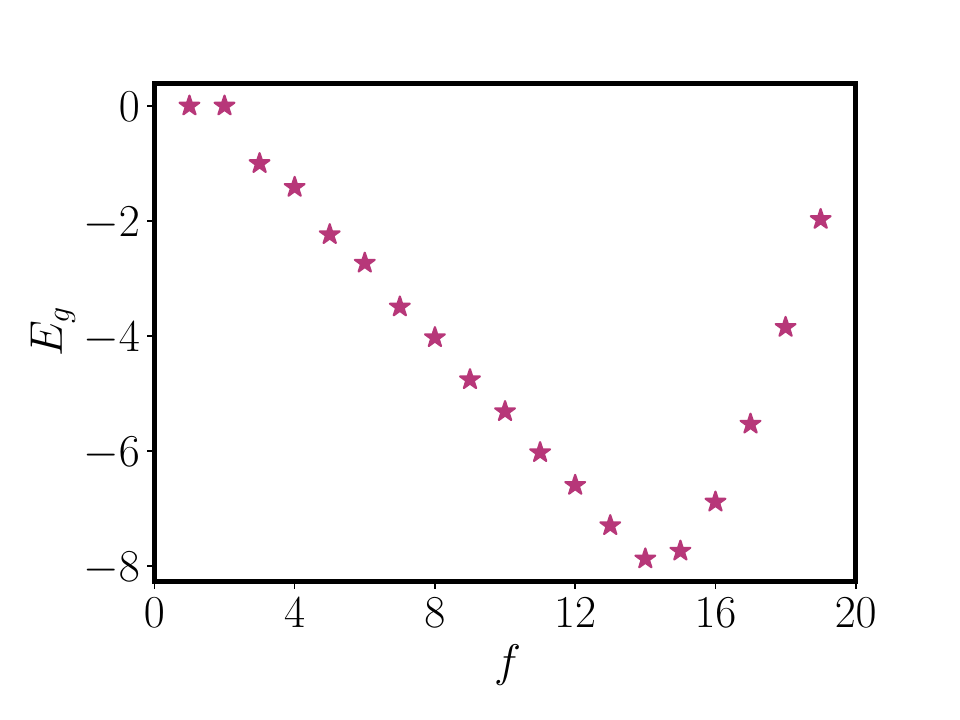}
\caption{Plot showing the ground state energy $E_{GS}$ versus the filling $N_{f}$ for the PXXP$+$PYYP model for $L = 20$ with OBC. We see that the ground state appears at $N_{f}=14$, which means the filling fraction is $\rho=0.7$. This is close to the 
value of $0.699$ which is predicted analytically in the thermodynamic limit.} \label{fig16}
\end{figure}

We will now argue that PXXP$+$PYYP falls under the class of constraint-integrable model~\cite{pozsgay_2024,zadnik_2021}, in contrast to the East model which is non-integrable~\cite{weak2,brighi_2023}. To prove the integrability of this model, we map
each fragment to a non-interacting tight-binding model of spinless fermions
with nearest-neighbor hopping on a system with OBC. This can accomplished by noting 
that facilitated hopping in this model always makes a pair of $10$ hop as a unit. Accordingly, there are effectively two degrees of freedom available during the transitions, which are $10$'s and $1$'s. This allows us to map $10$ and $1$ to a hole and a particle, given by $\ket{0}$ and $\ket{1}$, respectively. Following this mapping, 
we see that the model within each connected fragment simply reduces to a nearest-neighbor tight-binding model of spinless fermions, where the energy spectrum $E=\sum_{n}E_{n},$ 
where the single-particle energies are given by
\bea E_{n} ~=~ -2 \cos\left(\frac{\pi n}{N_{A}+N_{B}}\right), \label{en}
\eea
with $n=1,\cdots$, $N_{A}+N_{B}-1$. To find the total energy $E$, all possible combinations of $n$'s have to be taken into account, with the number of $n$'s
within a given fragment being given by $N_{B}-1$. Here, $N_{A}$ and $N_{B}$ define the number of $10$'s and $1$'s in the parts $\psi_{m}$ and $\psi_{R}$ of the root state;
alternatively, these give the total numbers of $0$'s and $1$'s in the mapped model. 
We also note that the rightmost $\ket{1}$ never participates in the hopping process for this newly mapped tight-binding model of length $L'=N_{A}+N_{B}$. This fact, along with the OBC, makes the single-particle dispersion different from a tight-binding model with PBC. The single-particle dispersion in Eq.~\eqref{en} can be obtained by observing that 
the single-particle wave function at site $j$ in state $n$ is given by $\psi_{n} (j) =\sin\left(\frac{\pi\,n j}{N_{A}+N_{B}}\right)$, which follows from the hard-wall boundary condition $\psi_n (0)=\psi_n (L'=N_{A}+N_{B})=0$. We remark here that the integrability of the PXXP$+$PYYP model has been discussed recently using the 
Bethe Ansatz in Ref. \onlinecite{pozsgay_2024}; however, our mapping of this model to a nearest-neighbor tight-binding model of non-interacting fermions within each fragment 
explains the integrability very easily.

Having discussed the integrability of the PXXP$+$PYYP model, we will now compute the filling fraction at which the ground state of the model appears. An analytical computation 
presented in Appendix \ref{appF} shows that in the thermodynamic limit,
the filling fraction at which the ground state lies is given by $\rho \simeq \pi/4.493 = 0.699$, and the ground state energy is $E_{GS}\simeq 0.435 L$. We numerically validate this result by studying the ground state of the PXXP$+$PYYP model for $L=20$ as a function of the filling $f$. This shows that the ground state lies at a filling, $f=14$. This implies $\rho=0.7$ which is very close to the analytically predicted value 
of $\rho \simeq 0.699$. This agrees with our numerical results; see Fig.~\ref{fig16}.
Surprisingly, we see that the ground state in this model lies at a filling fraction which is less than the filling fraction $0.724$ where this model 
has the largest fragment in the thermodynamic limit. This anomalous behavior is in
contrast to what we found in the East model; there the ground state lies in the largest fragment, and the filling fraction for the ground state shifts away half-filling with increasing system size.

\subsection{Experimentally realizable models}

We will discuss here how to experimentally realize the PXXP$+$PYYP model involving terms 
with four consecutive sites by considering a particular limit of the $t-V$ model 
(a model of spinless fermions which has a
nearest-neighbor hopping and a 
nearest-neighbor density-density interaction with strength $V$), which has 
a specific pattern of on-site potentials pattern~\cite{aditya_2024}. We consider the Hamiltonian
\bea H_4 &=& \sum_{j} ~[~ c_{j}^{\dagger}c_{j+1} ~+~ c_{j+1}^\dagger c_j \non \\
&& ~~+~ V_j ~(2n_{j}-1) ~(2n_{j+1}-1)+~~\mu_{j}(2n_{j}-1)],\non\\ \label{ham4} \eea
where $V_j$ varies with $j$ with period four, namely, $V_{4j+1} = V_1$, $V_{4j+2} = V_2$,
$V_{4j+3} = V_3$, and $V_{4j+4} = V_4$. Similarly, $\mu_{j}$ also has a spatial periodicity of four sites  such that $\mu_{4j+1}=\mu_{1}$, $\mu_{4j+2}=\mu_{2}$, $\mu_{4j+3}=\mu_{3}$ and $\mu_{4j+4}=\mu_{4}$. We have set the strength of the 
nearest-neighbor hopping to be unity for convenience.
% We choose $V_1 = V = - V_3$, and $V_2 = V' = - V_4$. 
We now consider all possible 
correlated-hopping processes involving four consecutive sites, namely, 
$1101 \leftrightarrow 1011$, $0100 \leftrightarrow 0010$, $1100 \leftrightarrow 1010$ and $0101 \leftrightarrow 0011$, for all the possible patterns of the interactions and on-site potentials.
%which are associated to each other through translation by one site. 
We will compare the energy cost on both sides of the different transitions mentioned
above for
$V_{1}, ~V_{2}, ~V_{3}, ~V_{4} \rightarrow\infty$, and $\mu_{1},~\mu_{2},~\mu_{3},~\mu_{4}\to\infty$, as shown in Table \ref{Table VII};
this will allow us to find an effective Hamiltonian in the limit of large interaction 
strengths and on-site potentials. The table shows that the energy costs on the
left and right sides of the correlated-hopping process $1101 \leftrightarrow 1011$ are equal to each other provided $V_1 = - V_3 = V_2 = V_4 = V$ with $\mu_3=\mu_4=-\mu_1=-\mu_2=V$. However, these parameter values do not support the other three correlated-hopping processes in the $V \to \infty$ limit as the energy costs on the two sides
of these processes differ by infinitely large amounts. Hence we see that an experimentally realizable $t-V$ model with a special pattern of 
large nearest-neighbor interactions and on-site potentials $\mu_{j}$ 
reduces to the PXXP$+$PYYP model.
%in the limit of large interactions and on-site potentials.

\begin{widetext}
\begin{center}
\begin{table}[htb]
\begin{tabular}{|c|c|c|c|c|} 
\hline
Pattern of&Pattern of & Correlated-hopping & Energy cost for left hand side of 
& Energy cost for right hand side of \\
$V_j$ & $\mu_j$ & process & the process in the third column & the process in the 
third column \\
\hline
$V_{1}\,V_{2}\,V_{3}\,V_{4}$&$\mu_{1}\,\mu_{2}\,\mu_{3}\,\mu_{4}$&$1101 ~\leftrightarrow~ 1011$& $V_{1}-V_{2}-V_{3}+\mu_1 +\mu_2-\mu_3+\mu_4$&$-~ 
V_{1}-V_{2}+V_{3}+\mu_1 -\mu_2+\mu_3+\mu_4$\\
\hline
$V_{1}\,V_{2}\,V_{3}\,V_{4}$&$\mu_{1}\,\mu_{2}\,\mu_{3}\,\mu_{4}$&$0100 ~\leftrightarrow ~0010$& $-~ V_{1}-
V_{2}+V_{3}-\mu_1 +\mu_2-\mu_3-\mu_4$&$V_{1}-V_{2}-V_{3}-\mu_1 -\mu_2+\mu_3-\mu_4$\\
\hline
$V_{1}\,V_{2}\,V_{3}\,V_{4}$&$\mu_{1}\,\mu_{2}\,\mu_{3}\,\mu_{4}$&$1100 ~\leftrightarrow ~1010$& $V_{1}-V_{2}+V_{3}+\mu_1 +\mu_2-\mu_3-\mu_4$&$-~ 
V_{1}-V_{2}-V_{3}+\mu_1 -\mu_2+\mu_3-\mu_4$\\
\hline
$V_{1}\,V_{2}\,V_{3}\,V_{4}$&$\mu_{1}\,\mu_{2}\,\mu_{3}\,\mu_{4}$&$0101 ~\leftrightarrow ~0011$& $-~ V_{1}-V_{2}-
V_{3}-\mu_1 +\mu_2-\mu_3+\mu_4$&$V_{1}-V_{2}+V_{3}-\mu_1 -\mu_2+\mu_3+\mu_4$\\
\hline

$V_{2}\,V_{3}\,V_{4}\,V_{1}$&$\mu_{2}\,\mu_{3}\,\mu_{4}\,\mu_{1}$&$1101 ~\leftrightarrow ~1011$& $V_{2}-V_{3}-V_{4}+\mu_2 +\mu_3-\mu_4+\mu_1$&$-~ 
V_{2}-V_{3}+V_{4}+\mu_2 -\mu_3+\mu_4+\mu_1$\\
\hline
$V_{2}\,V_{3}\,V_{4}\,V_{1}$&$\mu_{2}\,\mu_{3}\,\mu_{4}\,\mu_{1}$&$0100 ~\leftrightarrow ~0010$& $-~ V_{2}-
V_{3}+V_{4}-\mu_2 +\mu_3-\mu_4-\mu_1$&$V_{2}-V_{3}-V_{4}-\mu_2 -\mu_3+\mu_4-\mu_1$\\
\hline
$V_{2}\,V_{3}\,V_{4}\,V_{1}$&$\mu_{2}\,\mu_{3}\,\mu_{4}\,\mu_{1}$&$1100 ~\leftrightarrow ~1010$& $V_{2}-V_{3}+V_{4}+\mu_2 +\mu_3-\mu_4-\mu_1$&$-~ 
V_{2}-V_{3}-V_{4}+\mu_2 -\mu_3+\mu_4-\mu_1$\\
\hline
$V_{2}\,V_{3}\,V_{4}\,V_{1}$&$\mu_{2}\,\mu_{3}\,\mu_{4}\,\mu_{1}$&$0101 ~\leftrightarrow ~0011$& $-~ V_{2}-V_{3}-
V_{4}-\mu_2 +\mu_3-\mu_4+\mu_1$&$V_{2}-V_{3}+V_{4}-\mu_2-\mu_3+\mu_4+\mu_1$\\
\hline

$V_{3}\,V_{4}\,V_{1}\,V_{2}$&$\mu_{3}\,\mu_{4}\,\mu_{1}\,\mu_{2}$&$1101 ~\leftrightarrow~ 1011$& $V_{3}-V_{4}-V_{1}+\mu_3 +\mu_4-\mu_1+\mu_2$&$-~ 
V_{3}-V_{4}+V_{1}+\mu_3 -\mu_4+\mu_1+\mu_2$\\
\hline
$V_{3}\,V_{4}\,V_{1}\,V_{2}$&$\mu_{3}\,\mu_{4}\,\mu_{1}\,\mu_{2}$&$0100 ~\leftrightarrow ~0010$& $-~ V_{3}-
V_{4}+V_{1}-\mu_3 +\mu_4-\mu_1-\mu_2$&$V_{3}-V_{4}-V_{1}-\mu_3 -\mu_4+\mu_1-\mu_2$\\
\hline
$V_{3}\,V_{4}\,V_{1}\,V_{2}$&$\mu_{3}\,\mu_{4}\,\mu_{1}\,\mu_{2}$&$1100 ~\leftrightarrow ~1010$& $V_{3}-V_{4}+V_{1}+\mu_3 +\mu_4-\mu_1-\mu_2$&$-~ 
V_{3}-V_{4}-V_{1}+\mu_3 -\mu_4+\mu_1-\mu_2$\\
\hline
$V_{3}\,V_{4}\,V_{1}\,V_{2}$&$\mu_{3}\,\mu_{4}\,\mu_{1}\,\mu_{2}$&$0101 ~\leftrightarrow ~0011$& $-~ V_{3}-V_{4}-
V_{1}-\mu_3 +\mu_4-\mu_1+\mu_2$&$V_{3}-V_{4}+V_{1}-\mu_3 -\mu_4+\mu_1+\mu_2$\\
\hline
$V_{4}\,V_{1}\,V_{2}\,V_{3}$&$\mu_{4}\,\mu_{1}\,\mu_{2}\,\mu_{3}$&$1101 ~\leftrightarrow ~1011$& $V_{4}-V_{1}-V_{2}+\mu_4 +\mu_1-\mu_2+\mu_3$&$-~ 
V_{4}-V_{1}+V_{2}+\mu_4 -\mu_1+\mu_2+\mu_3$\\
\hline
$V_{4}\,V_{1}\,V_{2}\,V_{3}$&$\mu_{4}\,\mu_{1}\,\mu_{2}\,\mu_{3}$&$0100 ~\leftrightarrow ~0010$& $-~ V_{4}-
V_{1}+V_{2}-\mu_4 +\mu_1-\mu_2-\mu_3$&$V_{4}-V_{1}-V_{2}-\mu_4 -\mu_1+\mu_2-\mu_3$\\
\hline
$V_{4}\,V_{1}\,V_{2}\,V_{3}$&$\mu_{4}\,\mu_{1}\,\mu_{2}\,\mu_{3}$&$1100 ~\leftrightarrow ~1010$& $V_{4}-V_{1}+V_{2}+\mu_4+\mu_1-\mu_2-\mu_3$&$-~ 
V_{4}-V_{1}-V_{2}+\mu_4 -\mu_1+\mu_2-\mu_3$\\
\hline
$V_{4}\,V_{1}\,V_{2}\,V_{3}$&$\mu_{4}\,\mu_{1}\,\mu_{2}\,\mu_{3}$&$0101 ~\leftrightarrow ~0011$& $-~ V_{4}-V_{1}-
V_{2}-\mu_4 +\mu_1-\mu_2+\mu_3$&$V_{4}-V_{1}+V_{2}-\mu_4 -\mu_1+\mu_2+\mu_3$\\
\hline
\end{tabular}
\caption{Energy costs arising from the $V_j$ and $\mu_j$ terms in Eq.~\eqref{ham4} for the 
left and right hand sides of various correlated-hopping processes for all four sets of
patterns of $V_j$ and $\mu_{j}$ which are connected to one another through translation by 
one site, in the limit $V_{1}$, $V_{2}$, $V_{3}$, $V_{4}\rightarrow\infty$ and $\mu_1,\mu_2,\mu_3,\mu_4\to\infty$.}
%*** redo the table using only $V , V'$, better still, put $ V=V'$ also. **}
\label{Table VII}
\end{table}
\end{center}
\end{widetext}

\section{Discussion}
\label{Dis}

We will now summarize the main results obtained in this paper. We have studied a particle-number conserving quantum East model~\cite{brighi_2023,weak2}, which only allows nearest-neighbor hoppings between sites $j$ and $j+1$ when the neighboring site $j-1$ is occupied. This model is known to exhibit a filling-dependent weak-to-strong 
fragmentation transition~\cite{weak1,weak2}. To unravel the fragmentation structure, we have presented an alternative formalism based on the identification of the canonical representation of the root states which define each fragment~\cite{fredkin1,fredkin2}. This root identification procedure enables us to uncover various aspects of the fragmented Hilbert space in this model, namely, the total number of fragments, the number of frozen eigenstates, and the dimension of the largest fragment, using the transfer matrix and enumerative combinatorics methods~\cite{gen,aditya_2024,menon_1997,menon_1995,barma_1994}. In addition, the growth of all the fragments in this model can be captured by using the appropriate Dyck word sequences~\cite{catalan}. Our analysis further revealed that the filling fraction for the appearance of the ground state and of the largest fragment shifts away from half-filling as $\sqrt{L}/2$ asymptotically; this is a consequence of the distinctive fragmentation structure of this model in terms of Dyck sequence~\cite{catalan}.

Next, we have studied the static and dynamic signatures of the lack of thermalization~\cite{rev1, HSF2_moudgalya_2020,aditya_2024} which occurs in this model 
due to the HSF. Our analysis of energy level spacing correlation~\cite{huse,atas} reveals that although the ETH~\cite{ETH1,ETH2,ETH3,ETH4,rev1,rev2,rev3} is strongly violated within the full Hilbert space, a weaker version of the Krylov-restricted ETH~\cite{rev1} is satisfied within all sufficiently large fragments. The weaker violation of the ETH within the different fragments is supported by the presence of subspace-restricted quantum many-body scars~\cite{HSF2_moudgalya_2020,subscar1,subscar2,brighi_2023}, which are a consequence of the recursive quantum HSF~\cite{brighi_2023}. We then examined various measures of the ETH violation~\cite{rev1,rev2} in the scar subspaces, such as long-time revivals in the Loschmidt amplitude, abrupt jumps in the overlap amplitudes for the scar states, and some of the few-body observables not behaving as smooth functions of the energy. We also studied
the anomalous long-time behaviors of the bulk and boundary autocorrelators starting from a random initial state~\cite{HSF2_moudgalya_2020,stat_loc_2020,hart_2024}; these are a
hallmarks of typical fragmented systems. We found that the two boundaries in the East model behave in distinct manners due to the lack of spatial inversion symmetry in this model.

We have then studied the effects of nearest-neighbor density-density interactions
with strength $V$ on the East model~\cite{brighi_2023}; this perturbation keeps the fragmentation structure of the model intact. The introduction of a finite $V$ has a number of effects on the ground state and on the ETH properties of the system. To be specific, increasing the value of $V$ gradually moves the ground state closer to half-filling and also induces a spectral transition from the GOE statistics (weakly ETH violating~\cite{brighi_2023,HSF2_moudgalya_2020}) to the Poisson statistics (statistical bubble localized phase~\cite{SBL1,SBL2}) within sufficiently large fragments. The infinite-$V$ limit gives a significantly 
different model, namely, it turns into a constraint-integrable model~\cite{zadnik_2021,pozsgay_2024} called the PXXP$+$PYYP model. The integrability of this model can be readily seen by mapping it to a non-interacting neatest-neighbor tight-binding model of spinless fermions within each fragment. We have then unravelled the distinct fragmentation structure and anomalous ground state properties with the help of this mapped model. Finally, we have concluded by proposing an experimentally realizable special kind of $t-V$ model with a specific pattern of nearest-neighbor interactions and on-site potentials~$\mu$~\cite{aditya_2024} which reduces to the PXXP$+$PYYP model in a particular limit.

We end by proposing possible future avenues of research. It would 
be worthwhile to completely characterize the classical fragmentation structure of the families of longer-range facilitated East models~\cite{brighi_2023}, which have not been explored much until now. A more detailed analysis of the transport properties 
which takes the classical fragmentation structure into account would also be useful ~\cite{brighi_2023,weak2,barma_1994}. The behaviors of the
bulk and boundary autocorrelation functions for a random thermal state in the thermodynamic limit also demands a more careful analysis in order to understand the nature of the freezing transition~\cite{hart_2024}.
Finally, it would be useful to study the effects of disorder~\cite{tomasi_2019,Herviou} and dissipation~\cite{Li_2023} in this model.  

We expect that our results can be experimentally verified in cold-atom 
platforms~\cite{HSFexp1,HSFexp2} where one-dimensional systems of spinless fermions 
with spatially periodic potentials and strong interactions can be realized.
Recently, thermalization in some particular fragments of a model with HSF
has been observed in a Rydberg atom system in one dimension~\cite{HSFexp3}. Another observation of HSF has been noted in a superconducting processor in a system demonstrating Stark many-body localization~\cite{HSFexp4}.

\vspace{0.6cm}

\centerline{\bf Acknowledgments}
\vspace{0.4cm}
S.A. and D.S. thank Sanjay Moudgalya for helpful discussions. S.A. thanks MHRD, India for financial support through a PMRF. D.S. thanks SERB, India for funding through Project No. JBR/2020/000043.

\appendix

\section{Calculation of total number of fragments in the East model}
\label{appA}

\begin{figure}[h!]
\centering
\includegraphics[width=0.96\hsize]{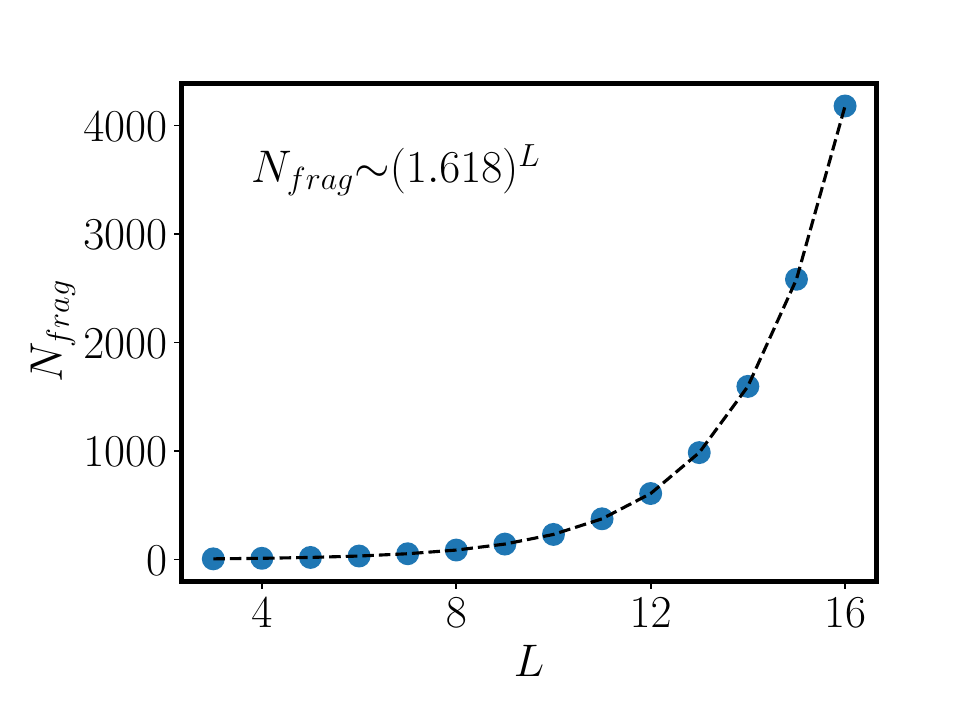}
\caption{Plot of the total number of fragments $N_{frag}$ versus $L$, for a system
with OBC. The numerical fitting shows that $N_{frag}$ grows as $1.618^L$.} 
\label{fig1} \end{figure}

We will compute here the total number of fragments $N_{frag}$ in the East model using the transfer matrix method~\cite{gen,aditya_2024,menon_1995,menon_1997,barma_1994}. We begin
by observing that since our model comprises terms involving three consecutive sites, it is necessary to construct a $4\times 4$ transfer matrix to compute the total number of fragments. We denote the matrix as $T_1 (c_{i},c_{j})$, where $c_{i}$ denotes the configurations $11$, $10$, $01$ and $00$ on two consecutive sites. In addition, as discussed earlier, a unique identification of the root state specifying each fragment requires us to move all occurrences of $10$'s to the left end of the system by successively changing $110$'s to $101$'s. This implies that any occurrences of $110$ 
should be disallowed by the transfer matrix while counting the total number of fragments. After imposing this condition, the transfer matrix is found to be
\bea T_1 =
\begin{bmatrix}
1 & 0 & 0 & 0 \\
0 & 0 & 1 & 1 \\
1 & 1 & 0 & 0 \\
0 & 0 & 1 & 1
\end{bmatrix},
\label{trans1} \eea
and the matrix $(T_1)^{L-2}$ leads to a 
construction of the root states from right to left for 
a system with $L$ sites.
The eigenvalues of $T_1$ are given by $\tau = (\sqrt{5}+1)/2 \simeq
1.618$, $-1/\tau \simeq -0.618$, 1 and 0. The total number of fragments, therefore, 
grows exponentially with the system size as $1.618^L$. We have confirmed this growth numerically as shown in Fig. \eqref{fig1}.

\section{Counting of frozen eigenstates in the East model}
\label{appB}

We will now count the total number of frozen eigenstates in the East model using a transfer matrix method~\cite{gen,aditya_2024,menon_1995,menon_1997,barma_1994}. Following a similar procedure as in Appendix \ref{appA}, we construct another $4\times4$ transfer matrix, $T_2 (c_{i},c_{j})$. However, the rule for constructing $T_2$ 
is different from the rule for $T_1$. This is because for frozen eigenstates,
both $110$'s and $110$'s should be disallowed by the transfer matrix.
Upon imposing this rule, the transfer matrix is found to be
\begin{comment}
hence they do not evolve at all under time evolution. Examples of such states are
$1 1 1$ and $0 0 0$. We want to know how the number of frozen state grows with the system size $L$. This can be found using another transfer matrix $T'$. Keeping
in mind the East model constraint that only $110 \leftrightarrow 101$ is allowed, 
the appropriate transfer matrix for frozen states is given, in the basis $\ket{11},
\ket{10}, \ket{01}, \ket{00}$, by
\end{comment}
\bea
T_2 =
\begin{bmatrix}
1 & 0 & 0 & 0 \\
0 & 0 & 0 & 1 \\
1 & 1 & 0 & 0 \\
0 & 0 & 1 & 1
\end{bmatrix}.
\label{trans2}
\eea
The eigenvalues of Eq. \eqref{trans2} are given by $1.466, -0.2328 \pm 0.7926i$ and $1$. The number
of frozen states, $N_{froz}(L)$ in our model, thus grows with system size as $1.466^L$ in the large-$L$ limit. For the first few values of $L$, $N_{froz}$ can also be 
computed by using the relation $N_{froz}(L) = \sum_{i,j} X_{i,j}$, where 
$X = T_2^{L-2}$; these values are shown in Table \ref{Tablefrzo}. We also have
$N_{froz}(L)=1,2,4$ for $L=0,1$, and $2$, respectively. We note that $N_{froz}$ 
given in Table \ref{Tablefrzo} for OBC follows the recursion relation $N_{froz}(L) = N_{froz}(L-1) + N_{froz}(L-3) + 1$ for $L \ge 3$. 

It is also possible to calculate the number of frozen eigenstates for a system
with PBC after incorporating the additional constraint that the four consecutive sites $(L-1,L,1,2)$ should not contain any occurrences of $110$'s and $101$'s. Imposing this rule, $N_{froz}$ with PBC is given by 
\bea N_{froz}&\,=\,&X(1,1)+X(1,3)+X(1,4)+X(2,4)\non\\
&&+ ~X(3,2)+X(3,4)+X(4,2)+X(4,3)\non\\
&&+ ~X(4,4), \eea
where $X=T_2^{L-2}$. The analytically obtained values of $N_{froz}$ for PBC for the first few values of $L$ are shown in Table \ref{Tablefrzo}; these agree with the results obtained
by numerical enumeration.

\begin{table}[h!]
\centering
\begin{tabular}{|c|c|c|c|c|c|c|c|c|c|c|c|}
\hline
$L$  & 3 & 4 & 5 & 6 & 7 & 8 & 9 & 10 \\
\hline
\(N_{froz}^{OBC}\) & 6 & 9 & 14 & 21 & 31 & 46 & 68 & 100 \\
\hline
\(N_{froz}^{PBC}\) & 5 & 6 & 7 & 11 & 16 & 22 & 32 & 47  \\
\hline
\end{tabular}
\caption{$N_{froz}$ versus $L$ with OBC and PBC as obtained analytically. These
agree with our numerical results.}
\label{Tablefrzo}
\end{table}

\begin{comment}

\begin{table}[h]
    \centering
    \begin{tabular}{|c|c|c|c|c|c|c|c|c|c|}
        \hline
        $L$ & 3 & 4 & 5 & 6 & 7 & 8 & 9 & 10 \\
        \hline
       
        \hline
    \end{tabular}
    \caption{$N_{frzo}$ versus $L$ with PBC obtained using the transfer matrix method, which is in agreement with our numerically obtained results.}
    \label{tab:Number of frozen states using PBC}
\end{table}

\begin{figure}[H]
    \centering
    \includegraphics[width=0.6\textwidth]{number_frozen_state_with_system_size_L_East_model_with_pbc.pdf}
    \caption{Plot of the number of frozen states growth with the system size $L$ for PBC. From the fitted curve we see the growth is $1.463^L$ which is close to the OBC result.}
 \label{fig3}
\end{figure}
\end{comment}

\section{Recursive quantum scar states with zero entanglement at a non-central cut
in the East model}
\label{appC}

\begin{figure}
\centering
\includegraphics[width=0.8\hsize]{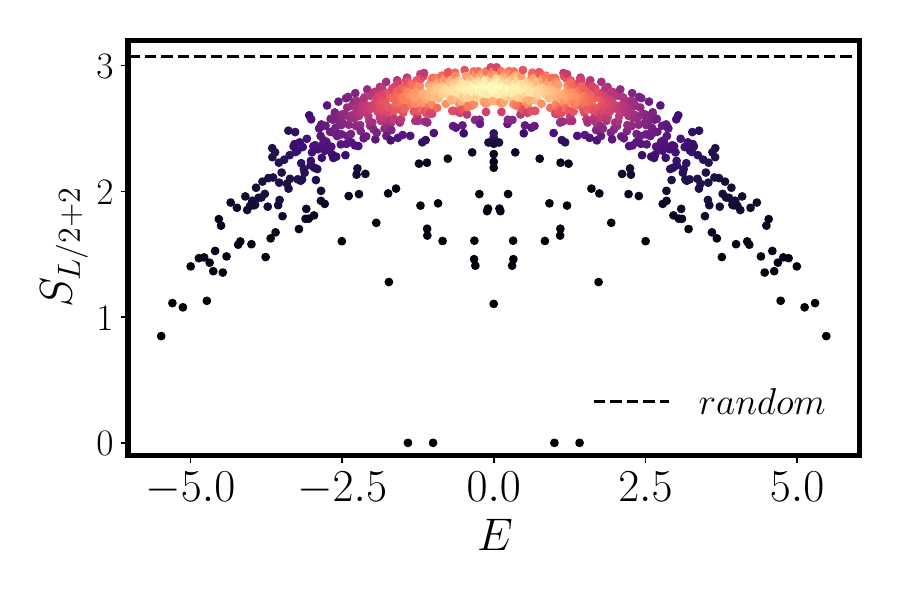}%
\caption{Plot showing the entanglement spectrum versus $E$ for the fragment generated from the root state $10101010101011$ for $L=14$ and $N_{f}=L/2+1$; the entanglement cut is made at $l=9$ instead of at the center. We see that there are exactly four zero entanglement states in the middle of the spectrum.
%originating due to recursive quantum HSF within this fragment. 
These states do not have zero entanglement if the cut is made at the center of
the system.}
\label{QHSFent}
\end{figure}

\begin{figure}
\centering
\subfigure[]{\includegraphics[width=0.5\hsize]{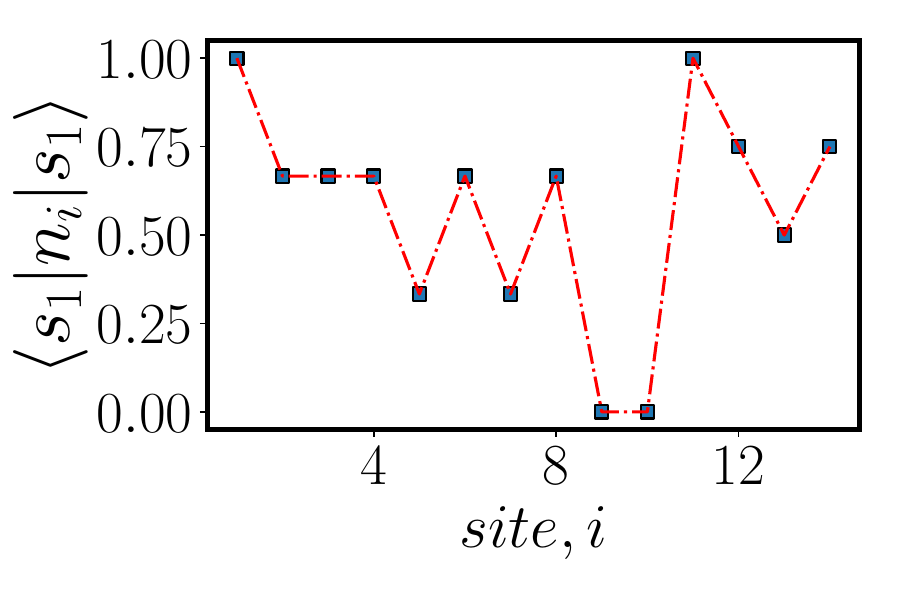}}%
\subfigure[]{\includegraphics[width=0.5\hsize]{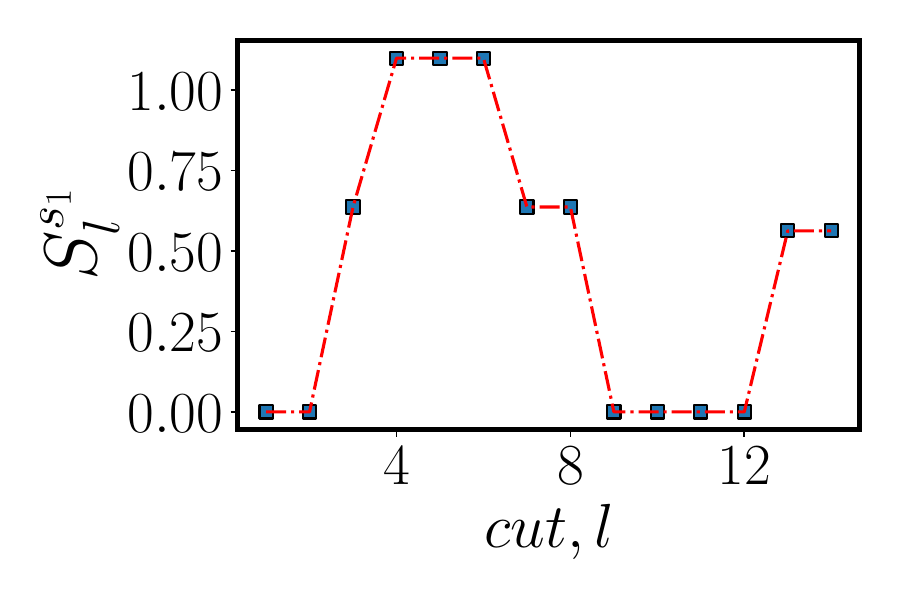}}\\
\subfigure[]{\includegraphics[width=0.5\hsize]{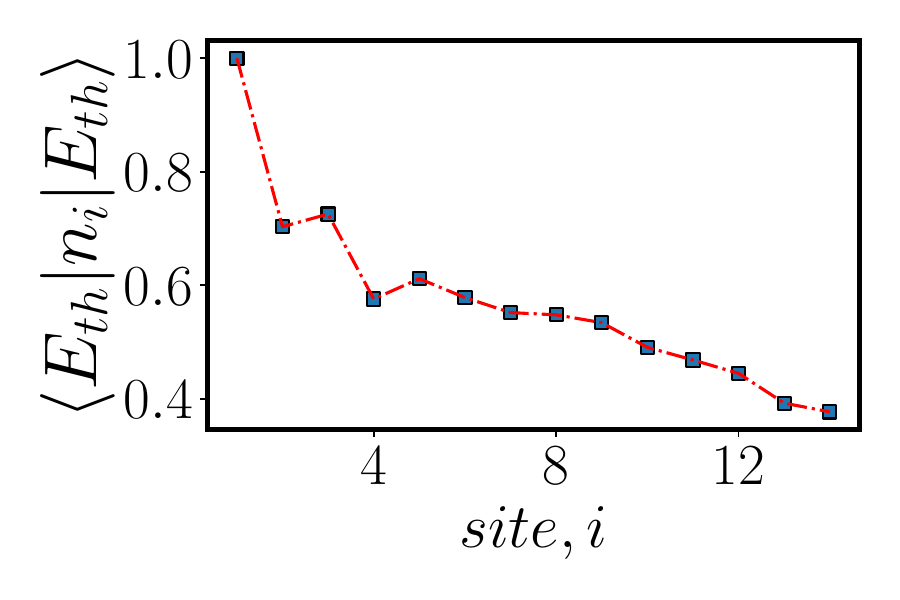}}%
\subfigure[]{\includegraphics[width=0.5\hsize]{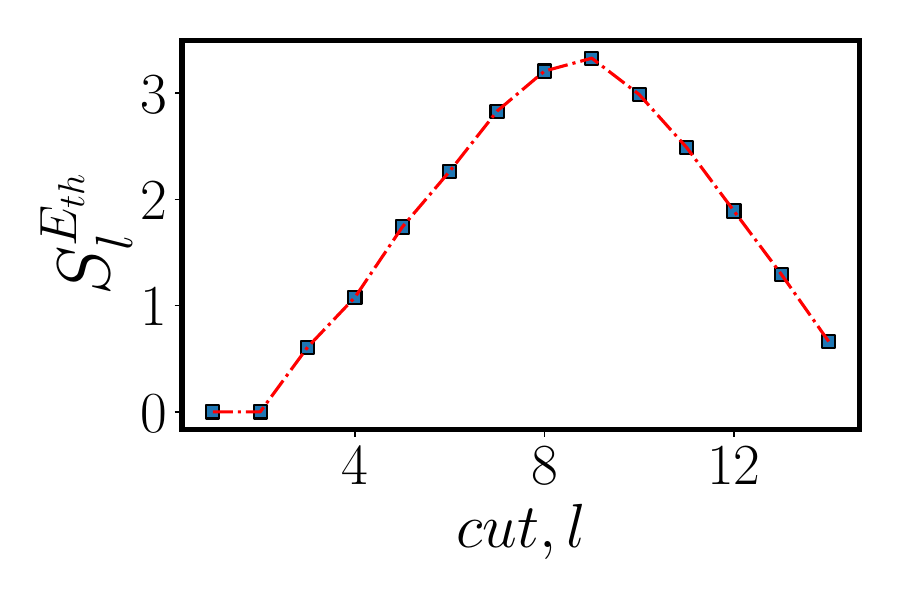}}%
\caption{(a) and (c): Variations of the expectation values of $n_{i}$ versus the site number $i$ in a scar state $\ket{s_1}$ and in a thermal state $\ket{E_{th}}$, both
lying in fragment originating from the root state $10101010101011$ for $L=14$ and $N_{f}=L/2+1$. (b) and (d): Variation of the entanglement entropy $S_{l}$ versus $l$ for $\ket{s_{1}}$ and $\ket{E_{th}}$, respectively, for the same fragment as in plots (a) and (c). The density profile in plot (a) confirms that $\ket{s_{1}}$ has a product state form at sites $9$ and $10$. Further, the leftmost site of $\ket{\psi_{R}}$ is frozen at $n_{j}=1$. These observations imply that the system will have zero entanglement for 
multiple locations of cuts, as we see in plot (b). On the other hand, a thermal state within this fragment shown in (c) demonstrates a distinct behavior since no part of the eigenstate has a product form. We observe that the minimum of the entanglement for $\ket{s_{1}}$ and the maximum of the entanglement for $\ket{E_{th}}$ appear when 
the cut is made at $l=9$; these features are exactly opposite of each other. }
\label{QHSFobs}
\end{figure}

We will now show that in some fragments, the recursive quantum scar states~\cite{brighi_2023} can have zero entanglement at a cut which is not at the
center of the system. In do so, we investigate a classical fragment originating from a root state $1010101011$ for $L=14$ and $N_{f}=L/2+1$. Our analysis shows that this fragment contains some exact quantum scar states. However, these states have zero entanglement, not at a central cut, but rather at a cut at $l=L/2+2$, as shown in Fig. \ref{QHSFent}. We have found four such zero-entangled states after following a similar method as discussed in the case of the largest fragment. We first note that there are two scar states with the entangled left-restricted part, 
\beq \ket{\phi_{L}'} ~=~ \frac{1}{\sqrt{3}} (\ket{11100110}-\ket{11010101}+\ket{10111001}), \eeq
while for the remaining two scar states, we have
\bea \ket{\phi_{L}'} &=& \frac{1}{2} (\ket{101111100}-\ket{110111010} \non \\
&& ~~~~- \ket{111100101} +\ket{111010110}). \eea
As in the previous case, $\ket{\phi_{R}}$ remains dynamically separated from $\ket{\phi_{L}'}$ by a $\ket{00}$. For the first two scar states, the Hamiltonian within the right-restricted part describes a tight-binding model with $L=4$ and $N_{f}=3$ with OBC. For the other two scare states, we get similar effective model with $L=3$ and $N_{f}=2$. In both cases, a single hole hops in the background of several particles, and
the leftmost site, denoted $l$, of the right-restricted part remains frozen 
with $n_{l}=1$. Hence, the single-particle energies in the first case are $E_{p}=-2\cos(\pi p/4)$ with $p=1,2,3$, and in the second case, $E_{p}=-2\cos(\pi 3/ p)$ with $p=1,2$; these are equal to $\pm 1.414, 0$ in the first case, and $\pm 1$ in the
second case. In Fig. \ref{QHSFobs} (a), we show the density profile of a scar state versus the site number within this fragment.
%as exhibited in Fig. \ref{QHSFent}. 
As Fig. \ref{QHSFobs} (a) shows, the scar state has the form of a product state in the number basis ($\ket{00}$) on the 9-th and 10-th sites; this gives zero entanglement 
if the system is cut at $l=9$ or $10$, as illustrated in Fig. \ref{QHSFobs} (b). This state also has $S_{l}=0$ for a cute at the 11-th cut since the leftmost site of the right-restricted part, $\ket{\phi_{R}}$, is frozen at $n_{i=11}=1$. Next, we compare the scar states with a thermal state; where the density profile as a function of the
site $i$ and the entanglement $S_{l}$ versus the location $l$ of the cute of a thermal state are shown in Fig. \ref{QHSFobs} (c-d), respectively. In the latter case, we notice that the density profile continuously decreases with increasing site number. However, $S_{l}$ first increases with $l$, reaches a maximum at $l=9$, and then decreases with $l$, as we see in Fig. \ref{QHSFobs} (d); this cannot be explained in a simple way
by the density profile illustrated in Fig. \ref{QHSFobs} (c).

\section{Counting the total number of fragments in the PXXP$+$PYYP model}
\label{appD}

\begin{figure}[h!]
\centering
\includegraphics[width=0.85\columnwidth]{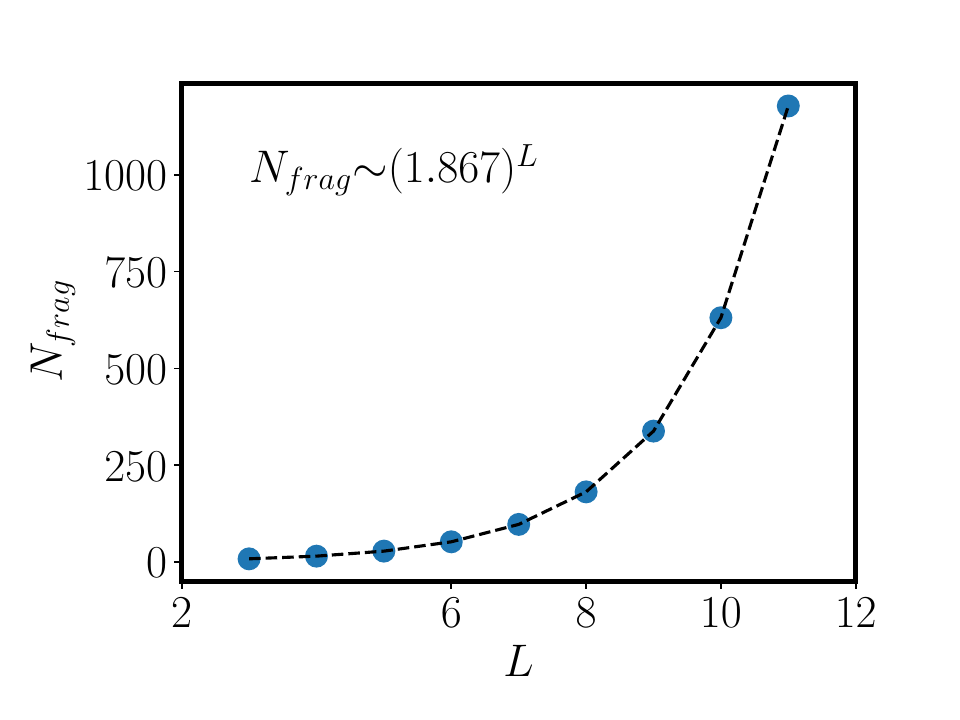}
\caption{Plot showing $N_{frag}$ versus $L$ for a system with OBC. The numerical fitting shows that $N_{frag}$ grows as $1.867^L$, which agrees with our analytically obtained result.} \label{fig13} \end{figure}

We will now discuss how to compute the total number of fragments $N_{frag}(L)$ in the PXXP$+$PYYP model using a transfer matrix method~\cite{gen,aditya_2024,menon_1997,menon_1995,barma_1994}.
Since this model allows transitions involving four consecutive sites, we need to 
construct a $8\times 8$ transfer matrix, represented by $T_3 (c_{i},c_{j})$ for our counting, $c_{i}$ denotes the configurations $111$, $110$, $101$, $011$, $100$, $010$, $001$ and $000$ on three consecutive sites. Next, the root identification in this model requires successive removals of all occurrences of $1101$'s to uniquely define the fragments. Hence the transfer matrix should 
not allow the configuration $1101$ to appear~\cite{gen,aditya_2024,menon_1997,menon_1995,barma_1994}. We then find that the transfer matrix must have the form
\begin{comment}Accordingly, 
To see how the number of fragments grows with the system size $L$ we can again use a transfer matrix method. For this four-site model, we need a $8 \times 8$ transfer matrix. We construct this by taking the basis of states $\ket{111}, \ket{110}, \ket{101}, \ket{011}, \ket{100}, \ket{010}, \ket{001},  \ket{000}$ to define the rows and columns. Now the crucial point to note is that we will not allow the configuration $1101$
to occur anywhere in a product of transfer matrices. We then have
\end{comment}
\bea T_3 =
\begin{bmatrix}
1 & 1 & 0 & 0 & 0 & 0 & 0 & 0 \\
0 & 0 & 0 & 0 & 1 & 0 & 0 & 0 \\
0 & 0 & 0 & 1 & 0 & 1 & 0 & 0 \\
1 & 1 & 0 & 0 & 0 & 0 & 0 & 0 \\
0 & 0 & 0 & 0 & 0 & 0 & 1 & 1 \\
0 & 0 & 1 & 0 & 1 & 0 & 0 & 0 \\
0 & 0 & 0 & 1 & 0 & 1 & 0 & 0 \\
0 & 0 & 0 & 0 & 0 & 0 & 1 & 1 \\
\end{bmatrix}.
\label{tranPXXP}
\eea
The eigenvalues of $T_3$ are given by $1.867$, $-0.867$, $0.5 \pm 0.6067i$, and zero 
appearing four times. The total number of fragments, $N_{frag}$ therefore grows as $1.867^{L}$ for large $L$. This agrees with our numerical results; see Fig.~\ref{fig13}. We see that $N_{frag}$ grows much faster in this model than in the East model. This can be understood from the fact the the infinite-$V$ limit 
disallows some processes which the East model allows. Hence the degree of fragmentation in this model is larger in the infinite-$V$ model.

\begin{comment}
Hence, as the system size $L$ becomes large,
the total number of fragments for this model grows exponentially as $1.8667^L$. A very important observation following is that this model is more constrained than the East model; hence the degree of fragmentation is larger, and the number of fragments grows faster here. 
We have also checked this growth of the total number of fragments numerically and
we find that the result matches the one found analytically (see Fig. \ref{fig13}).
\end{comment}

\section{Counting of frozen eigenstates in the PXXP$+$PYYP model}
\label{appE}

We will now compute the number of frozen fragments~\cite{rev1} in this model using the transfer matrix method~\cite{gen,aditya_2024,barma_1994,menon_1997,menon_1995}. We will discuss this calculation for a system with OBC.
%, which is relatively easier compared to a system with PBC. 
The counting procedure again requires us to construct a $8\times 8$ transfer matrix, similar to the construction shown in the calculation of the total number of fragments. However, the imposed rules to be imposed here are different from the previous one, namely, any occurrences of $1101$ and $1011$ must be disallowed
as there can be transitions between these two; frozen states are dynamically inactive
by definition. This rule leads to a transfer matrix $T_4 (c_i,c_{j})$ of the form
\bea T_4 =
\begin{bmatrix}
1 & 1 & 0 & 0 & 0 & 0 & 0 & 0 \\
0 & 0 & 0 & 0 & 1 & 0 & 0 & 0 \\
0 & 0 & 0 & 0 & 0 & 1 & 0 & 0 \\
1 & 1 & 0 & 0 & 0 & 0 & 0 & 0 \\
0 & 0 & 0 & 0 & 0 & 0 & 1 & 1 \\
0 & 0 & 1 & 0 & 1 & 0 & 0 & 0 \\
0 & 0 & 0 & 1 & 0 & 1 & 0 & 0 \\
0 & 0 & 0 & 0 & 0 & 0 & 1 & 1 \\
\end{bmatrix},\label{transfrozPXXP}
\eea
where $c_{i}$ denotes the configurations $111$, $110$, $101$, $011$, $100$, $010$, $001$ and $000$ on three consecutive sites, as before.
The eigenvalues of $T_4$ are given by $1.812$, $1$, $0.344 \pm \, 0.864i$, $-0.750 \pm 0.277i$, and zero appearing twice.. Hence, the total number of frozen states $N_{froz}$ grows as $1.8124^{L}$ for large $L$. $N_{froz}$ can be analytically computed by using
the fact that $ N_{froz} = \sum_{i,j = 1}^8 Y_{i j}$, where $Y=T_4^{L-3}$, as shown 
in Table \ref{tablefrozPXXP} for a few values of $L$. We have verified that the analytical results agree with the results obtained by
numerical enumeration. In addition, we have $N_{froz}= 1, 2, 4$ for $L = 0, 1$ and $2$, respectively. We also note that $N_{froz}$ follows a different recursion relation than the East model, which can be seen from Table \ref{tablefrozPXXP}, as
\bea N_{froz}( L ) &=& N_{froz}( L-1 ) + N_{froz}( L -2 )\non\\&& +N_{froz}( L-4 ) +
N_{froz}( L-5 ), \eea
for $L\geq 5$.

\begin{table}[h]
\centering
\begin{tabular}{|c|c|c|c|c|c|c|c|c|c|c|c|c|c|c|c|c|}
\hline
$L$ & 3 & 4 & 5 & 6 & 7 & 8 & 9 & 10 & 11 \\
\hline
\(N_{froz}\) & 8 & 14 & 25 & 45 & 82 & 149 & 270 & 489 & 886  \\
\hline
\end{tabular}
\caption{$N_{froz}$ versus $L$ as obtained analytically.} 
\label{tablefrozPXXP}
\end{table}

\section{Computation of filling fraction for the ground state of the PXXP$+$PYYP model}
\label{appF}

We will now compute the filling fraction at which the ground state of the PXXP$+$PYYP
model appears. We will do this by using the mapping to a non-interacting tight-binding model. Using the single-particle dispersion given in Eq. \eqref{en}, the ground state energy for a specific fragment can be shown to be
\bea
    E_{GS} = - 2 ~\sum_{n = 1}^{N_B -1} \cos \left(\frac{\pi n}{N_A + N_B}\right).\label{Egs}
\eea
Since we are interested in the ground state energy in the thermodynamic limit, i.e., $N_{A},N_{B}\to\infty$, we can convert the summation in Eq. \eqref{Egs} to an 
integration after defining a variable $x=n/(N_{A}+N_{B})$. This yields
\bea
E_{GS} &=& - 2 (N_A + N_B) \int_0^\frac{N_B}{N_A + N_B} dx ~ \cos{(\pi x)}\non\\
& =& - \frac{2}{\pi} (N_A + N_B) \,\, \sin \left( \frac{\pi N_B}{N_A + N_B}\right).\label{simEgs}
\eea

\begin{comment}
\begin{equation}
    E_{g} = -2 (N_A + N_B) \sum_{n = 1}^{N_B } \frac{1}{N_A + N_B}  \cos{(\frac{\pi n}{N_A + N_B})}.
\end{equation}
Let us define $x = \frac{n}{(N_A + N_B)}$, so that $\Delta n = 1$ and $\Delta x = \frac{1}{(N_A + N_B)}$.
\bea
        E_{GS} &=& - 2 (N_A + N_B) \int_0^\frac{N_B}{N_A + N_B} dx \cos{(\pi x)}\non\\
        & \,\,\,\,\, = - \frac{2}{\pi} (N_A + N_B) \,\, \sin{(\pi \frac{N_B}{(N_A + N_B)})}\\
        & \,\,\,\,\,\, =  - \frac{2}{\pi} L \rho \,\, \sin{(\frac{L ( 2 \rho - 1) \pi}{\rho L})}\\
        & \,\,\,\,\,\, = - \frac{2}{\pi} L \rho \,\, \sin{(2 \pi - \frac{\pi}{\rho})}\\
        & \,\,\,\,\,\, =  \frac{2}{\pi} L \rho \,\, \sin{( \frac{\pi}{\rho})}.
   \eea
\end{comment}

We now note that the PXXP$+$PYYP model with a length $L=2N_{A}+N_{B}$ and a filling 
fraction $\rho=(N_{A}+N_{B})/L$ reduces to a tight-binding model with a length $L'=N_{A}+N_{B}$ and a filling fraction $\rho'=N_{B}/L'$. These relations imply
that $\frac{1}{2}\leq\rho\leq 1$ in the original model, and $0\leq\rho'\leq 1$ in the mapped model. Hence, rewriting $N_{A}$ and $N_{B}$ in terms of $L$ and $\rho$, Eq. \eqref{simEgs} can be recast as
\bea E_{GS}=\frac{2L\rho}{\pi} \, \sin\left( \frac{\pi}{\rho}\right). \eea
We can now define a variable $y=\pi/\rho$, which lies between $\pi$ and $2\pi$; we 
then have $E_{GS} = \frac{2L}{y}\sin y$. Extremizing $E_{GS}$ with respect to $y$, we numerically find that the value of $y$ at which $E_{GS}$ has its minimum is $y_{0}= 4.493$. Hence, the filling fraction at which the ground state lies in the 
PXXP$+$PYYP model is $\rho \simeq \pi/4.493 =0.699$, and the ground state 
energy at this filling fraction is $E_{GS} \simeq 0.435 ~L$.

\section{Robustness of quantum scars under various perturbations}
\label{appG}

\begin{figure*}
\subfigure[]{\includegraphics[width=0.33\linewidth]{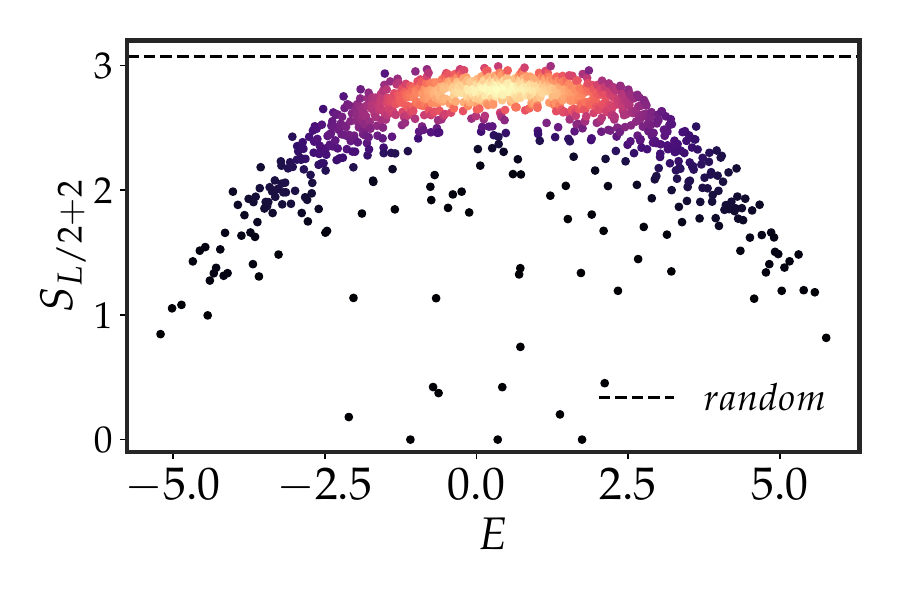}}%
\subfigure[]{\includegraphics[width=0.33\linewidth]{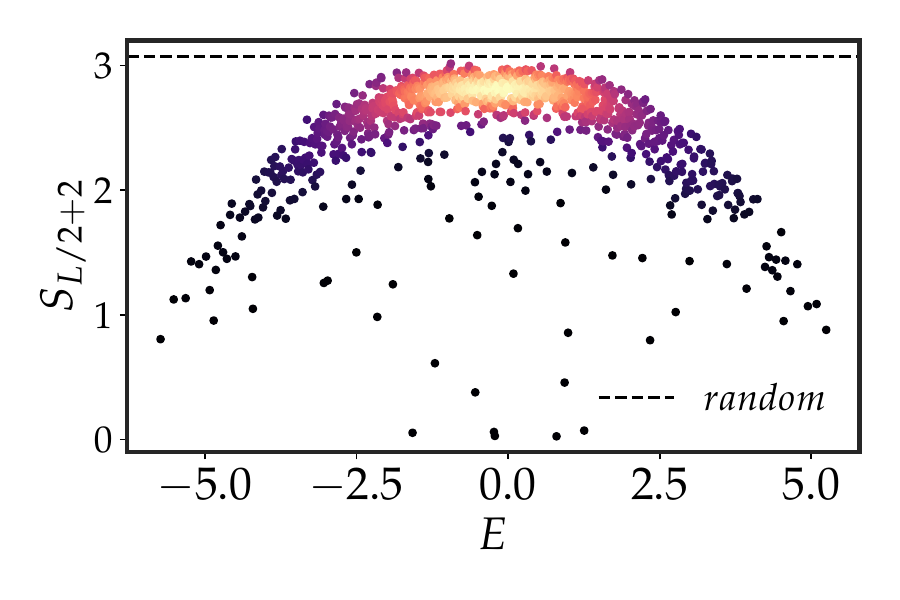}}%
\subfigure[]{\includegraphics[width=0.33\linewidth]{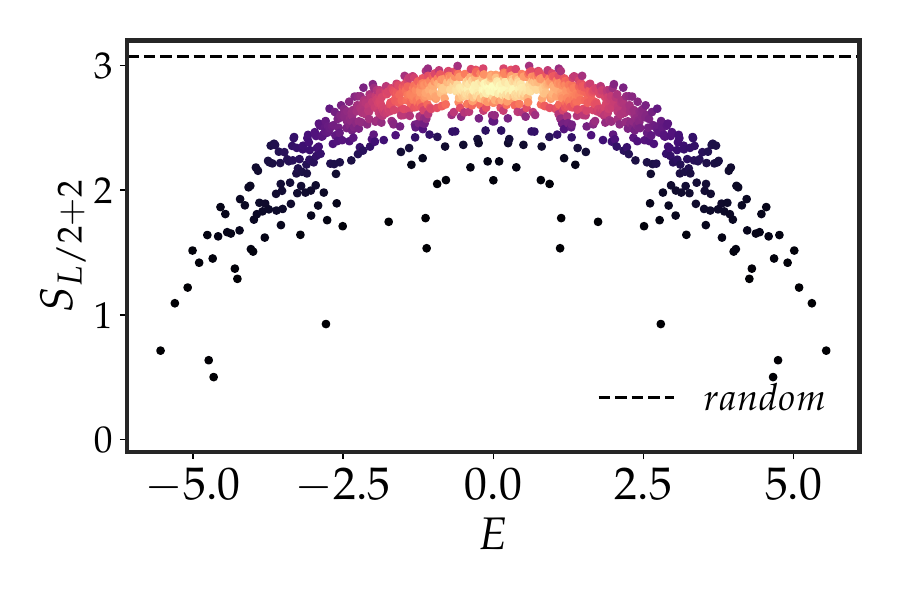}}\\
\subfigure[]{\includegraphics[width=0.3\linewidth]{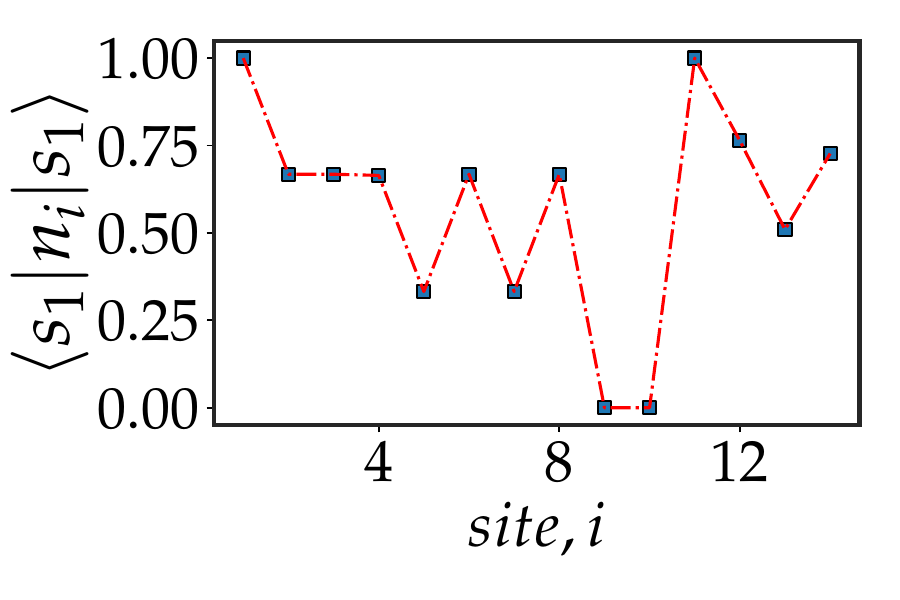}} \hspace*{.6cm}
\subfigure[]{\includegraphics[width=0.3\linewidth]{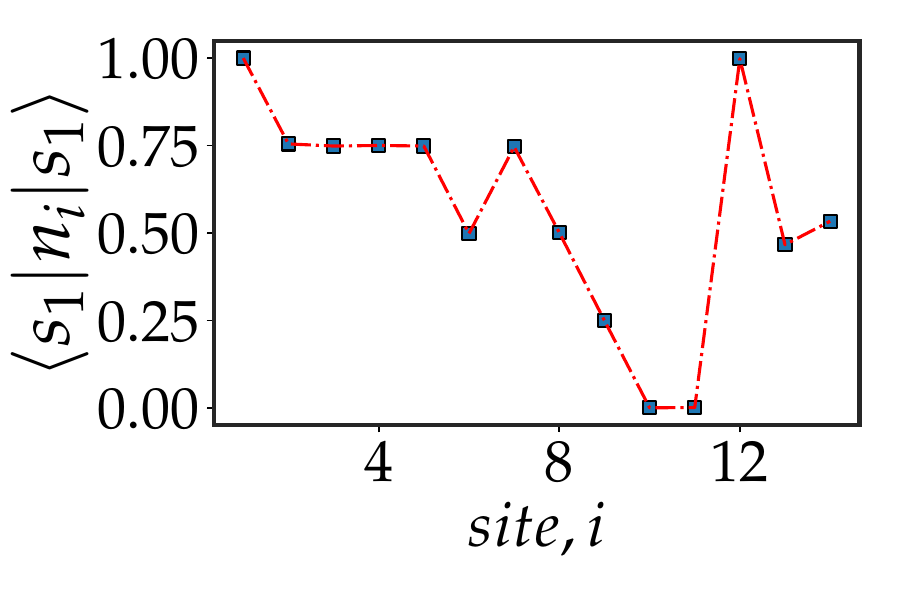}}%
\caption{(a-c) Plots showing the entanglement spectrum, $S_{L/2+2}$ versus $E$ within the largest fragment originating from the root state $10101010101011$ for $L=14$ and $N_{f}=L/2+1$, in the presence of three kinds of perturbations: (a) a nearest-neighbor density-density interaction $V=0.1$, (b) a random on-site potential $\mu_{j}$ taken from a uniform distribution $[-0.1,0.1]$ with zero mean, and (c) a random hopping of the
form $1+\delta J_{j}$, where $\delta J_{j}$ are random numbers taken from the uniform distribution $[-0.1,0.1]$ with zero mean, respectively. Plot (a): we observe that having $V=0.1$ reduces the number of zero-entanglement states to three from
the four states found for the East model ($V=0$) as shown in Fig.~\ref{QHSFent}. 
However, the density profile for one of the scar states shown in plot (d) turns out to be similar to one of the scar states of the East model as shown in Fig. \ref{QHSFobs} (a). Plot
(b): the entanglement of the scar states is not exactly zero when there is a random
on-site potential. Namely, we find that the value of $S_{L/2+2}$ is of order $10^{-2}$, 
unlike the states with exactly zero entanglement (numerically of order $10^{-15}$)
found for the East model. Further, these scar states do not have a perfect $\ket{001}$ form on the 9-th, 10-th and 11-th sites, as shown in plot (e).
%; however, they have $\langle n_{j}\rangle$ very close to zero at these two sites. 
Plot (c): The scar states quickly vanish when a small random hopping is introduced.}
\label{perturb} \end{figure*}

We will now examine how robust the quantum many-body scars~\cite{brighi_2023} are in the presence of various perturbations, specifically, a nearest-neighbor density-density interaction with strength $V$, a random on-site potential taken from a uniformly distributed random numbers $\mu_{j}$ with zero mean, and a random hopping $\delta J_{j}$ on top of a uniform hopping $J=1$ which is also taken from a uniform random distribution with zero
mean. We note that all three perturbations keep the fragmentation structure of the East model unchanged. In Figs. \ref{perturb} (a-c), we show the entanglement entropy versus the energy within the classical fragment generated from the root state $10101010101011$ for $L=14$ and $N_{f}=L/2+1$ for a cut at the site $L/2 + 2$, in the presence of (a) $V=0.1$, (b) $\mu_{j}=0.1$, and  (c) $\delta J_{j}=0.1$, respectively. 
% To analyze the stability of the scar states, we demonstrate the entanglement spectrum $S_{L/2+2}$ versus $E$ within this fragment. 
In Fig. \ref{perturb} (a), with a small value of $V$, we see that the number of zero-entanglement states reduces to three from four for the unperturbed East model, 
% which has been observed in the absence of any other perturbations, 
as shown in Fig. \ref{QHSFent}. We show the density profile of one of these scar states in Fig. \ref{perturb} (d). We find that this is very similar to what is seen in the
East model in Fig.~\ref{QHSFobs}. For example, we see a state of the form $\ket{001}$ 
on the 9-th, 10-th and 11-th sites in both cases. In Fig. \ref{perturb} (b), with a
small random potential, we see that there are some scar states with small but non-zero entanglements. Specifically, we 
find that the entanglements are of order $10^{-2}$, in contrast to the four
scar states with exactly zero-entanglement (of order $10^{-15}$) found in the unperturbed East model; see Fig.~\ref{QHSFent}.
The density profile of one of these scar states also looks a little different
from the scar states of the East model. For instance, the state does not
exactly have a $\ket{001}$ form at the 9-th, 10-th and 11-the sites as we 
see in Fig. \ref{perturb} (e).
%; this further causes deviation in entanglement from the exact zero value. 
%Also, in both of these cases, the energy spectrum does not appear in $E$ and $-E$ pairs as these perturbations remain invariant under the sublattice symmetry transformation, whereas the unperturbed part transforms as $H_{unper}\rightarrow -H_{unper}$. 
In Fig. \ref{perturb} (c), we observe that the scar states completely vanish when a
small random hopping $\delta J_{j}$ is introduced; this is a very distinct behavior 
from the first two kinds of perturbations.
%One should note that this term transforms in a similar manner to the unperturbed case under the action of the sublattice symmetry operator.

\end{document}